\documentclass[12pt,prd,superscriptaddress,nofootinbib,tightenlines,preprintnumbers]{revtex4}
\pdfoutput=1
\usepackage[english]{babel}
\usepackage{slashed}
\usepackage{amsmath}
\usepackage[]{units}
\usepackage{multirow}
\usepackage{color}
\usepackage{enumitem}
\usepackage{hyperref}
\usepackage[all]{hypcap}
\usepackage{array}
\usepackage{float}
\usepackage{graphicx} 
\usepackage{amsfonts}
\usepackage{placeins}

\usepackage[caption=false]{subfig}

\usepackage[dvipsnames]{xcolor}

\newenvironment{psmallmatrix}
  {\left(\begin{smallmatrix}}
  {\end{smallmatrix}\right)}

\newcommand{\DM}{\text{DM}}
\newcommand{\dd}{\mathrm{d}}
\newcommand{\FF}{F^m}
\newcommand{\SSS}{S^r}

\begin{document}
\preprint{ULB-TH/17-13}

\title{Prospects for discovering a neutrino line induced\\ by dark matter annihilation}

\author{Chaimae El Aisati}
\email{Chaimae.El.Aisati@ulb.ac.be}
\author{Camilo Garcia-Cely}
\email{Camilo.Alfredo.Garcia.Cely@ulb.ac.be}
\author{Thomas Hambye}
\email{thambye@ulb.ac.be}
\author{Laurent Vanderheyden}
\email{laurent.vanderheyden@ulb.ac.be}

\affiliation{Service de Physique Th\'eorique - Universit\'e Libre de Bruxelles, Boulevard du Triomphe, CP225, 1050 Brussels, Belgium}


\begin{abstract}

In the near future, neutrino telescopes are expected to improve their sensitivity to the flux of monochromatic neutrinos produced by dark matter (DM) in our galaxy. 
This is illustrated by a new limit   on the corresponding cross section that we derive from public IceCube data. 
In this context,  we study which DM models  could produce an observable flux of monochromatic neutrinos from DM annihilations.
To this end, we proceed in two steps.
First, within a set of simple and minimal assumptions  concerning the properties of the DM particle, we determine the  models that could give rise to a significant annihilation into monochromatic neutrinos at the freeze-out epoch. 
The list of models turns out to be very limited as a result of various cons\-traints, in parti\-cular direct detection and neutrino masses at loop level. 
Given the fact that, even if largely improved, the sensitivities will be far from reaching  the thermal annihilation cross section soon, a signal could only be observed if the annihilation into neutrinos today is boosted with respect to the freeze-out epoch. This is why, in a second step, we analyze  the possibility of having such a large enhancement from the Sommerfeld effect. 
For each scenario, we also compute the cross sections into other annihilation products and confront our results with experimental constraints. 
We find that, within our simple and minimal assumptions, the expectation to observe  monochromatic neutrinos is only possible in very specific scenarios. 
Some  will be  confirmed or excluded in the near future because they predict signals slightly below the current experimental sensitivities. 
We also  discuss how these  prospects change by relaxing our assumptions as well as by considering  other types of sharp spectral features. For the latter, we consider   boxed-shaped and bremsstrahlung spectra and provide the corresponding limits from IceCube data. 

\end{abstract}

\maketitle

\newpage
\tableofcontents
\newpage

\section{Introduction}

In indirect searches of dark matter (DM) with photons or neutrinos from the galactic center, the experimental sensitivities to monochromatic signals---or more generally to sharp spectral features---are better than the ones to spectra characterized by a continuum. This is mostly due to the fact that sharp features can better be discriminated from the smooth background.
The most recent searches for photon signals of this kind illustrate well this statement. For example, for a 100 GeV DM candidate in the Milky Way whose distribution is given by the Einasto profile, the Fermi-LAT collaboration reports an upper bound on the annihilation cross section into two photons given by
$\langle \sigma v\rangle_{\DM \DM \rightarrow \gamma \gamma} \lesssim  \unit[2\times10^{-28}]{cm^3/s}$~\cite{Ackermann:2015lka}. In contrast, the search for the continuum of photons produced by the charged primary particles arising from 100 GeV DM annihilations in our galaxy reports upper bounds about four orders of magnitude weaker: e.g., $\langle \sigma v\rangle_{ \DM \DM \rightarrow W W} \lesssim 2\times \unit[10^{-24}]{cm^3/s}  $ and $\langle \sigma v\rangle_{ \DM \DM \rightarrow b \bar{b}} \lesssim \unit[10^{-24}]{cm^3/s} $~\cite{Ackermann:2012qk}.
Similarly, for a 10 TeV DM candidate, 
the HESS collaboration reports $\langle \sigma v\rangle_{ \DM \DM \rightarrow \gamma \gamma} \lesssim \unit[10^{-26}]{cm^3/s} $~\cite{Abramowski:2013ax}, whereas $\langle \sigma v\rangle_{ \DM \DM \rightarrow WW} \lesssim  \unit[2\times10^{-25}]{cm^3/s}  $ and $\langle \sigma v\rangle_{ \DM \DM \rightarrow b \bar{b}} \lesssim  \unit[10^{-25}]{cm^3/s}$~\cite{Abdallah:2016ygi}. The same situation holds  if the photons arise from the decay of a DM particle. Even if the sensitivity is better for monochromatic photons, this does not mean that the search for continuum signals should not be pursued, since in many models the cross section into monochromatic photons is suppressed with respect to the production of other particles.

For neutrinos the situation is similar, up to a few differences which crucially depend on whether one considers decaying or annihilating DM. For decays, and for  DM masses above a few TeV and up to $\sim100$~TeV, the sensitivities on the DM decay width into monochromatic neutrinos are close to those on the decay width into monochromatic photons~\cite{Aisati:2015vma}, rendering neutrino line searches especially interesting and competitive. At higher masses, there are no interesting bounds on DM decay into photons---in contrast to those provided by neutrino telescopes up to $\unit[10^5]{TeV}$---and below few TeV, the neutrino sensitivity quickly becomes orders of magnitude weaker than for photons.
For DM annihilations the situation is different, as the sensitivities reported for photons are orders of magnitude better than for neutrinos, unless the DM mass $m_{\DM}$ lies above 70~TeV.
For instance, for $m_{\DM}=100$~GeV (10 TeV), the current upper bounds from galactic DM searches by IceCube~\cite{Aartsen:2017ulx, Aartsen:2016pfc} give $\langle \sigma v\rangle_{\DM \DM \rightarrow \nu \bar{\nu}}\lesssim \unit[0.5\times10^{-23}]{cm^3/s}$ ($\lesssim \unit[4\times10^{-23}]{cm^3/s}$) (assuming a NFW DM profile). These limits are orders of magnitude weaker than those given above for monochromatic photons.
Also, neutrino telescopes currently only probe cross sections much larger than the thermal value $\langle \sigma v \rangle_{fo}\sim \unit[3\times10^{-26}]{cm^3/s}$. 
This  does not  mean that one could not see a neutrino-line signal in the near future, but it does mean that one must look into specific models where the production of monochromatic photons is suppressed with respect to neutrinos. Moreover, the fact that the current  sensitivity is far from reaching the thermal value does not mean that we could not see a signal but it means that we need to boost the production of neutrinos today with respect to the freeze-out epoch.

One must stress that this discussion crucially depends on the improvement in sensitivity that neutrino telescopes will be able to attain in the near future. As discussed in Section \ref{sec:limits}, we can expect that these improvements will be important, even if still far from for reaching the thermal value. 
This of course highly motivates the analysis performed below.

In this context, the goal of this work is to quantitatively determine 
if neutrino telescopes can observe a signal from DM annihilations in the near future.
 To answer this question one first needs to determine the simple models that give rise to a significant annihilation into monochromatic neutrinos at the freeze-out epoch, see Refs.~\cite{Lindner:2010rr,Farzan:2011ck,Arina:2015zoa} for related studies. 
Then, among these models, we must determine which ones lead to an enhancement of the annihilation rate  today. Here, we do not consider the possibility of  large astrophysical boosts, such as considered in Ref.~\cite{Arina:2015zoa}.

Since doing this in full generality is not feasible\footnote{For a systematic study in the case of one-loop DM annihilations into two photons, see Ref.~\cite{Garcia-Cely:2016hsk}.}, we proceed in the following way.
First, in Section \ref{sec:classification}, we state a series of general assumptions concerning the simplicity of models under consideration. Besides the fact that we limit ourselves to DM particles with spin zero or one-half,  
these assumptions are: (i) DM is made of a single particle annihilating into neutrinos through the exchange of a mediator,  
(ii) the $SU(2)_L$ multiplets of the DM and the mediator are not larger than triplets, (iii) the relic density of the DM particle stems from the thermal freeze-out mechanism. 
Having made these simple assumptions, we then determine the models which  give rise to a significant annihilation into neutrinos at the freeze-out epoch. This  takes into account the following  constraints: 1)  DM annihilations must proceed via the s-wave (so that  they are not suppressed today  
in DM halos), 2) DM direct detection and 3)  the  annihilation must not lead to too large neutrino masses.
We will see that the list of models satisfying these constraints is very short. This list is compared to a similar one determined in Ref.~\cite{Lindner:2010rr}. In particular, we show that the criterion of not inducing too large neutrino masses at one loop
considerably reduces the number of viable cases.

Subsequently, we discuss in Section~\ref{sec:SE} and \ref{sec:SEnoEW} the possibility that the annihilation into neutrinos today is boosted with respect to the freeze-out epoch, so that an observable flux of monochromatic neutrinos can be generated.
Section~\ref{sec:SE} focuses on models which can benefit from an electroweak Sommerfeld effect, whereas Section~\ref{sec:SEnoEW} considers the possibility of another source of Sommerfeld enhancement. We will compare the predicted cross section to the sensitivities of current telescopes. Given the fact that, in general, these models do not produce only neutrinos but also other cosmic rays---charged leptons in particular---, we also determine if these neutrino fluxes are compatible with the bounds existing on the production of these other cosmic rays.  In Section \ref{sec:flavor} we look at the possibility to further distinguish the various scenarios from the flavor composition of these fluxes. 

These results naturally bring us to Section~\ref{sec:caveats}, where we discuss the possibility to produce observable monochromatic neutrino fluxes by going beyond the simple assumptions made in the previous sections. 
Before concluding in Section~\ref{sec:conclusions}, we 
discuss in Section~\ref{sec:beyondlines} the possibility of neutrino sharp spectral features other than lines: box-shaped and bremsstrahlung spectra. Appendices~\ref{sec:AppSE} and \ref{sec:AppSEIDM}  are devoted to details on the Sommerfeld effect for Dirac triplets and scalar doublets, respectively.

\section{Limits on lines}
\label{sec:limits}

The possibility that DM could produce neutrinos has been explored
previously in multiple contexts. Since neutrinos travel undeflected in astrophysical environments, observing them with sufficiently good angular resolution allows to infer where they were produced. Therefore, the non-observation of neutrinos  coming from regions of the sky where DM is expected in large amounts constrains the properties of DM. In this work, we are interested in observing the neutrinos emitted through the annihilation of DM in DM halos. The relevant quantity probed in this way is the DM annihilation cross section into neutrinos.\footnote{We will not look at the possibility to observe neutrinos that could have been emitted from DM annihilations inside the Sun.
This typically sets upper limits on the spin-dependent DM-proton or DM-electron cross section, which controls the DM capture rate in the Sun (see Ref.~\cite{Garani:2017jcj} and references therein), and not directly on the DM annihilation cross section we are interested in here.} 
Before looking at explicit models that could lead to an observable monochromatic signal, we shortly discuss 
in this section the state of the art of the experimental constraints on this cross section.

\begin{figure}[t]
\includegraphics[width=0.65\textwidth]{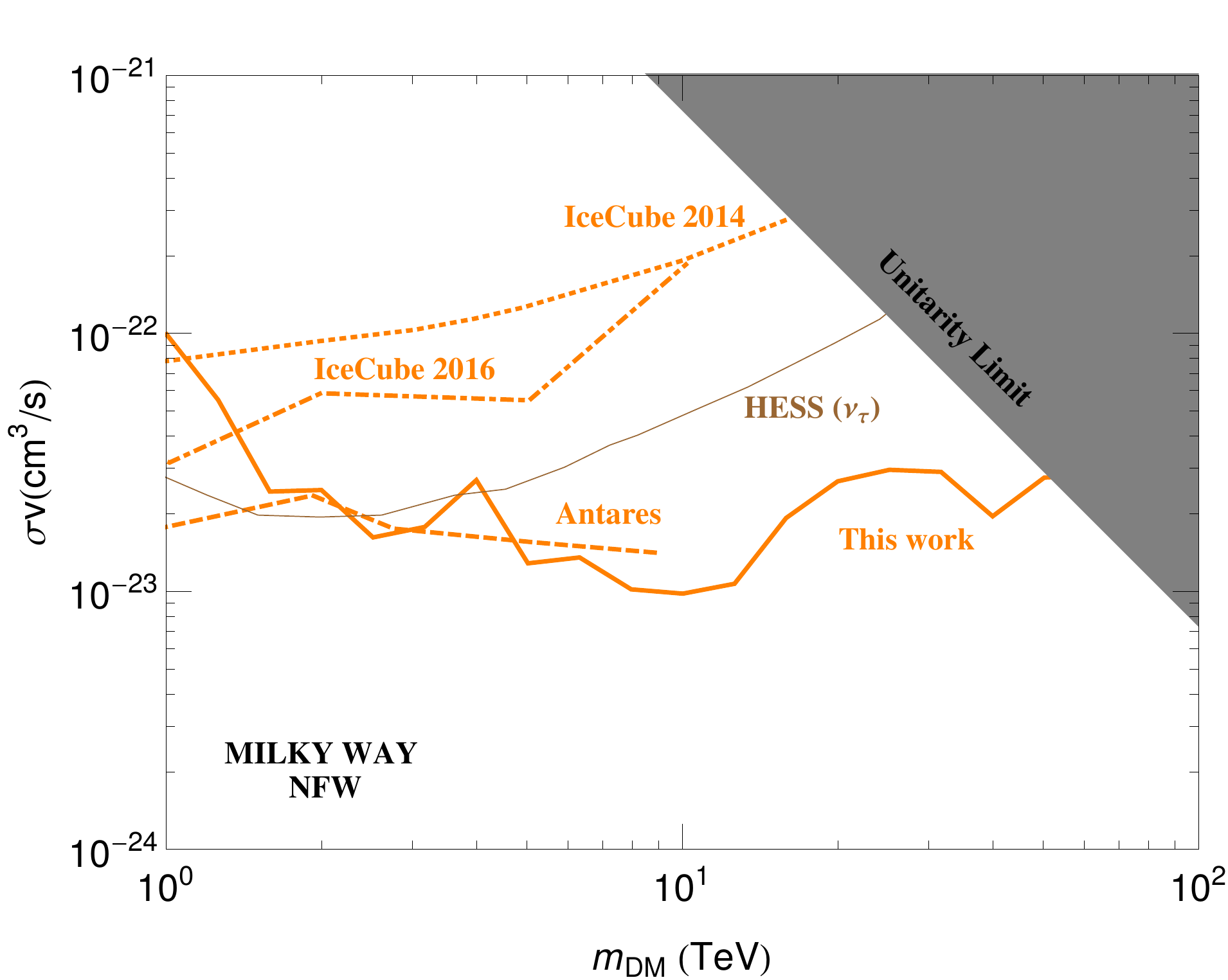}
\caption{Upper bounds on $\sigma v$, the total DM annihilation cross section into neutrinos, as a function of the DM mass $m_\DM$. Given are the 
experimental 90\% CL limits from IceCube 2014~\cite{Aartsen:2014hva}, from IceCube 2016~\cite{Aartsen:2016pfc}, from Antares 2015~\cite{Adrian-Martinez:2015wey}, and those we derived at 95\% CL from a 2-year public data sample of IceCube \cite{Aartsen:2014muf} following the approach considered in section II.C of Ref.~\cite{Aisati:2015vma}.
Also given are the unitarity limit in our galaxy (excluding the shaded area) and the indirect ``Hess ($\nu_\tau$)" limit from the production of secondary gamma-rays through radiative corrections \cite{Queiroz:2016zwd}. Here we assume self-conjugate DM. {For complex or Dirac DM, the limits are a factor of two weaker.}
}
\label{fig:dNdxline}
\end{figure} 

Fig.~\ref{fig:dNdxline} presents a compendium of the existing upper limits on the annihilation cross section $\sigma v$, as a function of DM masses $m_\DM$ ranging from 1 to 100 TeV.
The cross section $\sigma v$ given on the vertical axis is defined as the sum of the cross sections into the various neutrino flavor final states, $\sigma v \equiv \sum_{\alpha,\beta=e, \mu, \tau} \sigma v_{\nu_\alpha \overline{\nu}_\beta}$. 
This definition is convenient because it is directly linked to the total neutrino flux, which does not depend on how neutrinos oscillate on their way to Earth.
All the limits in this figure are given assuming an equal production of $\nu_e$, $\nu_\mu$ and $\nu_\tau$ at the source, and thus in the detector too. Under this assumption, the limits on the total cross section differ by a factor of 3 from those into a single neutrino flavor, often reported in the literature.\footnote{Going from the limit on an individual flavor cross section to the one on the total cross section is model dependent as it requires to know the flavor structure at production. However, in practice, it is a rather good approximation to apply such a factor of 3 in between, because neutrino oscillations approximately lead to a flavor-democratic flux in the detector, see Section~\ref{sec:flavor}.}
The dotted and dashed-dotted curves come from two different analyses of different samples by the IceCube collaboration, respectively ~\cite{Aartsen:2014hva}  and~\cite{Aartsen:2016pfc}, at the 90 \% CL. 
To make the comparison as fair as possible with all the other limits, we have rescaled this limits obtained with an Einasto profile to account for a NFW profile with a local density of 0.39 GeV/cm$^3$.
The Antares collaboration has also reported 90 \% CL limits on the presence of a flux of $\nu_\mu$ and $\overline{\nu}_\mu$ from the galactic center (dashed curve). 

Also given in Fig.~\ref{fig:dNdxline} is an indirect limit on the cross section $\sigma v$ which has been derived under certain assumptions at 95\% CL in Ref.~\cite{Queiroz:2016zwd}. This limit has been obtained by looking at the gamma-ray continuum emitted by $\nu_\alpha \overline{\nu}_\alpha$ final states (with $\alpha = e, \mu, \tau$) as a result of electroweak corrections, and by confronting the corresponding spectra to Fermi and HESS data. We selected the most stringent limits (brown curve labeled as HESS ($\nu_\tau$)) and multiplied it by a factor of 3 since it constrains one third of our cross section (assuming, as said above, a flavor universal flux at production).
For completeness, we also give the unitarity limit on the total annihilation cross section (upper right corner of Fig.~\ref{fig:dNdxline}, assuming a DM relative  velocity of $2\times 10^{-3} c$ in our galaxy~\cite{Bertone:2010zza}), which 
holds as a model-independent upper bound on $\sigma v$.

\begin{figure}[t]
\centering
\includegraphics[width=0.5\textwidth]{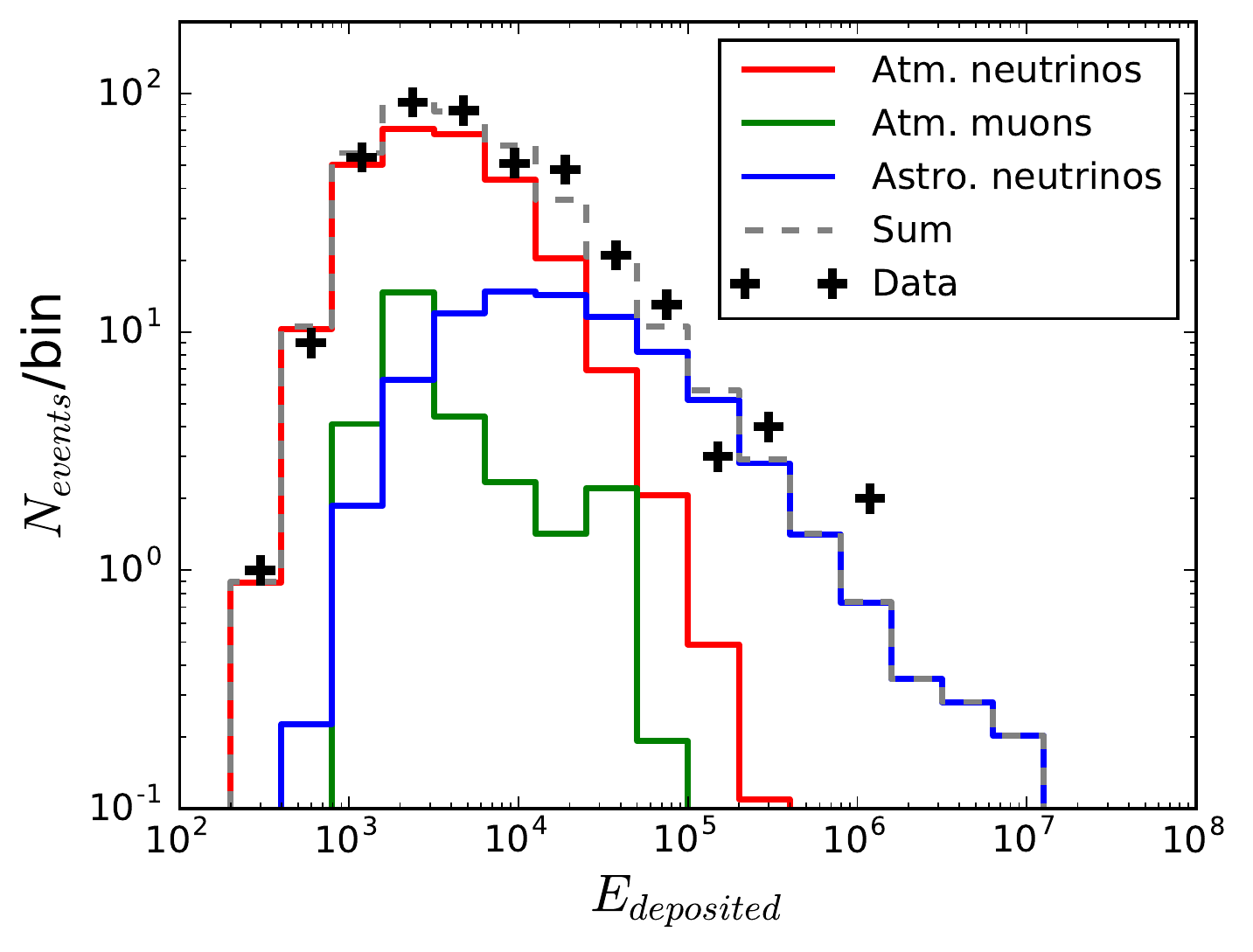}
\caption{Neutrino spectrum analysed as a function of the deposited energy $E_\text{deposited}$ at IceCube. The black crosses show the data, while the green, red and blue curves respectively denote the atmospheric muons, the atmospheric neutrinos, and astrophysical neutrinos (all as derived in \cite{Aartsen:2014muf}).}
\label{fig:eventdistribution2014IC}
\end{figure}

Last, but not least, Fig.~\ref{fig:dNdxline} also shows 95 \% CL limits (solid line) that we have derived by applying to  DM annihilation the approach which has been presented for decaying DM in  section II.C of Ref.~\cite{Aisati:2015vma}. In contrast with all the other limits presented in Fig.~\ref{fig:dNdxline}, this approach is line-feature oriented, meaning that the potential presence of a monochromatic signal was constrained by studying the (deposited) energy distribution of the data and of the backgrounds. These distributions are displayed in Fig.~\ref{fig:eventdistribution2014IC}, where the crosses, the green, the red and the blue curves respectively denote the 2-year data sample of Ref.~\cite{Aartsen:2014muf}, the atmospheric muons, the atmospheric neutrinos, and astrophysical neutrinos. Concretely, we define $N^i_\text{lim}$, the upper bound on the number of events observed in each bin $i$, through Neyman's construction \cite{Neyman:1937uhy},
\begin{equation}
\sum_{k=0}^{N^i_\text{obs}} \dfrac{(N^i_\text{limit})^k}{k!} e^{-N^i_\text{limit}} = 1 - q, 
\end{equation}
where $q$ denotes the confidence level (in our case, 95\% CL). From there, the limit on $\langle \sigma v \rangle$ is defined as the maximum value such that the putative DM-induced signal---which is proportional to $\langle \sigma v \rangle$---plus the background predictions (gray curve in Fig.~\ref{fig:eventdistribution2014IC}) do not overshoot $N^i_\text{lim}$ in any bin.
  The improvements achieved in this way (see Fig.~\ref{fig:dNdxline}) illustrate well the fact that important progresses at neutrino telescopes are possible. Moreover, further improvements are expected in the future 
 for at least two reasons. First, only 
small data samples have been analyzed until now as far as DM searches are concerned.
In comparison, IceCube has now accumulated 6 years of data with the complete detector. Second, spectral features must properly be searched for by optimizing the study of both the energy and angular distributions of the data, which has never been done  to the best of our knowledge.

All in all, from these arguments, it would not be necessarily surprising that an experiment like IceCube could reach, in a near future,  a sensitivity  on $\sigma v$ as good as few times $\unit[10^{-25}]{cm^3/s}$.

\section{Models leading to  large annihilation cross sections into monochromatic neutrinos}
\label{sec:classification}

This section is devoted to the determination of models which lead to s-wave annihilation into neutrinos and are in agreement with direct detection and neutrino mass constraints. As mentioned in the introduction, we will perform this determination within a basic setup based on simple assumptions. Besides postulating  that DM is made of real scalar, complex scalar, Majorana or Dirac particles\footnote{Spin-1 DM will not be considered here.}, we make the following three assumptions:
\begin{itemize}
\item[(i)] \emph{Single DM component and one mediator}.  We assume that DM is made of a single particle species and, in order for DM to  annihilate into neutrinos, we assume a single mediator  beyond the SM.  From the various discussions below, it will be clear why one needs an extra mediator for DM to annihilate into neutrinos, i.e. why annihilations mediated  by SM particles cannot lead to an observable neutrino flux. We will see that this is related to the fact that annihilations mediated by SM particles proceed through a $Z$ boson in the s-channel.

\item[(ii)] \emph{Size of multiplets.} We assume that the DM and  the mediator particles belong to  $SU(2)_L$ multiplets not larger than a triplet. Notice that  the mediator, like the DM, must be electrically neutral so that the annihilation can proceed at tree level
. Thus, multiplets in our discussion can be singlets ``S", doublets with hypercharge 1 ``D" , triplets with hypercharge 0 ``$T_0$", or triplets with hypercharge 2 ``$T_2$". 
We impose this requirement on the size of the multiplet simply for the sake of not having too many models. In Section~\ref{sec:caveats}, we discuss the consequences of giving up this assumption.

\item[(iii)] \emph{Thermal DM production}. Furthermore, in this basic setup we assume that DM  is thermally produced as a result of the various 
(co-)annihilation channels in the model. {In general, we will see that,} in order to get an observable neutrino signal, it is important that  the annihilation  into neutrinos plays a non-negligible role in the freeze-out process.

\end{itemize}

Given these assumptions, we must now consider the following constraints:

\begin{enumerate}
\item \emph{s-wave annihilation}.  In order to have a neutrino flux sufficiently large,   one is  forced to consider  s-wave annihilations. Any other partial  wave would be suppressed  today and could not lead to any observable flux of monochromatic neutrinos. If no Sommerfeld effect is present, the suppression arises as powers of the velocity.  Otherwise, the suppression is due to powers of the coupling constant associated to the Sommerfeld effect~\cite{Cassel:2009wt}.\footnote{As pointed out recently, when certain selection rules apply, the suppression due to the coupling constant might be avoided~\cite{Das:2016ced}. We will not consider such scenarios in this work.} 

For an \underline{annihilation into $\nu \bar{\nu}$} such a requirement eliminates many scenarios. Since the neutrino is left-handed, the antineutrino right-handed, and since they have negligible masses, the final state has, in this case, one unit of  total angular momentum along the direction of motion of the neutrinos in the annihilation frame. Accordingly, the initial state must have $J=1$. Thus, if we consider only annihilations taking place via the s-wave, the total spin of the initial state must equal one, $S=1$. This is a very strong restriction because it excludes scalar or Majorona DM, which may only form pairs with $S=0$. As a result, this annihilation channel requires Dirac DM. In particular, this excludes the possibility of splitting  Dirac particles into Majorana fermions, which would be required if the DM multiplet had hypercharge, as discussed below when addressing the direct detection constraint.

In contrast, for an \underline{annihilation into $\nu\nu$}, since both neutrinos are left-handed and have a negligible mass, the final state has zero total angular momentum along the direction of motion. This and the fact that there is no CP-symmetry constraint  on this channel ($\nu\nu$ is not a CP eigenstate) leave the possibility of  having s-wave annihilations open.

\item \emph{Direct detection}. Direct detection constraints are an important concern for all DM candidates coupling to the $Z$ boson, i.e., for all DM multiplets with non-vanishing hypercharge. 
As well known, the exchange of a $Z$ boson between DM and nucleons induces a scattering process far too fast to be compatible with direct detection bounds. This can only be circumvented by splitting the  field coupling to the Z boson into two self-conjugate fields of opposite CP parity, in such a way that the mass difference of the latter renders the direct detection process kinematically forbidden.
For a scalar  DM doublet, $\phi_D$, 
this can be done through the interaction Lagrangian
\begin{equation}
{\cal L}=-\frac{1}{2}\lambda_5 [(H^\dagger \phi_D)^2 +h.c.], 
\label{eq:lambda5def}
\end{equation}
where $H$ is  the SM doublet. Concretely, from Eq.~\eqref{eq:DD_IDM} of Appendix~\ref{sec:AppSEIDM} we have
\begin{equation}
|\lambda_5| \gtrsim 3 \cdot 10^{-6} \left( \frac{m_\text{DM}}{\unit[1]{TeV}}  \right) \left(\frac{\delta m}{\unit[100]{keV}}\right)
\end{equation}
with $\delta m\sim 100$~keV being the typical minimum mass splitting that is needed to suppress the $Z$-mediated direct detection process~\cite{TuckerSmith:2001hy}.
For other candidates with non-vanishing hypercharge, this mass splitting can only be induced through non-renormalizable interactions or from the adjunction of extra scalar multiplets. As examples\footnote{For  other examples in the context of left-right symmetric DM models, see Ref.~\cite{Garcia-Cely:2015quu}}, for  a scalar triplet with $Y=2$,  a second scalar triplet with the same hypercharge  allows to write an interaction term similar to  Eq.~\eqref{eq:lambda5def}. For a fermionic vector-like doublet or  fermionic vector-like triplet with $Y=2$, the chiral components can be split into two Majorana fermions from higher dimensional operators involving several SM scalar doublets (or by coupling pairs of these fermions to a scalar field with $Y=2$ or $Y=4$). They lead,  through electroweak symmetry breaking, to different Majorana masses for the chiral components. Note that, in addition to assuming as above a DM multiplet and a mediator multiplet, these cases require an extra multiplet whose exchange or vev leads to the mass splittings.
 
For the \underline{$\nu\bar{\nu}$ channel}, this criterion eliminates all Dirac DM candidates with non-vanishing hypercharge because splitting the mass of the neutral components converts DM into Majorana fermions, which annihilate into $\nu\bar{\nu}$ via its p-wave. Thus, for this channel only hyperchargeless singlet or triplet Dirac candidates are viable.

Note that this direct-detection constraint also explains  why at tree level  one needs a mediator beyond the SM  to produce monochromatic neutrinos,  as stated in assumption (i). Without any new mediator,  the annihilation into neutrinos at tree level can only proceed through a $Z$ in the s-channel, which is not possible  due to the present direct-detection constraint.

\begin{figure}
\includegraphics[width=0.27\textwidth]{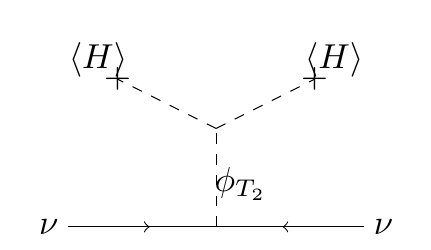}
\caption{ 
Tree-level diagram  leading to neutrino masses for models whose mediator is a scalar triplet with $Y=2$.}
\label{fig:diagram-tree}
\end{figure}

\begin{figure}
\includegraphics[width=0.27\textwidth]{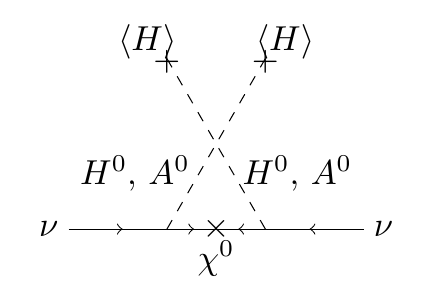}
\includegraphics[width=0.27\textwidth]{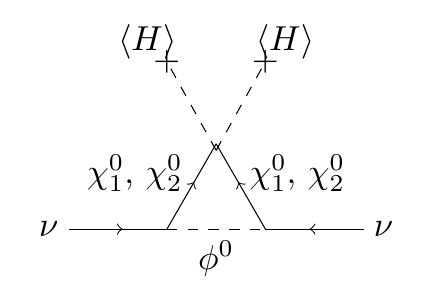}\\
\includegraphics[width=0.27\textwidth]{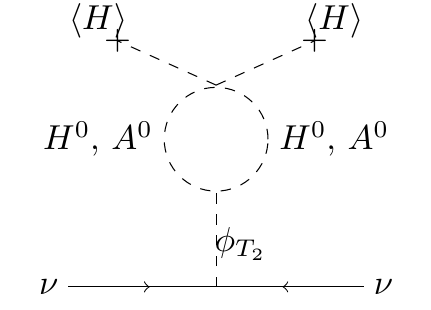}
\includegraphics[width=0.27\textwidth]{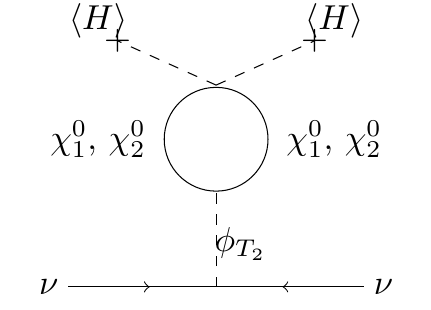}
\caption{Neutrino-mass diagrams at one loop induced by annihilations of doublet DM. The diagrams of the first row are finite. Those of the second row diverge and renormalize the tree-level process of Fig.~\ref{fig:diagram-tree}}.  

\label{fig:diagrams2}
\end{figure}

\item \emph{Neutrino masses}.  One issue which is fully relevant for the \underline{$\nu\nu$ channel}  concerns the possibility that the same diagrams leading to annihilation into neutrinos could give rise to neutrino-mass Weinberg operators with too large coefficients. This is not necessarily a problem at tree level\footnote{\label{footnote:6}A priori, this tree-level concern only applies to the case of a scalar triplet with $Y=2$ in the  s-channel (the models we will call $\SSS_1$ and $\FF_1$ in Table \ref{table:models} below). In this case, the neutrino-mass constraint requires the vev of this mediator to be small, implying that its coupling to two SM Higgs doublets is small as well, see Fig.~\ref{fig:diagram-tree}. However, since this vev does not enter in the annihilation process, both issues are parametrically separated. Note that to assume a flavour setup with large Yukawa couplings whose contributions cancel each other in the neutrino mass matrix (such as the ones based on an approximatly conserving lepton number $U(1)_L$ symmetry, see Refs. \cite{Shaposhnikov:2006nn},\cite{Pilaftsis:2003gt} and \cite{Branco:1988ex}) does not help here. This will not boost the $\nu\nu$ annihilation channel, only the $\nu\bar{\nu}$ one which remains p-wave suppressed.} but it is a matter of concern at the loop level and
actually excludes quite a number of models.
For instance, the models involving a DM scalar doublet annihilating into the $\nu\nu$ final state in the t-channel are excluded because if they satisfy the direct detection constraint, they necessarily induce too large neutrino masses through the first diagram of Fig.~\ref{fig:diagrams2}.
This stems from the fact that, in this case and as said above, the direct detection constraint requires a minimum value for the coupling responsible for splitting the CP-even and CP-odd neutral scalars in the doublet. In the presence of such an interaction, from the annihilation channel into $\nu\nu$, one can directly form a loop by attaching both DM particles together. Given the lower bound on $\lambda_5$, one can see  that an observable flux of neutrinos leads to a too large neutrino mass contribution (of order $\sim 100$~keV at least). As shown in Ref.~\cite{Andreas:2009hj} for the production of monochromatic neutrinos, such problem occurs in the scotogenic model~\cite{Ma:2006km} (which is model $\SSS_3$ in Table.~\ref{table:models}).

Such an incompatibility between an observable flux of neutrinos and neutrino-mass constraints basically holds for all models which involve a source of lepton number violation\footnote{In other words, this incompatibility applies to all models which lead to a one-loop neutrino-mass diagram involving 
two arrow flips: one from the Majorana character of the fermion particle in the diagram   and one occurring for the scalar particle, as induced for example by a $\lambda_5$ interaction.   
}.
 Nevertheless, note that for the models which proceed to an annihilation into neutrinos in the s-channel---last two diagrams of Fig.~\ref{fig:diagrams2}---this exclusion is not strict. For these two cases (which we call $S_1^r$ and $F_1^m$ below), and unlike for the first two diagrams of  Fig.~\ref{fig:diagrams2}, the loop renormalizes  the interaction associated to the diagram of Fig.~\ref{fig:diagram-tree}. Thus even if, just in the same way as for the first two diagrams, the finite part of this loop gives far too large neutrino masses, in principle this large finite part could be canceled by an appropriate counter-term in the tree-level diagram, leading to sufficiently small neutrino masses. In the following, we will consider these models as not absolutely excluded and discuss their phenomenology accordingly.
Thus, except for these two fine-tuned models, all the models that could be probed by neutrino telescopes do not involve lepton number violation associated to the DM and mediator particles. In other words, apart from $\SSS_1$ and $\FF_1$, models that involve a source of lepton-number violation necessarily lead to too large neutrino masses at loop level.  
\end{enumerate}

\begin{table}[t]
\begin{tabular}{|c|c|c|c|c|c|c|c|c|}\hline
Annihilation &  \multicolumn{2}{c|}{\multirow{2}{*}{DM}} & \multicolumn{2}{c|}{\multirow{2}{*}{Mediator}}  & $m_\nu$  OK  & Suppressed  & \multirow{2}{*}{$\ell^+\ell^-$} & \multirow{2}{*}{Model} \\
Channel &  \multicolumn{2}{c|}{} &  \multicolumn{2}{c|}{}  & at 1-loop? & by $v_\text{EW}/m_\DM$?  & &\\\hline\hline
\multirow{4}{*}{$\overline{\text{DM}}\text{DM}\to \overline{\nu}\nu$} &
\multirow{4}{*}{Dirac} & $T_0$ & s-chann. vector& $S$ &   \multirow{4}{*}{Yes} &\multirow{4}{*}{No{}} & \multirow{4}{*}{=} &\hyperref[model:F1]{\textcolor{red}{$F_1$}}\\\cline{3-5}\cline{9-9}
&& $T_0$ & t-chann. scalar& $D$ &  & && \hyperref[model:F2]{\textcolor{red}{$F_2$}} \\\cline{3-5}\cline{9-9} 
&& $S$ & s-chann. vector& $S$  & & && \hyperref[model:F3F4]{\textcolor{red}{$F_3$}} \\\cline{3-5}\cline{9-9} 
&& $S$ & t-chann. scalar& $D$   & & && \hyperref[model:F3F4]{\textcolor{red}{$F_4$}} \\\hline

\multirow{19}{*}{$\text{DM}\text{DM}\to \nu\nu$} & \multirow{8}{*}{Real Scalar } &$D$ & s-chann. scalar & $T_2$  & $\pm$ & No{}  & \multirow{18}{*}{/}& \hyperref[model:SSS1]{\textcolor{red}{$\SSS_1$}} \\\cline{3-7}\cline{9-9}
& & $S$ & \multirow{7}{*}{ t-chann. Majorana }& $D$   &\multirow{6}{*}{No} &  Yes{} && $\SSS_2$   \\\cline{3-3}\cline{5-5}\cline{7-7}\cline{9-9}
&& $D$ & & $S$& &No{}  && $\SSS_3$    \\\cline{3-3}\cline{5-5}\cline{7-7}\cline{9-9}
&& $D$ & & $T_0$ && No{} && $\SSS_4$    \\\cline{3-3}\cline{5-5}\cline{7-7}\cline{9-9}
&& $D$ &  & $T_2$  && Yes{} && $\SSS_5$   \\\cline{3-3}\cline{5-5}\cline{7-7}\cline{9-9}
 
&& $T_0$ & & $D$ &  &  Yes{} &&
 $\SSS_6$  \\\cline{3-3}\cline{5-5}\cline{7-7}\cline{9-9}

&& $T_2$ &  & $D$ && Yes{} && $\SSS_7$ \\  
\cline{2-7}\cline{9-9}
&\multirow{8}{*}{ Majorana }& $D$ & s-chann. scalar & $T_2$&  $\pm$ & No{}  && \hyperref[model:FF1]{\textcolor{red}{$\FF_1$}}\\\cline{3-7}\cline{9-9}
 &  & $S$ & \multirow{7}{*}{t-chann. scalar}& $D$    &\multirow{6}{*}{No} &  Yes{} && $\FF_2$   \\\cline{3-3}\cline{5-5}\cline{7-7}\cline{9-9}
&& $D$ & & $S$& &No{}  && $\FF_3$    \\\cline{3-3}\cline{5-5}\cline{7-7}\cline{9-9}
&& $D$ & & $T_0$ & & No{} && $\FF_4$    \\\cline{3-3}\cline{5-5}\cline{7-7}\cline{9-9}
&& $D$ &  & $T_2$ &  & Yes{} && $\FF_5$   \\\cline{3-3}\cline{5-5}\cline{7-7}\cline{9-9}
 
&& $T_0$ & & $D$ &&   Yes{} &&
 $\FF_6$  \\\cline{3-3}\cline{5-5}\cline{7-7}\cline{9-9}

&& $T_2$ &  & $D$&& Yes{} && $\FF_7$    \\\cline{2-7}\cline{9-9}

&\multirow{2}{*}{Complex Scalar} &  $S$&\multirow{2}{*}{ t-chann. Majorana} & $D$ &\multirow{2}{*}{Yes} & \multirow{2}{*}{Yes{}}  && \hyperref[model:S1S2]{\textcolor{red}{$S_1$}}\\\cline{3-3}\cline{5-5}\cline{9-9}
&& $T_0$ &  & $D$& &   &&  \hyperref[model:S1S2]{\textcolor{red}{$S_2$}}\\\cline{2-7}\cline{9-9}
&
\multirow{2}{*}{Dirac} &  $S$ &\multirow{2}{*}{t-chann. scalar} & $D$ & \multirow{2}{*}{Yes} & \multirow{2}{*}{Yes{}}  && $F_4$\\\cline{3-3}\cline{5-5}\cline{9-9}
&& $T_0$ &  & $D$ & &   &&  $F_2$\\\hline
\end{tabular}
\caption{
DM models which pass the direct detection constraint and the requirement of s-wave annihilation into neutrinos. 
 $S$, $D$, $T_0$ and $T_2$ hold for a field which is a singlet, doublet, $Y=0$ triplet and $Y=2$ triplet, respectively. 
The models leading to too large neutrino masses have a ``No'' in the  column  labeled $m_\nu$, whereas the two models 
 which satisfy the one-loop neutrino mass-constraint after fine tuning are  indicated with ``$\pm$ ". Those labeled as  ``Yes'' in the column $m_\nu$ do not break lepton number and do not generate neutrino masses. The models whose annihilation into neutrinos requires an electroweak vev insertion (with a rate  suppressed by at least $(v_\text{EW}/m_\DM)^4$)  are indicated in the next column.  These models cannot {perturbatively} lead to a large annihilation cross section into neutrinos if $m_\text{DM}\gg v_\text{EW}$. 
The column  $\ell^+\ell^-$  specifies whether the model leads  to annihilations into charged lepton pairs. The ``="  sign means that electroweak symmetry leads to equal rates in $\nu\bar{\nu}$ and in $\ell^+\ell^-$ (up to corrections proportional to the electroweak vev $v_\text{EW}$). 
The ``/" sign holds for the case when associated charged lepton production is not present at tree level. The labels in red correspond to the models that are extensively discussed in the next sections.  
}
\label{table:models}
\end{table}

Table \ref{table:models} summarizes the whole discussion above by showing
all the models that proceed through s-wave annihilation---explaining  why there are no Majorana and scalar DM scenarios with the final state $\nu\bar{\nu}$---and that are in agreement with the $Z$-exchange direction detection constraint---explaining why there is no case with  Dirac and complex scalar DM having non-zero hypercharge.
This singles out 9 scalar and 11 fermionic DM models.
Table \ref{table:models} also shows which of these models are excluded by the neutrino mass constraint, i.e.~all   scenarios with the real scalar or Majorona DM annihilating into $\nu\nu$, except, as explained above, two models proceeding in the s-channel ($S_1^r$ and $F_1^r$). Note that there is a one-to-one correspondence between the list of models with the real scalar DM and that of  Majorana DM. This is  a consequence of the fact that 
scalar or Majorana DM pairs always have total spin $S=0$ and therefore obey  the various constraints in the same way.

Table~\ref{table:models} also displays which  models involve an annihilation into neutrinos with a rate proportional to powers of $v_\text{EW}/m_\DM$. Notice that these models are those that violate hypercharge. The neutrino flux predicted by them is naturally suppressed above the electroweak scale. { One can}  show that for multi-TeV DM, and due to this suppression, these models {do not} lead to a sufficiently large neutrino flux in a perturbative way\footnote{{If the couplings are perturbative, the annihilation into neutrinos is negligible in the freeze-out process for $m_\DM \gg v_\text{EW}$. Thus, to have an observable monochromatic flux today, one needs a very large boost.  We will not elaborate further on that.}}. This will be useful in our discussion of the Sommerfeld effect in the next two sections, because if the Sommerfeld enhancement is induced by electroweak interactions, it will only be large when $m_\DM\gg v_\text{EW}$, i.e.~in the multi-TeV range.

Another property listed in Table~\ref{table:models} has to do with the annihilations involving charged leptons. Whenever the final state into neutrinos is present, $SU(2)_L$ invariance predicts another process into charged leptons (not necessarily with the same initial state), whose rate is equal to that of neutrinos, up to terms proportional to $v_\text{EW}$. More concretely, the swapping  of a neutrino for an electron is a particular transformation induced by the second generator of the $SU(2)_L$ algebra
\begin{align}
\begin{pmatrix}
\nu\\ \ell^-
\end{pmatrix}
 &\to 
\begin{pmatrix}
\ell^-\\ -\nu
\end{pmatrix} = e^{ i \pi \, T_2}
\begin{pmatrix}
\nu\\ \ell^-
\end{pmatrix}
 \,.
\end{align}
For the $\nu\nu$ final state, the charged-lepton counterpart is $\ell^-\ell^-$, which is not an available channel for DM annihilations. In contrast, for the $\nu\overline{\nu}$ final state, the corresponding charged-lepton pair is  $\ell^-\ell^+$. For the models in Table~\ref{table:models} that give rise  to the final state $\nu\overline{\nu}$, the action of $e^{ i \pi \, T_2}$ leaves the initial state invariant. This is obvious for singlets, while for triplets with $Y=0$, we have 
\begin{align}
\begin{pmatrix}
\psi^+\\ \psi^0\\\psi^-
\end{pmatrix}
 &\to 
\begin{pmatrix}
\psi^-\\ -\psi^0\\\psi^+
\end{pmatrix} = e^{ i \pi \, T_2}
\begin{pmatrix}
\psi^+\\ \psi^0\\\psi^-
\end{pmatrix}
 \,.
\end{align}
Hence,  for the models in Table~\ref{table:models} that lead to $\nu\overline{\nu}$, we always have DM annihilations into  $\ell^-\ell^+$ with the same rate (again, up to terms proportional to the electroweak vev). We will further discuss the implications of this  in Sections~\ref{sec:results} and ~\ref{sec:bremsstrahlung}.

Combining all constraints only leaves 8 models which can be split as 4+2+2. The first four are the fermion DM  models, labeled as $F_{1-4}$, which readily satisfy all constraints. These models require a Dirac fermion DM multiplet with vanishing hypercharge, that is, a singlet or a triplet. All four models involve an annihilation into $\nu\bar{\nu}$, two of them ($F_{1,3}$) proceed via the s-channel exchange of a $Z'$ spin-1 boson, whereas the other two ($F_{2,4}$) proceed via the t-channel  exchange of a scalar doublet. Models $F_2$ and $F_4$ also involve an annihilation into $\nu\nu$, but for DM masses well above the electroweak scale, it is suppressed by powers of $v_\text{EW}/m_\text{DM}$ and is negligible with respect to the annihilation into $\nu\bar{\nu}$.
The next  two models are the fine-tuned models which require canceling out neutrino-mass diagrams. These are the model $\SSS_1$, based on a scalar real scalar DM particle, and the model $\FF_1$, based on a Majorana DM fermion, both belonging to a doublet.  Finally, 
the last two models  can only satisfy  all constraints  perturbatively for DM masses below the $\sim$TeV scale. Both are based on a complex scalar DM setup out of a hyperchargeless singlet ($S_1$) or triplet ($S_2$), both with a fermion doublet mediator.

We would like to remark that a list of models leading to DM annihilations into monochromatic neutrinos has also been determined  in Ref.~\cite{Lindner:2010rr}. Even though there are many similarities, both approaches are qualitatively different, in particular due to the fact that we focus from the start on a neutrino flux at the reach of neutrino telescopes.
A small difference is that Table~\ref{table:models} disregards from the start the models leading to p-wave or d-wave annihilation as well as the models involving possible right-handed neutrinos evolving into SM neutrinos via mixing in the final state (because as already said in Ref.~\cite{Lindner:2010rr} these processes are suppressed by the mixing). Another difference is that we include the perturbativity constraint for DM scenarios above few TeV.
Also,  Table \ref{table:models} is more detailed because it specifies if fermions are Dirac or Majorana particles. Hence, a single instance in the tables of Ref.~\cite{Lindner:2010rr} corresponds to several models in our classification. 
Moreover, $Z'$ mediators  were not considered in Ref.~\cite{Lindner:2010rr}.
Besides the above, the main and most important difference is that the possibility that these models could lead at one loop to too large neutrino masses is not considered in Ref.~\cite{Lindner:2010rr}. 
As we have seen above, such a constraint  excludes many models.

In summary, the requirement of large s-wave annihilation into neutrinos at freeze-out, combined with direct-detection and neutrino-mass constraints, lead to
a very limited number of possible DM candidates and simple models which, apart from two fine-tuned models and two sub-TeV scenarios, all involve a hyperchargeless Dirac DM fermion, conserve lepton number and lead to the $\nu\bar{\nu}$ final state.

\newpage 
\section{Models with electroweak Sommerfeld Enhancement}

\label{sec:SE}

So far, we have determined the models which could lead to a large (s-wave) annihilation rate into neutrinos, that is to say, not several orders of magnitude smaller than the freeze-out thermal value. However, as discussed in Section~\ref{sec:limits}, even this value is not expected to lead to an observable flux of monochromatic neutrinos, at least for the very near future. Thus, in order to observe a flux  one would need an enhancement of the cross section into neutrinos today with respect to the freeze-out epoch. An enhancement expected to arise for certain DM candidates comes from the Sommerfeld effect.

As is well known, in order to have Sommerfeld enhancement one needs the DM  to couple to a lighter mediator that can be exchanged between the particles of the initial state in the annihilation process. Unless an extra light mediator of this kind is added, this is not operative for the singlet DM cases. Moreover,  a large electroweak Sommerfeld enhancement is only possible for DM masses much larger than the electroweak boson masses, that is to say, above few TeV. Thus, out of the eight models of the previous section, only four can lead to an electroweak Sommerfeld boost. These are $F_{1,2}$, both involving a hyperchargeless Dirac triplet and annihilation into $\nu \bar{\nu}$, and the doublet models $\SSS_1$ and $\FF_1$.
In the following, we determine the Sommerfeld enhancement for these models. In Section~\ref{sec:SEnoEW}, we also briefly consider models $F_{3,4}$  and $S_{1,2}$ which all require non-electroweak  Sommerfeld mediator.

Let us start with models $F_{1,2}$, where the DM candidate belongs  to a Dirac triplet with no hypercharge, that we write as  $\psi= (\psi_1^+, \psi^0,  \psi_2^-)^T $. In order to protect its Dirac nature, we will assume, for instance, that $\psi$ is the lightest multiplet charged under a $U(1)'$ group.
Including Sommerfeld effects, the 
s-wave cross section for a pair of DM particles 
annihilating into a two-body final state $ab$ can be written in both models as
\begin{equation}
\sigma v\left(\psi^0 \overline{\psi^0}\to a b\right)  =  \,d^*\, \Gamma_{ab}\, d^T\,,
\label{eq:SEsigmav_text}
\end{equation}
where $d = (d_+, d_0, d_+)$ are the Sommerfeld factors and $\Gamma_{ab}$ is the corresponding annihilation matrix. The former encode the non-perturbative effect associated to exchange of SM gauge bosons in the initial state, whereas the latter describes the perturbative physics associated to the annihilation. 
In appendix~\ref{sec:AppSE}, we calculate the Sommerfeld effect  for a Dirac triplet.  In particular, we show that the Sommerfeld factors are unequivocally determined by fixing the mass and the relative velocity of the DM. The mass splittings between the neutral and charged components of the triplet also play a role,  but they are  fixed to 166 MeV by radiative corrections~\cite{Cirelli:2005uq}.

\subsection{Model $F_2$}
\label{model:F2}

\begin{figure*}[h]
\includegraphics[trim=0cm 0cm 0cm 0cm,clip,height=0.15\textheight]{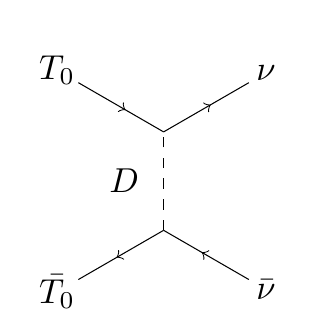}
\caption{ DM annihilation into neutrinos for model $F_2$. We use the notation of Table~\ref{table:models}.}
\end{figure*}

Here, in addition to the Dirac triplet, there is a heavier doublet  $\varphi_D= (\varphi^+, \varphi^0)^T $ that interacts with the SM lepton doublets $L_\alpha$ by means of 
\begin{eqnarray}
{\cal L}_D &\supset& y_\alpha\, \overline{\psi } P_L L_{\alpha} \varphi_D+ h.c.\,.
\label{eq:LF2}
\end{eqnarray}

In the following we calculate the cross section of this model for the $\psi^0 \bar{\psi}^0\to \nu\bar{\nu}$ channel.
Interestingly, 
the process $\psi^0 \psi^0\to \nu\nu$ also exists but is suppressed by two powers of  $v_\text{EW}/m_\DM$ because it violates hypercharge by two units. Since, in this particular section, we are interested in multi-TeV DM, neglecting this channel is justified and we only consider the final states with no lepton number.

The matrices $\Gamma_{\nu_\alpha \overline{\nu_\beta}}$ describing tree-level annihilations into neutrinos receive two contributions that interfere with each other. The first one is associated to the  creation of the neutrinos through the exchange of the mediator $\varphi_D$  and is therefore proportional to four Yukawa couplings, see Eq.~\eqref{eq:LF2}. The other contribution is given by SM gauge interactions creating  neutrinos through a $Z$ boson in the s-channel. This gives the annihilation matrices

\begin{align}
\label{eq:GammaF2Gaugev2}
\Gamma^{S=1}_{\nu_\alpha\bar{\nu}_\beta} =  \frac{|y_\alpha|^2 |y_\beta|^2 m_\DM^2}{96 \pi (m_\DM^2+m_D^2)^2} 
\begin{psmallmatrix}
  4   & 2 & 0\\
 2 & 1 &0\\
 0 & 0 & 0
\end{psmallmatrix}
+
\frac{|y_\alpha|^2 \alpha_2 \delta_{\alpha \beta} }{96 (m_\DM^2+m_D^2)} 
 \begin{psmallmatrix}
 8 & 2 & -4\\
 2 & 0 & -2\\
-4 & -2 & 0
\end{psmallmatrix}
 +  \frac {\alpha_2^2 \pi \delta_{\alpha \beta}}{ 24 m_\DM^2} 
\begin{psmallmatrix}
 1 & 0 & -1 \\
 0 & 0 & 0 \\
 -1 & 0 & 1
\end{psmallmatrix}\,,\\
\Gamma^{S=1}_{\ell_\alpha\bar{\ell}_\beta} = \frac{|y_\alpha|^2 |y_\beta|^2 m_\DM^2}{96 \pi (m_\DM^2+m_D^2)^2} 
\begin{psmallmatrix}
 0 & 0 & 0\\
 0 & 1 & 2\\
 0 & 2 & 4
\end{psmallmatrix}
+
\frac{|y_\alpha|^2 \alpha_2 \delta_{\alpha \beta} }{96 (m_\DM^2+m_D^2)}  
\begin{psmallmatrix}
 0 & -2 & -4\\
 -2 & 0 & 2\\
 -4 & 2 & 8
\end{psmallmatrix}
 +  \frac {\alpha_2^2 \pi  \delta_{\alpha \beta}}{ 24 m_\DM^2} 
\begin{psmallmatrix}
 1 & 0 & -1 \\
 0 & 0 & 0 \\
 -1 & 0 & 1
\end{psmallmatrix}\,.
\end{align}
In these matrices, we emphasize that the corresponding process only takes place when the initial state has total spin $S=1$.   According to Eq.~\eqref{eq:SEsigmav_text}, the cross section into neutrinos is 
\begin{eqnarray}
\sigma v \left(\psi^0 \overline{\psi^0}\to \nu_\alpha\bar{\nu}_\beta\right) &=&
3\sigma v^{S=1} \left(\psi^0 \overline{\psi^0}\to \nu_\alpha\bar{\nu}_\beta\right)=
 \sigma v_0 \big|d_0+2d_+\big|^2 \text{Br}_{\alpha\beta}\,,
\label{eq:neutrinos}
\end{eqnarray}
with
\begin{align}
\sigma v_0=\frac{9  \overline{y}^4\,m_\DM^2}{32 \pi (m_\DM^2+m_D^2)^2} \,&&\text{and}&&\text{Br}_{\alpha\beta}\bigg|_{F_2} &= \frac{|y_\alpha|^2|y_\beta|^2}{\left(\sum_{\alpha'}|y_{\alpha'}|^2\right)^2}\,,
\label{eq:BrF2}
\end{align}
where $\overline{y}^2 \equiv \sum_\alpha |y_\alpha|^2/3$. For annihilation into charged leptons, the cross section is the same. The Sommerfeld enhancement factors $d = (d_+, d_0, d_-)$ are calculated in Appendix~\ref{sec:AppSE}. Due to the fact  that $d_+$ and $d_-$ are equal,  the pure gauge part and the interference terms do not contribute to the annihilation cross section in Eq.~\eqref{eq:BrF2}, even though they do contribute to the annihilation matrix. 

The relic density is set by two classes of interactions: the one associated to Eq.~\eqref{eq:LF2} and the one involving only the electroweak interactions of the Dirac triplet. Each class involves various annihilation and co-annihilation processes with their corresponding Sommerfeld effect. Summing the contributions to the freeze-out of all these processes, the $26\%$ relic density constraint fixes the bare cross section $\sigma v_0$ (or equivalently $\bar{y}$ for a given $m_\DM$ and $m_D$). See the details in Appendix~\ref{sec:AppSE}.
For $y_\alpha=0$, the observed relic density is only obtained  for one single value of $m_\text{DM}$---fixed by electroweak interactions---which turns out to be    $\simeq\unit[2]{TeV}$.   Notice that this is a factor of $\sqrt{2}$ smaller than for Wino DM. For larger values of $m_\text{DM}$, the electroweak annihilation channels are too slow to account for the relic density, but the annihilation channels involving the $y_\alpha$ coupling can account for it. In this case, if the mass is well above $\unit[2]{TeV}$, the electroweak annihilation channels can safely be neglected,
and if the Sommerfeld effect at freeze-out is also neglected,
the relic density constraint gives $\sigma v_0 \approx 0.5\times 4.7\times\unit[10^{-26}]{cm^3/s}$. The factor 0.5 in front  is due to co-annihilations. This is just an approximation because the Sommerfeld effect can be sizable at the freeze-out, as it is the case for the quintuplet minimal DM~\cite{Cirelli:2007xd}. Based on the method introduced in Refs.~\cite{Cirelli:2007xd,Cirelli:2009uv}, and as explained in appendix~\ref{sec:AppSE}, we have made an estimate of this Sommerfeld effect in the early universe for Dirac triplets. We found that, for large DM masses, including such effect at freeze-out reduces the cross section by roughly 25\%. 
From now on, to determine $\sigma v_0$, we will use the estimate in Eq.~\eqref{eq:relicF2}.

The knowledge of $\sigma v_0$ allows us to predict unambiguously the annihilation cross section today in DM halos as a function of $m_\DM$ and of the DM velocity. To this end, we plug the value of $\sigma v_0$ in Eq.~\eqref{eq:neutrinos}, taken with the DM velocities corresponding to those in DM halos today. Fig.~\ref{fig:nuFluxF2}  gives the annihilation cross section we obtain in this way in the Milky Way and dwarf galaxies. These results are discussed below together with those of other models.

On top of the numerical results presented in  Fig.~\ref{fig:nuFluxF2}, it is also interesting to derive by simple means an order of magnitude estimate for the annihilations in DM halos. To this end, we can consider the electroweak symmetric limit, $m_\DM\gg v_\text{EW}$. In that limit, we find\footnote{\label{ft:potential}For $m_\DM\gg v_\text{EW}$, in the total-isospin basis, the most attractive potential is the one corresponding to the isospin singlet $V(r)=-2\alpha_2/r$, whose Sommerfeld factor is ${\cal S}(-4\alpha_2\pi/v) \approx 4\alpha_2\pi/v$. See appendix~\ref{sec:AppSE} for details.}  that $|d_0+2d_+|^2 \approx 4\alpha_2\pi/v$. Thus, the total annihilation cross section into neutrinos is roughly $\unit[3.6\times10^{-24}]{cm^3/s}$  
in the Milky Way (assuming $v=2\times10^{-3} c$~\cite{Bertone:2010zza}). 
In fact, in Fig.~\ref{fig:nuFluxF2} we observe that the  cross section numerically computed oscillates around this estimate  and approaches it at large masses.

\subsection{Model $F_1$}
\label{model:F1}

\begin{figure*}[h]
\includegraphics[trim=0cm 0.9cm 0cm 0.1cm,clip,height=0.14\textheight]{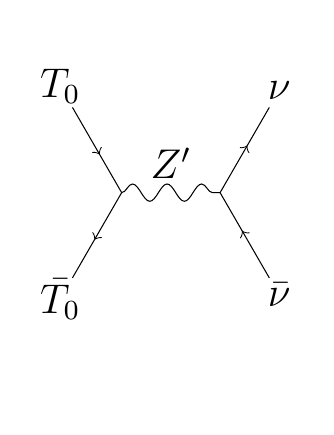}
\caption{ DM annihilation into neutrinos for model $F_1$. We use the notation of Table~\ref{table:models}.}
\end{figure*}

In this model, the Dirac triplet interacts with a heavier neutral vector boson $Z'$. The simplest realization of such a scenario assumes that  the $Z'$ is the gauge boson of a $U(1)'$ symmetry, and that it acquires its mass $m_{Z'}$ by means of the spontaneous breaking of the $U(1)'$ symmetry\footnote{In order to protect the Dirac nature of the DM field, we assume a charge assignment for the fields such that a Majorana mass term does not arise after symmetry breaking.} or via the Stueckelberg mechanism. Thus in model $F_1$ there is a direct link between the mediator of the annihilation into neutrinos and the Dirac nature of the DM multiplet. The coupling of the $Z'$ boson to the DM and lepton fields is given by
\begin{eqnarray}
{\cal L}_{Z'} \supset g_D Z'_\mu 
\left(\overline{\psi}\gamma^\mu\psi + Q \overline{L_\alpha}\gamma^\mu L_\alpha \right)\,.
\label{eq:LF1}
\end{eqnarray}
With this, the leptonic annihilation matrices are 
\begin{eqnarray}
\label{eq:GammaF1Gaugev2}
\Gamma^{S=1}_{\nu_\alpha\bar{\nu}_\beta} &=&\frac{\pi \delta_{\alpha\beta}}{24 m_\DM^2}\left(
 \alpha_2^2
\begin{psmallmatrix}
1 & 0 & -1 \\
0 & 0 & 0 \\
 -1 & 0 & 1
\end{psmallmatrix}
+
(2 Q)^2 \alpha_D^2 {\cal A} 
\begin{psmallmatrix}
1 & 1 & 1\\
 1 & 1 & 1\\
1 & 1 & 1
\end{psmallmatrix}
+
(2 Q) \alpha_D \alpha_2\left(1-\tfrac{m_{Z'}^2}{4m_\DM^2}\right){\cal A}
\begin{psmallmatrix}
2 & 1 & 0\\
 1 & 0 & -1\\
0 & -1 & -2
\end{psmallmatrix}
\right) \nonumber\\\
\\
\label{eq:GammaF1Gaugev3}
\Gamma^{S=1}_{\ell_\alpha\bar{\ell}_\beta} &=&\frac{\pi \delta_{\alpha\beta}}{24 m_\DM^2}\left(
 \alpha_2^2 
\begin{psmallmatrix}
1 & 0 & -1 \\
0 & 0 & 0 \\
 -1 & 0 & 1
\end{psmallmatrix}
+
(2 Q)^2 \alpha_D^2 {\cal A} 
\begin{psmallmatrix}
1 & 1 & 1\\
 1 & 1 & 1\\
1 & 1 & 1
\end{psmallmatrix}
-
(2 Q) \alpha_D \alpha_2\left(1-\tfrac{m_{Z'}^2}{4m_\DM^2}\right){\cal A}
\begin{psmallmatrix}
2 & 1 & 0\\
 1 & 0 & -1\\
0 & -1 & -2
\end{psmallmatrix}
\right) \nonumber\\\,
\end{eqnarray}
with $\alpha_D = g_D^2/4\pi$ and  ${\cal A} =1/\left(\left(1 - \tfrac{m_{Z'}^2}{4m_\DM^2}\right)^2+ \tfrac{\Gamma_{Z'}^2 m_{Z'}^2}{16m_\DM^4}\right)$. Here, we neglect the interference terms proportional to the width of the $Z'$ boson because they are higher order in $\alpha_D$.
The cross sections are then
\begin{equation}
\sigma v \left(\psi^0 \overline{\psi^0}\to \nu_\alpha\bar{\nu}_\beta\right) =
\sigma v \left(\psi^0 \overline{\psi^0}\to \ell_\alpha\bar{\ell}_\beta\right) =
 \sigma v_0 |d_0 +2d_+|^2 \text{Br}_{\alpha\beta}\,,
\end{equation}
with
\begin{align}
\sigma v_0= \frac{3 Q^2\pi \alpha_D^2  }{2 m_\DM^2}{\cal A} \,&&\text{and}&&\text{Br}_{\alpha\beta}\bigg|_{F_1} = \frac{1}{3}  \delta_{\alpha \beta}\,.
\label{eq:BrF1}
\end{align}

The calculation of the annihilation cross section today is analogous to the one of the previous scenario. We first determine  $\alpha_D Q$ in the early universe by means of the relic density constraint, see Eq.~$\eqref{eq:relicF1}$. Then, we calculate $\sigma v$ today in the  Milky Way and in dwarf galaxies taking into account the Sommerfeld effect. The results are shown in Fig.~\ref{fig:nuFluxF2}.

\subsection{Models $\SSS_1$  and $\FF_1$  }
\label{model:SSS1}

\begin{figure*}[h]
\includegraphics[trim=0cm 0.9cm 0cm 0.1cm,clip,height=0.14\textheight]{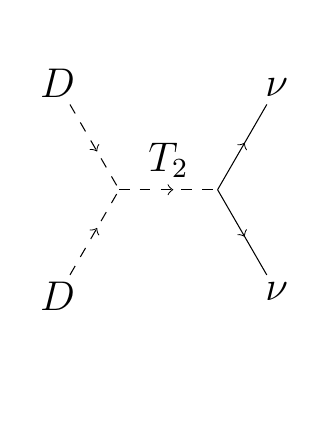}
\caption{ DM annihilation into neutrinos for model $\SSS_1$. We use the notation of Table~\ref{table:models}. For model $\FF_1$, the diagram is the same except that the DM particle is fermionic.}
\end{figure*} 
\FloatBarrier

As explained above, besides $F_{1}$ and $F_{2}$, there exist two other non-excluded models where DM has non-trivial electroweak quantum numbers (even if fine-tuned because they require cancellations to take place in the neutrino masses).
These are  $\SSS_1$  and $\FF_1$.
In the scenario $\SSS_1$, considered before in Refs.~\cite{Lindner:2010rr,Farzan:2011ck},
the DM belongs to a scalar inert doublet $\phi_D$ with the following relevant interactions 

\begin{equation}
{\cal L} = \mu\, \phi_D \phi_{T_2} \phi_D- Y^L_{\alpha\beta}\,\overline{L_\alpha} \phi_{T_2}L_\beta^c+ h.c.\,.\\
\label{eq:LSm7Fm7}
\end{equation}
The $Y^L$ coupling gives a type-II seesaw contribution to the neutrino masses equal to $m^\nu_{\alpha\beta}=Y^L_{\alpha\beta} v_T$, where
$v_{T}$ is the vev of the triplet (see Fig.~\ref{fig:diagram-tree}). Thus, if neutrino masses only stem from this source, the Yukawa coupling is given by 
$Y^L_{\alpha\beta}=m^\nu_{\alpha\beta}/v_T= \sum_i m_i U_{\alpha i}U_{\beta i}/v_T $, where $m_i$ are the neutrino masses, and $U$ the  Pontecorvo--Maki--Nakagawa--Sakata mixing matrix. 

For simplicity, we neglect the contribution of electroweak interactions to the annihilation matrices. (For doublet representations, such a contribution can only account for the relic density below the $\sim$ TeV scale, see .e.g.~\cite{Hambye:2009pw,ArkaniHamed:2006mb}).  The annihilation matrices take then the simple form
\begin{eqnarray}
\label{eq:GammaSm7Fm7}
\sum_{\alpha\beta}
\Gamma^{S=1}_{\nu_\alpha\nu_\beta} =
\frac{1}{2}\sigma v_0\,
\begin{psmallmatrix}
1&-1&0\\ 
-1&1&0 \\
0&0&0
\end{psmallmatrix} \,,
\end{eqnarray}
where~\cite{Lindner:2010rr}
\begin{equation}
\sigma v_0 =\frac{1}{8\pi}\frac{\mu^2\, \hbox{Tr}(Y^L Y^{L\dagger})}{\left(4m_\DM^2-m_T^2\right)^2 +\Gamma_T^2 m_T^2}
=\frac{1}{8\pi v_T^2}\frac{\mu^2\sum_{i}m_{i}^2}{\left(4m_\DM^2-m_T^2\right)^2 +\Gamma_T^2 m_T^2} 
\end{equation}
where the second equalities hold for the case where the $Y^L$ interactions are the only source of neutrino masses.

From this, it is clear that having a significant flux of neutrinos from DM annihilations into neutrinos is only possible when  $\mu\gg v_{T}$, as mentioned in footnote~\ref{footnote:6}. 
The total annihilation cross section into neutrinos is 
\begin{equation}
\sum_{\alpha\beta}\sigma v \left(H^0 H^0\to \nu_\alpha\nu_\beta\right) = \sigma v_0 |d_1-d_2|^2   \,. 
\end{equation}
Notice that, when the $Y^L$ interactions are the only source of neutrino masses, the monochromatic neutrinos are produced as mass eigenstates with branching ratios given by
\begin{equation}
\text{Br}_{ij}=\frac{m_i^2}{\sum_{i'} m_{i'}^2} \delta_{ij}, \hspace{1cm}i,j=1,2,3\,.
\label{eq:BrSr7}
\end{equation}
 
By carefully taking into account co-annihilations (see Appendix~\ref{sec:AppSEIDM})
we find $\sigma v_0\approx\frac{4}{3}\times\left(\unit[2.35\times 10^{-26}]{cm^3/s}\right)$ .
The total annihilation cross section into neutrinos is presented in Fig.~\ref{fig:nuFluxSm7Fm7}. There, we also show the corresponding cross sections into SM gauge bosons, which we obtain by  following Ref.~\cite{Garcia-Cely:2015khw}. Note that, unlike for the fermion DM models above, for the scalar model  $\SSS_1$, the mass splittings are not fixed by electroweak radiative corrections but are induced at  tree level by quartic scalar couplings whose value are not fixed. Accordingly, the results of Fig.~\ref{fig:nuFluxSm7Fm7} are  for two representative choices of mass splittings. In fact, to further illustrate the dependence with the mass splittings, we vary them and show the corresponding mass value at the first Sommerfeld peak in Fig.~\ref{fig:picks}. 

\label{model:FF1}
Model $\FF_1$ is  similar to model $\SSS_1$. Since in both cases DM belongs to a doublet, the Sommerfeld factors are identical for the same  mass differences among the particles in the doublet. In particular, the position of the first peak as a function of both mass splittings is the same as in Fig.~\ref{fig:picks}.  Nevertheless, as explained in Section \ref{sec:classification}, 
for fermionic DM the splittings necessarily arise from radiative corrections and/or non-renormalizable interactions. As a result, even though the neutrino cross section is very similar for models $\FF_1$ and  $\SSS_1$ (the left panel of Fig.~\ref{fig:nuFluxSm7Fm7}), the cosmic-ray counterpart (the right panel of Fig.~\ref{fig:nuFluxSm7Fm7}) is different. We will not elaborate further on that.

\newpage
\subsection{Discussion of the results and cosmic-ray constraints }
\label{sec:results}
\begin{figure*}[t]
\centering{Model $F_2$}\\
\includegraphics[width=0.495\textwidth]{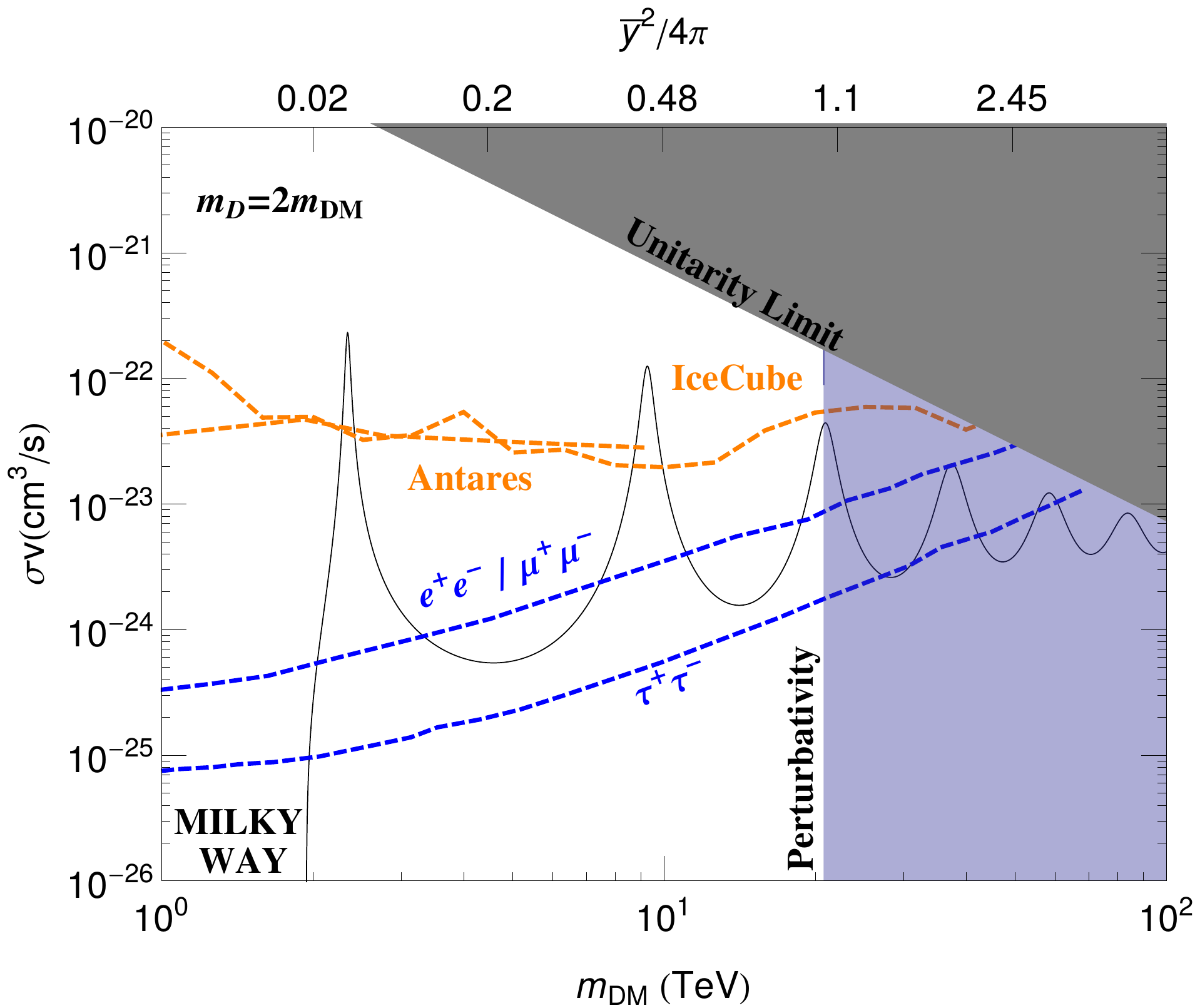}
\includegraphics[width=0.495\textwidth]{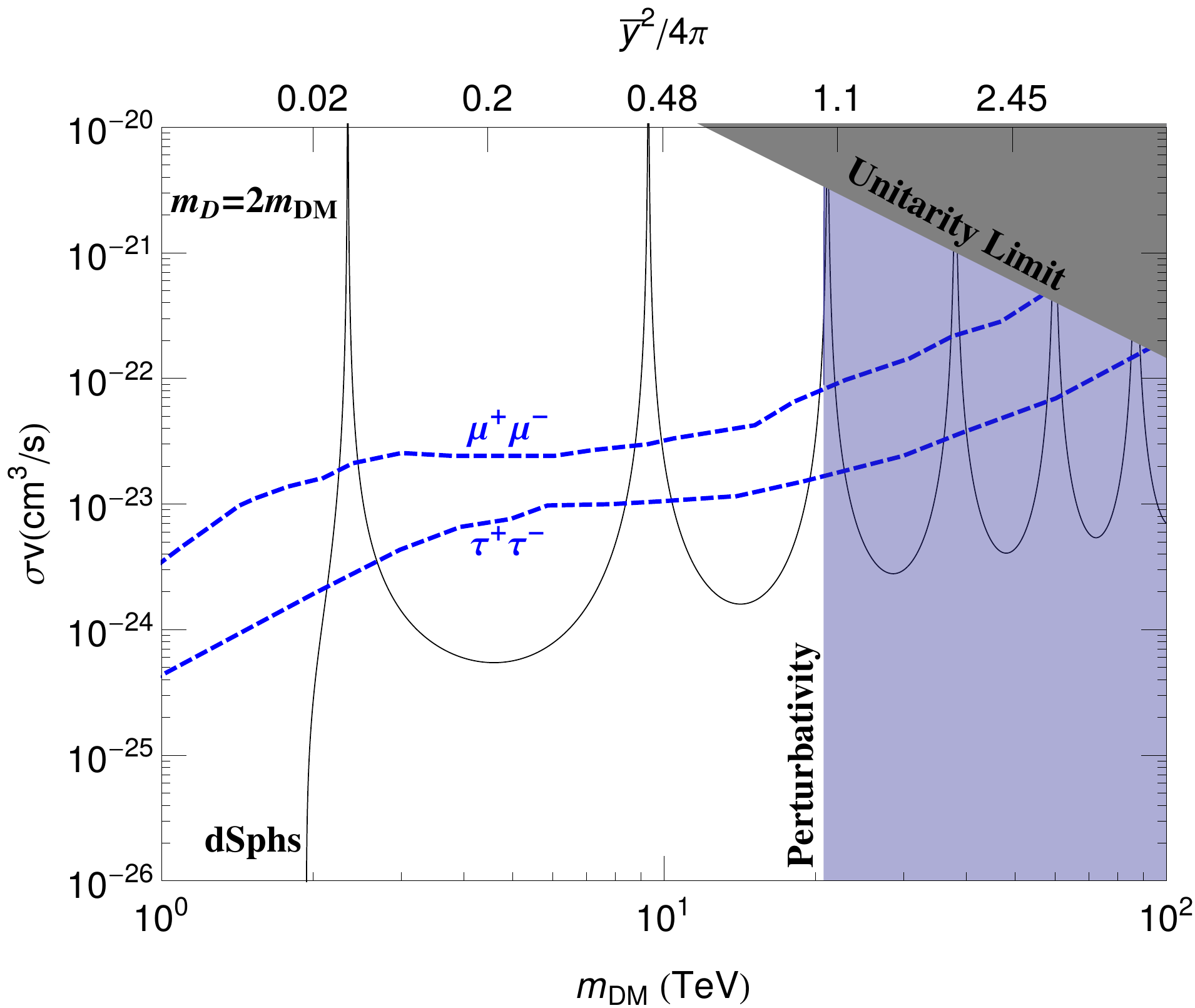}\\
\centering{\vspace{1cm}Model $F_1$}\\
\includegraphics[width=0.495\textwidth]{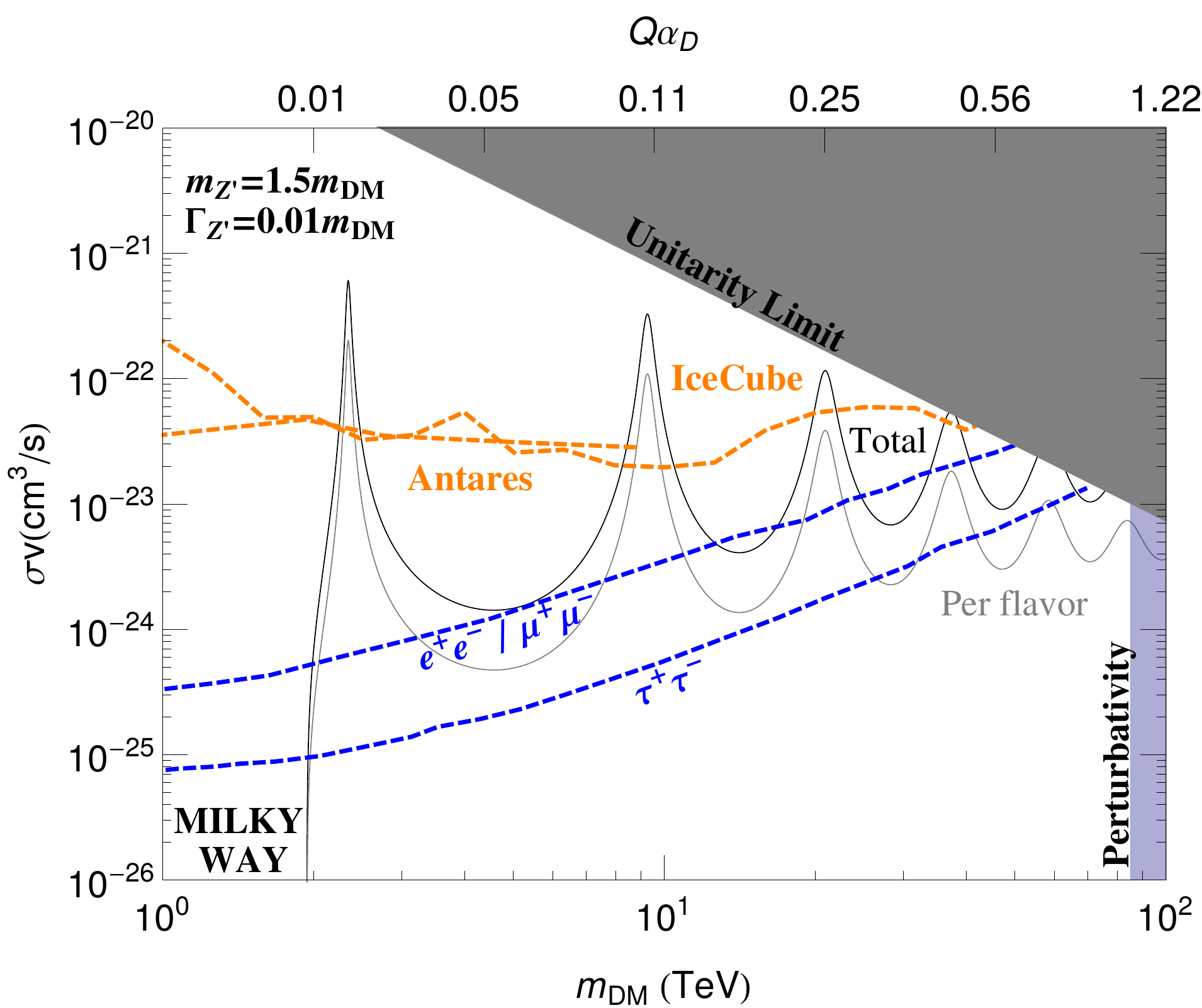}
\includegraphics[width=0.495\textwidth]{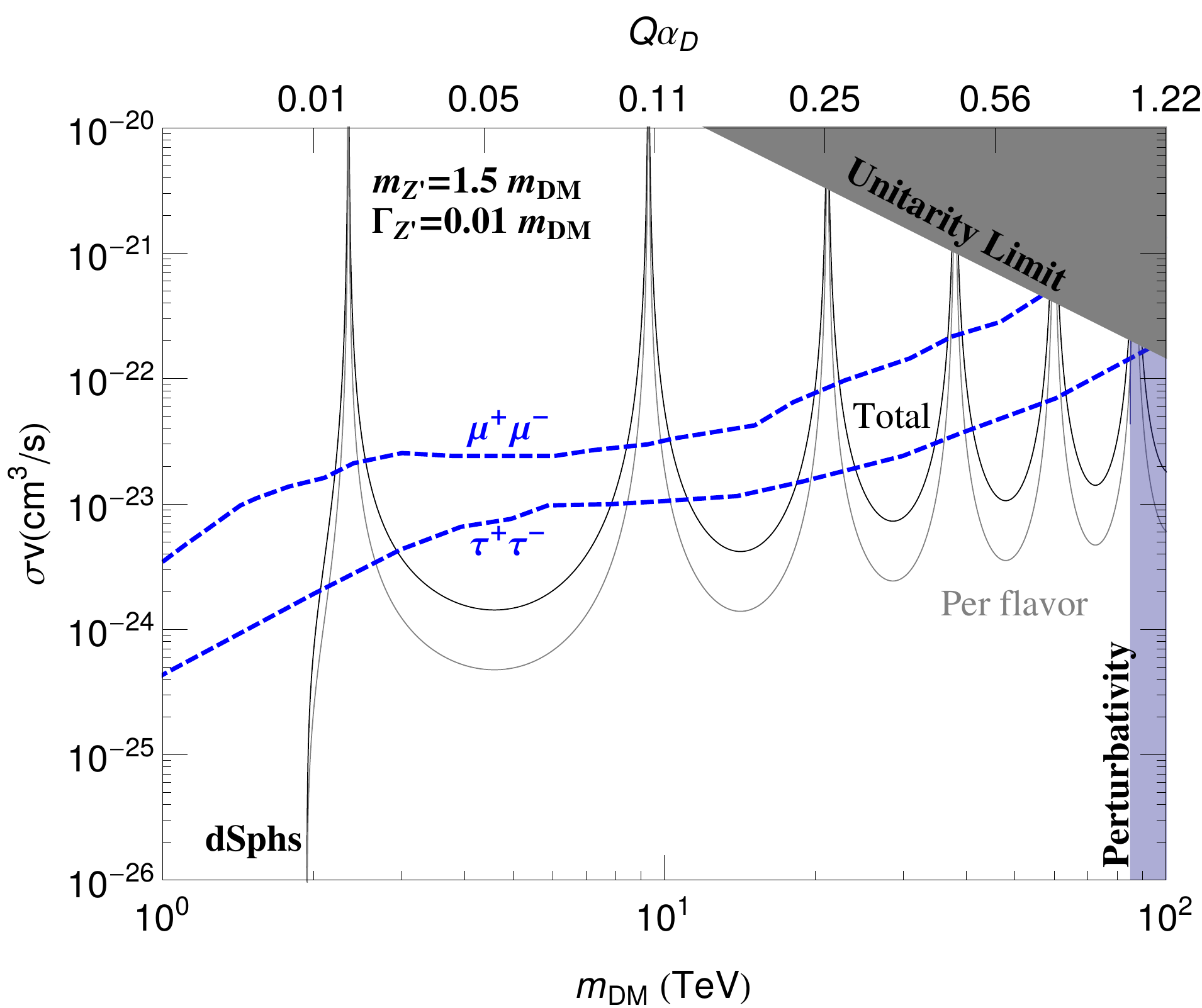}\\
\caption{The DM annihilation cross section into neutrinos of all flavors (black) for models $F_2$ (top) and $F_1$ (bottom) as a function of the DM mass. For these models, the cross sections into charged lepton are given by the same curves.  For model $F_1$, we also show the cross section into only one flavor (gray). 
{The parameters indicated in the upper left corners have been used to calculate the coupling (top axis) that leads to the observed relic density according to Eq.~\eqref{eq:relicF1} or \eqref{eq:relicF2}.}
The  purple region corresponds to  $\overline{y}^2> 4 \pi$ or $Q \alpha_D>1$.  \emph{Left panel:}  annihilations in the Milky Way, assuming a DM relative velocity of $v\approx 2\times10^{-3} c$~\cite{Bertone:2010zza}. We show the most stringent   limits  from IceCube and Antares data (see Fig.~\ref{fig:dNdxline}).   The limits for charged leptons are  from HESS~\cite{Abdallah:2016ygi} and correspond to ten years of observation of the galactic center. All the cross sections and experimental limits are given for a NFW profile. The unitarity limit is the classical result from Ref.~\cite{Griest:1989wd}. \emph{Right panel:} Same as the left panel, but for dwarf galaxies for which we assume $v\approx10^{-5} c$~\cite{Bertone:2010zza}. The limits on charged leptons consist of a combination of MAGIC and Fermi-LAT observations of dwarf galaxies~\cite{Ahnen:2016qkx} . All the limits in this figure have been rescaled to account for the fact that we are considering Dirac DM instead of the usual self-conjugate DM.} 
\label{fig:nuFluxF2}
\end{figure*}

\begin{itemize}
\item \emph{Models $F_1$ and $F_2$:}
As explained above, the neutrino cross sections for these models are totally fixed by the freeze-out through the annihilation into neutrinos and the fact that the Sommerfeld enhancement is fixed by electroweak physics for Dirac triplets. The results are given in Fig.~\ref{fig:nuFluxF2}. The cross section predicted by model $F_1$ is roughly 2.6 times that predicted by $F_2$ because of co-annihilation effects (this follows from Eqs.~\eqref{eq:relicF2} and ~\eqref{eq:relicF1}). As expected, below $\unit[2]{TeV}$ one can not expect an observable neutrino flux because in this case the electroweak annihilation is too fast, which leads to a relic density smaller than 26\%. Above $\unit[2]{TeV}$ instead, both figures show that the Sommerfeld enhancement is always large, even between the Sommerfeld peaks. The peaks show up at 
\begin{equation}
m_\DM=2.3\,,\,\,\,9.3\,,\,\,\, 21.0\,,\,\,\, 37.4\,,\,\,\, 58.4\text{   and }\unit[84.7]{TeV}\,.
\end{equation}
Between these peaks, the minima are obtained at
\begin{equation}
m_\DM=4.8\,,\,\,\,14.1\,,\,\,\,28.3\,,\,\,\,47.2\,,\,\,\,70.7\text{   and }\unit[98.4]{TeV}\,.
\end{equation}
At these minima the Sommerfeld-enhanced cross section regularly increases and approaches the values of $\unit[3.6\times10^{-24}]{cm^3/s}$ and $\unit[9.3\times10^{-24}]{cm^3/s}$  for models $F_2$ and $F_1$, respectively. 
Even though both models share the same Sommerfeld boost factor, due to perturbativity, only $F_1$ can reach DM masses above $\sim\unit[25]{TeV}$. This is because the perturbativity constraint on the coupling giving rise to the relic density  is weaker for  processes taking place via the s-channel, even if the enhancement due to the s-channel resonance is small (compare Eq.~\eqref{eq:GammaF2Gaugev2} and Eq.~\eqref{eq:GammaF1Gaugev2}).

Unless one lies around a peak, the cross sections reported in Fig.~\ref{fig:nuFluxF2}  are typically below the present neutrino-telescope limits. In fact, these cross sections could reasonably be probed in the future by them.

However,  one must still check whether these two models do not lead to too large fluxes of cosmic rays. The strongest cosmic-ray constraint is associated to charged leptons  whose cross section,  as mentioned in Table~\ref{table:models},
equals that of neutrinos. One can therefore directly compare the cross section obtained for neutrinos with the best existing limits on the production of charged leptons, which come from HESS~\cite{Abdallah:2016ygi}---for the galactic center---and a joint analysis by MAGIC and Fermi-LAT~\cite{Ahnen:2016qkx}, for dwarf galaxies.

Fig.~\ref{fig:nuFluxF2} shows that for DM masses below $\unit[25]{TeV}$, in model $F_2$ the Sommerfeld enhancement always leads to a total cross section in charged leptons larger than the current limit on taus. Therefore, in order to avoid that bound,  annihilations must proceed into muons or electrons, which are only allowed around the dips located at $m_\DM\sim \unit[4.8]{TeV}$  and  $\unit[14.1]{TeV}$. This can be achieved  by an appropriate choice of the Yukawa couplings in Eq.~\eqref{eq:LF2}.  

In contrast, model $F_1$  predicts equal branching ratios into $e^+e^-$, $\mu^+\mu^-$ and $\tau^+\tau^-$. Hence, the limit on the $\tau^+\tau^-$ channel must be compared with  the total  cross section divided by 3. After accounting for this, one observes that the tau limit lies a factor 2.5 (1.5) below the predicted cross section around the first (second) dip. 
Nevertheless, the $\tau^+\tau^-$ limit lies slightly above the prediction for the next dips, which---below the regions excluded by  perturbativity and unitarity---are located around $27$~TeV, $47.2$~TeV and $70.7$~TeV. 

\FloatBarrier

\begin{figure*}[t]
\centering{Model $\SSS_1$}\\
\includegraphics[width=0.495\textwidth]{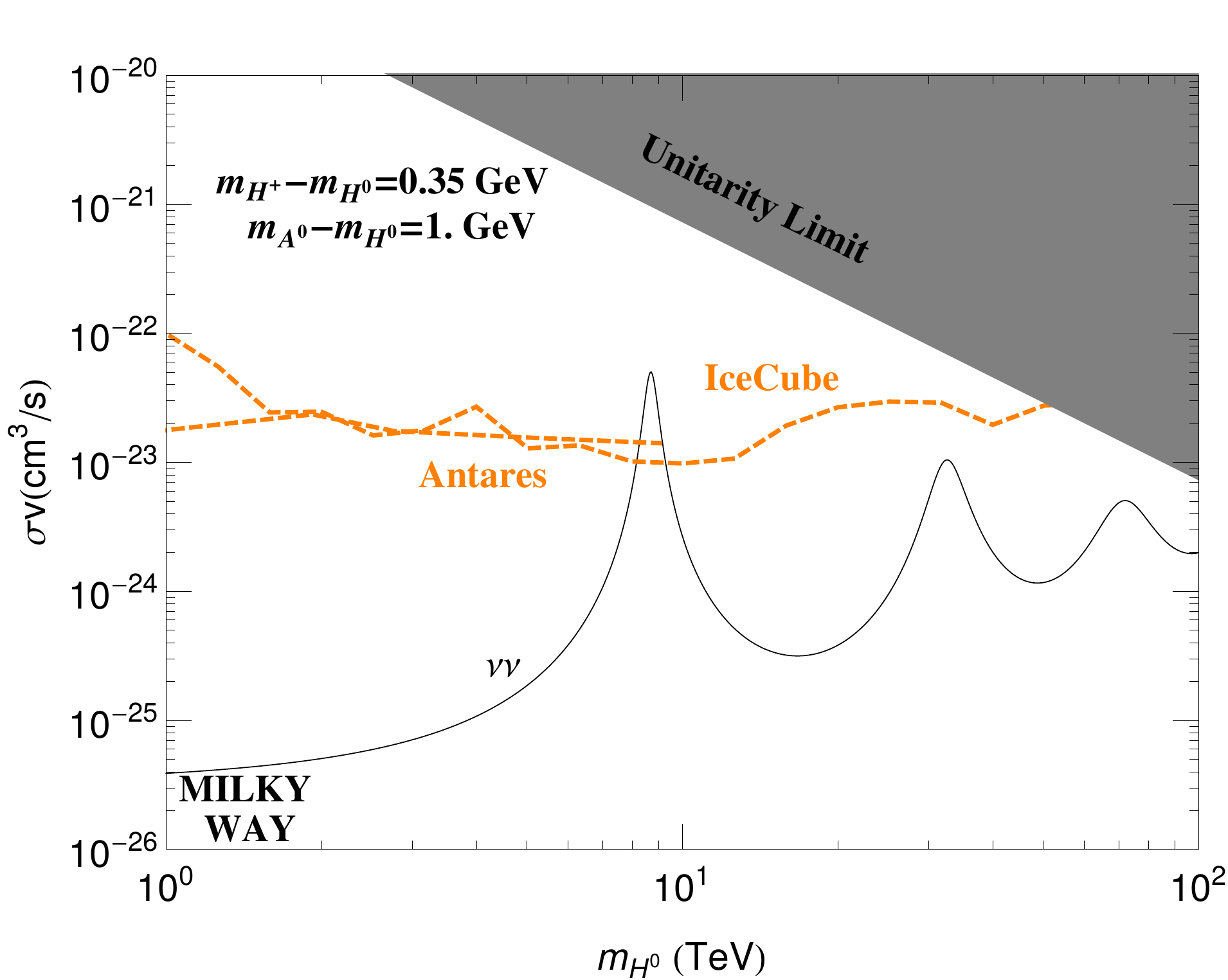}
\includegraphics[width=0.495\textwidth]{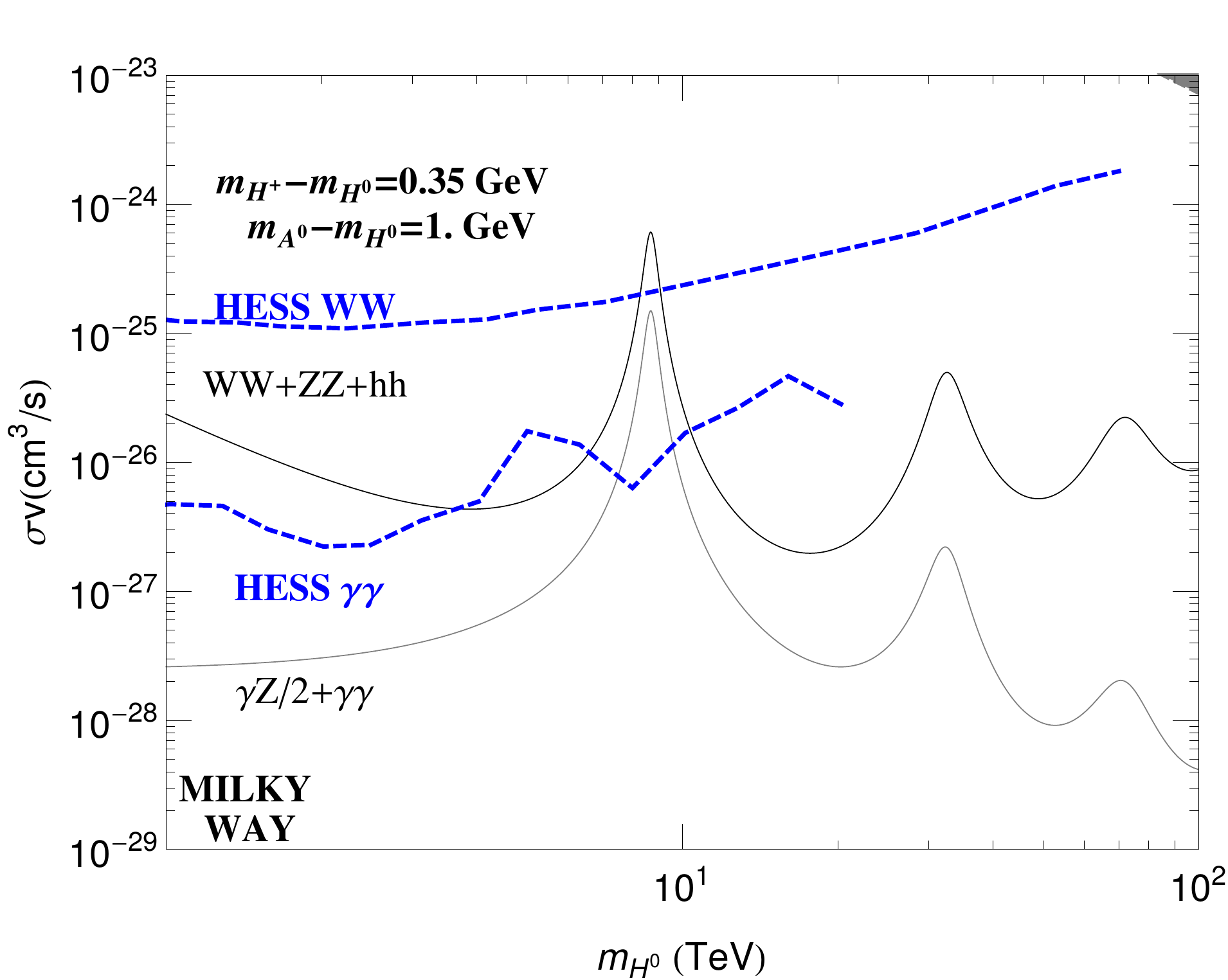}\\
\includegraphics[width=0.495\textwidth]{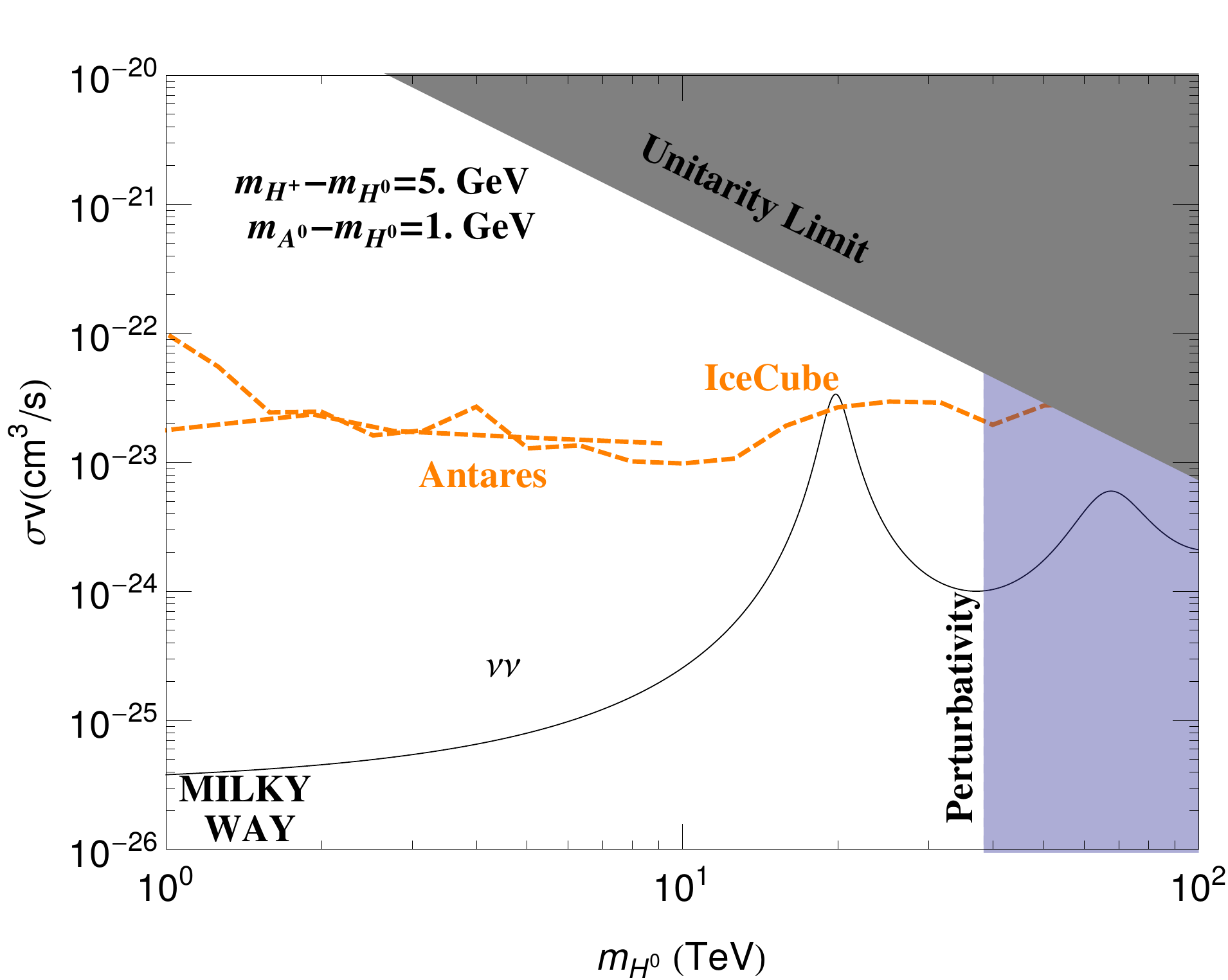}
\includegraphics[width=0.495\textwidth]{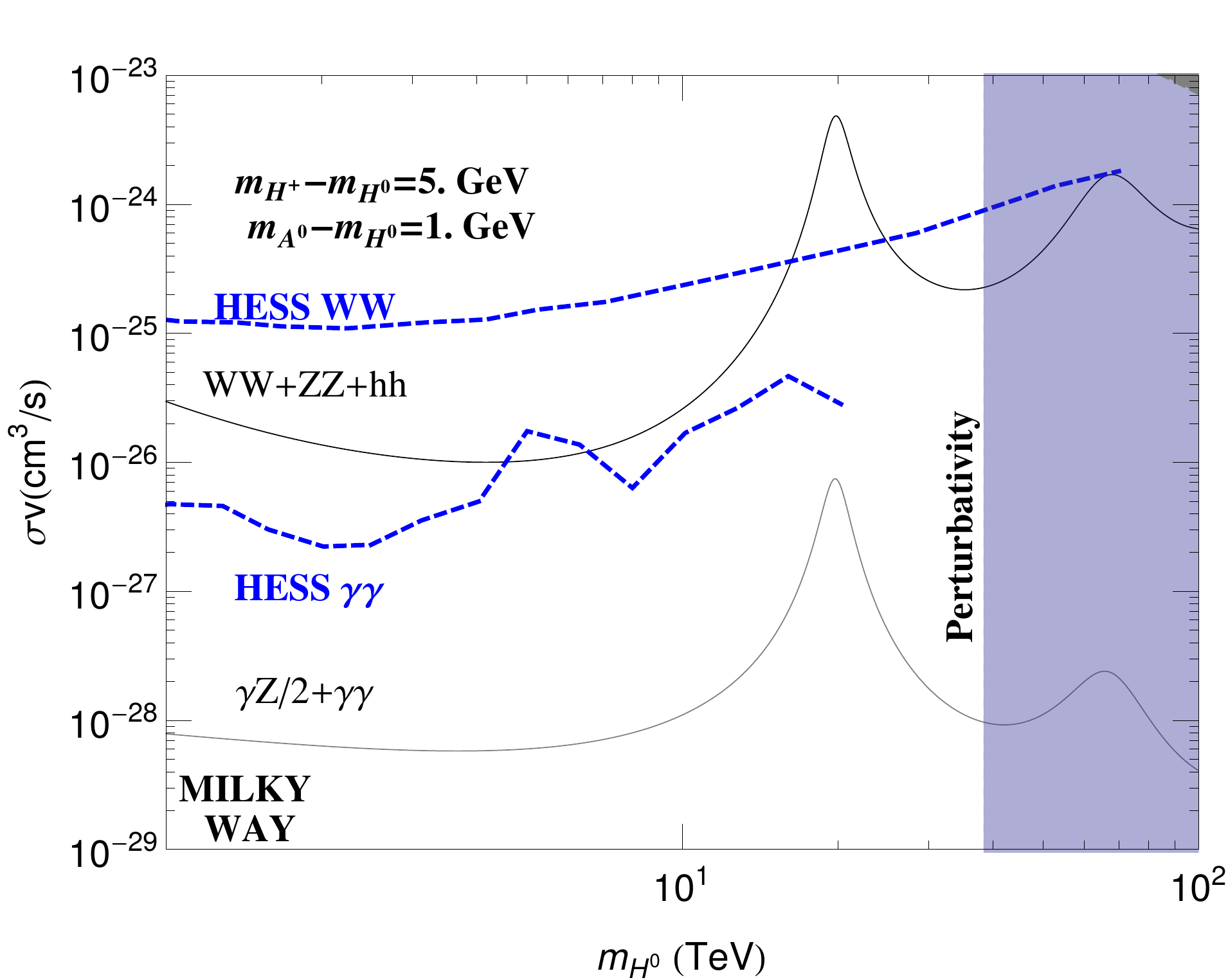}
\caption{\emph{Left panel:}
DM annihilation cross section into neutrinos of all flavors (black) as a function of $m_\DM$ for model $\SSS_1$.  For this model, there is \emph{no} annihilation into charged leptons.  We consider here annihilations in the Milky Way, assuming a DM relative velocity of $v\approx 2\times10^{-3} c$~\cite{Bertone:2010zza}. The IceCube and Antares limits are as in Fig.~\ref{fig:dNdxline}.
\emph{Right panel:} annihilation cross section into SM bosons for model $\SSS_1$.
Two  limits are shown, one from  Ref.~\cite{Abramowski:2013ax} which constrains the annihilation in monochromatic photons, and the other from Ref.~\cite{Abdallah:2016ygi} which constrains annihilation into a continuum of photons (assuming a WW final state). All curves in both panels are given for a NFW profile and for two representative choices of mass splittings between the components of the DM multiplet (see upper left corner). 
 In the purple region, the mass splittings are so large that perturbative quartic couplings can not generate them.
}
\label{fig:nuFluxSm7Fm7}
\end{figure*}

It must be stressed that our calculation of the Sommerfeld-enhanced cross sections  is expected to receive corrections from next-to-leading-order effects and Sudakov's logarithms  (see e.g. Ref.~\cite{Ovanesyan:2016vkk})  as well as from the finite width of the bound states determining the Sommerfeld peaks~\cite{Blum:2016nrz}. A precise determination of such corrections is beyond the scope of our work. However, judging by their impact on Wino DM,  we must mention they can be important and in some cases  our results might change by a factor of a few.  The reader should have this in mind  when interpreting
our results.

Thus, one concludes that for both models one could still have an observable neutrino line around all the dips.
This would require an improvement of the neutrino-line sensitivity by one order of magnitude  or more, depending on the dip under consideration. In all cases, the observation of an associated  flux of charged leptons should be around the corner. Alternatively,
an improvement of the limits\footnote{Preliminary limits from the HAWC collaboration on DM annihilations in dwarf galaxies further constrain the regions  around the Sommerfeld peaks  above 30 TeV\cite{Albert:2017vtb}.} on the production of charged leptons from DM annihilation by no more than a factor of a few would exclude these scenarios and thus definitely close the possibility of  observing a neutrino line. 

\begin{figure*}[t]
\includegraphics[width=0.7\textwidth]{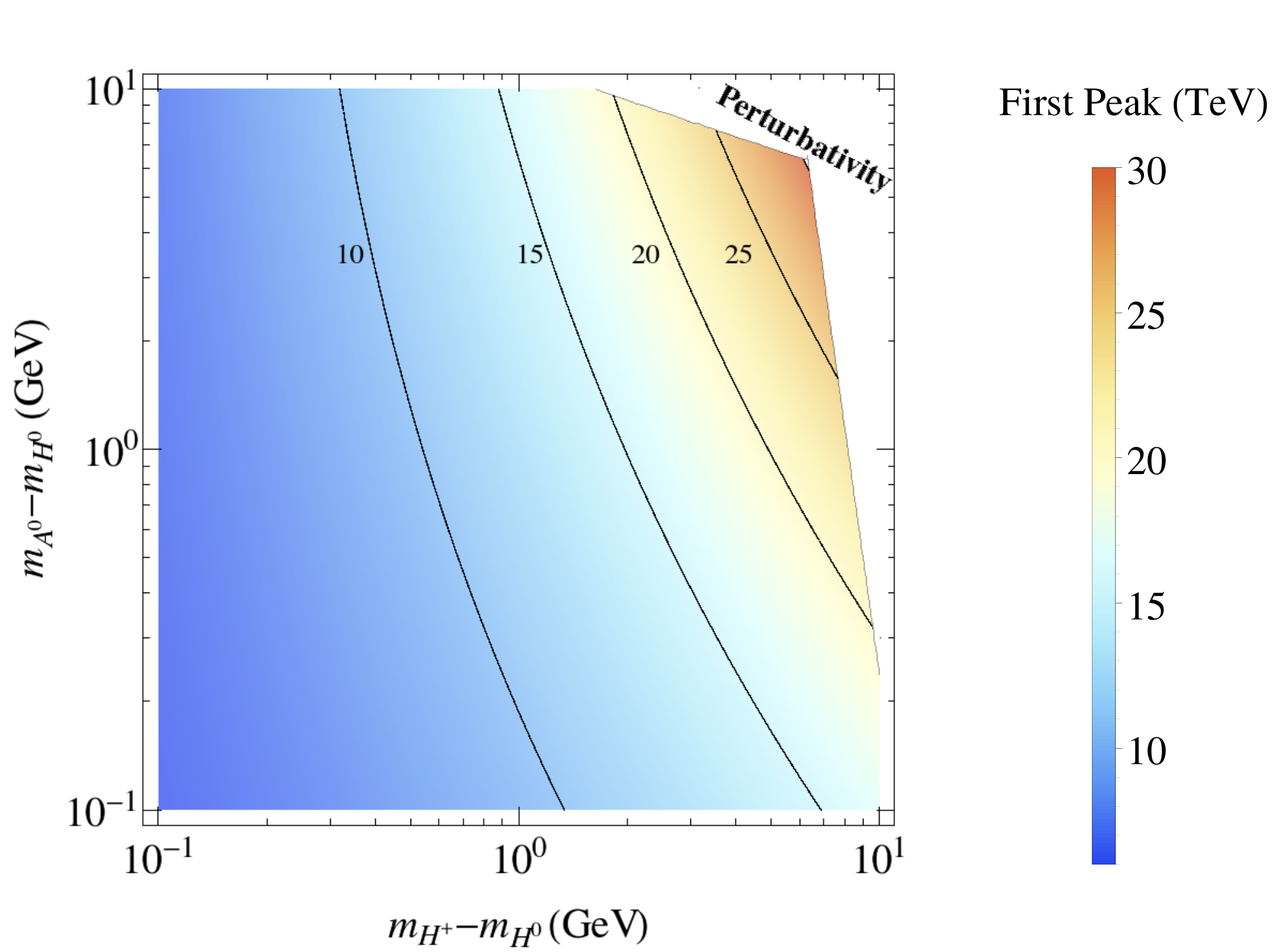}
\caption{Position of the first Sommerfeld peak for scalar doublet annihilations as a function of the mass splittings between its particles (model $S_1^r$). In the white region, the mass splitting are so large that perturbative quartic couplings can not generate them. 
}
\label{fig:picks}
\end{figure*}

\item \emph{Model $\SSS_1$:} Here the situation is different from the previous models in various ways. One difference is that  the neutrino signal has no charged-lepton counterpart. Never\-theless, the model predicts a gamma-ray flux which arises from DM annihilations into $WW$, $ZZ$ and $hh$---that produce a continuum photon spectra---as well as from DM annihilations into $Z\gamma$ and $\gamma\gamma$, that produce photon lines. These fluxes are severely constrained  by HESS observations of the galactic center~\cite{Abramowski:2013ax, Abdallah:2016ygi}. In particular, the limits on these channels exclude the neutrino signal around the first Sommerfeld peak, as illustrated in Fig.~\ref{fig:nuFluxSm7Fm7} for two different benchmarks. Details on the calculation of the cross sections are discussed in Appendix~\ref{sec:AppSEIDM}.

Another  aspect concerns  the masses of the scalar doublet particles. As already mentioned above, in contrast to Dirac triplets,  the mass splittings between the DM multiplet components are not fixed, but change as a function of  the quartic couplings (see Eq.~\eqref{eq:IDMmasses}). As a result, the position of the Sommerfeld peaks are  not fixed either. This is illustrated in Fig.~\ref{fig:picks}, where we show the position of the first peaks as a function of the mass splittings in the doublet. In particular, large mass splittings lead to a higher mass at the first Sommerfeld peak.

Note that the HESS limits become weaker for large masses and, for the case of lines, do not exist  above $\unit[20]{TeV}$. Thus, to avoid the gamma-ray constraints that exclude the neutrino signals around the first peak, one could think about considering large mass splittings, since in this case the first peak is located at a larger DM mass.  Nonetheless, this is not possible for two reasons. First, the mass splittings are bounded from above by the perturbativity condition on the quartic couplings. For instance, Eq.~\eqref{eq:DD_IDM} gives  $(m_{A^0}-m_{H^0}) <  |\lambda_5| v_\text{EW}^2/2m_{H^0}$. The area which is not accessible perturbatively is shown in blank in Fig.~\ref{fig:picks} and in purple in the bottom panel of Fig.~\ref{fig:nuFluxSm7Fm7}. Second, large splittings require large quartic couplings for  high DM  masses, which  induce fast annihilations into $WW$, $ZZ$ and $hh$. Again, this is shown in the bottom-right panel of Fig.~\ref{fig:nuFluxSm7Fm7}, where a mass splitting between the DM and the charged components of $\unit[5]{GeV}$ leads to relatively large annihilation cross sections into SM gauge bosons. Having mentioned this, it must be clear that a neutrino signal is excluded around first peak, no matter the mass splittings considered. 

The previous discussion suggests to consider small mass splittings and to study the neutrino signals around the second or the third  peaks. Numerically, we find the following approximated expressions for the position of the second and third peaks
\begin{align}
m_2 = 3.7 m_1&&\text{and}&& m_3 = 8.3 m_1\,,
\end{align}
where $m_n$ is the position of the n$^\text{th}$ peak. For instance, for the second Sommerfeld peak in the top right panel of Fig.~\ref{fig:nuFluxSm7Fm7}, the constraints from gamma-ray observations are not strong and an improvement of one order of magnitude by IceCube could probe the 
corresponding  neutrino signal in the near future. 

\end{itemize}

\FloatBarrier

\section{Models with an additional mediator inducing a Sommerfeld Enhancement}
\label{sec:SEnoEW}

\subsection{Models $F_3$ and $F_4$ }
\label{model:F3F4}

\begin{figure*}[h]
\includegraphics[trim=0cm 0.9cm 0cm 0.1cm,clip,height=0.14\textheight]{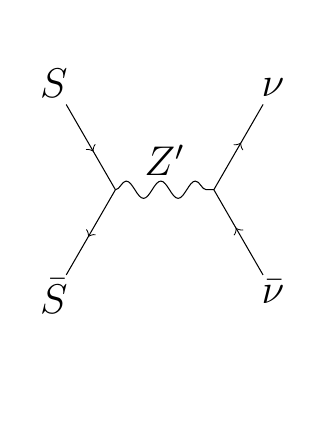}
\includegraphics[trim=0cm 0.0cm 0cm 0.0cm,clip,height=0.14\textheight]{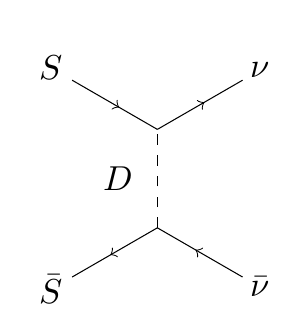}
\caption{ DM annihilation into neutrinos for model $F_3$ (left) and model $F_4$ (right). We use the notation of Table~\ref{table:models}.}. 
\end{figure*} 

Model $F_3$ consists of a DM Dirac singlet $\psi$ coupled to a $Z'$ mediator, see Eq.~\eqref{eq:LF1}. Model $F_4$ consists of a DM Dirac singlet $\psi$  coupled to a doublet mediator $\varphi_D$, see  Eq.~\eqref{eq:LF2}. A   Sommerfeld enhancement requires another mediator with a mass much smaller than $m_\DM$ . This can most easily be done by introducing a real scalar particle $\phi$ and coupling it to the DM by means of a Lagrangian
\begin{equation}
{\cal L} =  (4\pi \alpha_\text{SE})^{1/2} \,\phi\,\overline{\psi} \psi \,.
\end{equation}

 In that case, the additional annihilation channel $\overline{\text{DM}} \text{DM} \to \phi \phi$ is p-wave suppressed\footnote{This is simply angular momentum and CP conservation. On the one hand, the final state composed of two identical scalars has even total angular momentum and CP$=1$.  On the other hand, the initial state has  CP=$(-1)^{S+1}$, which implies that the total spin of the Dirac particles is $S=1$ and consequently that the initial orbital angular momentum must be odd, so that the total angular momentum is even.}. Consequently,  the freeze-out process in the early universe is determined, to a very good approximation, by DM annihilations into neutrinos and charged leptons. This leads to the following cross sections in DM halos

\begin{equation}
\sigma v \left(\DM \overline{\DM}\to \nu_\alpha\bar{\nu}_\beta\right) =
\sigma v \left(\DM \overline{\DM}\to \ell_\alpha\bar{\ell}_\beta\right) \approx
  \text{BF}\,\times \text{Br}_{\alpha\beta} \,\left(\unit[2.35\times10^{-26}]{cm^3/s}\right)\,,
\label{eq:relicF3F4}
\end{equation}
where
\begin{align}
\label{eq:BrF3F4}
\text{Br}_{\alpha\beta}\bigg|_{F_3} &= \frac{1}{3}  \delta_{\alpha \beta}\,,&\text{Br}_{\alpha\beta}\bigg|_{F_4} &= \frac{|y_\alpha|^2|y_\beta|^2}{\left(\sum_{\alpha'}|y_{\alpha'}|^2\right)^2}\,,
\end{align}
and with BF being the boost factor due to Sommerfeld enhancement. Notice that  for these models, there are no co-annihilations and the Sommerfeld effect  in the early universe is small~\cite{Tulin:2013teo}. Consequently, the total annihilation cross section at freeze-out is roughly $\unit[4.7\times10^{-26}]{cm^3/s}$~\cite{Steigman:2012nb} . In addition, we calculate BF by using the analytic expressions obtained for the Hulth\'en potential~\cite{Cassel:2009wt}, which approximates the Yukawa potential induced by the scalar exchange  reasonably well. 

\begin{figure*}[t]
\centering{Model $F_4$ }\\
\includegraphics[width=0.495\textwidth]{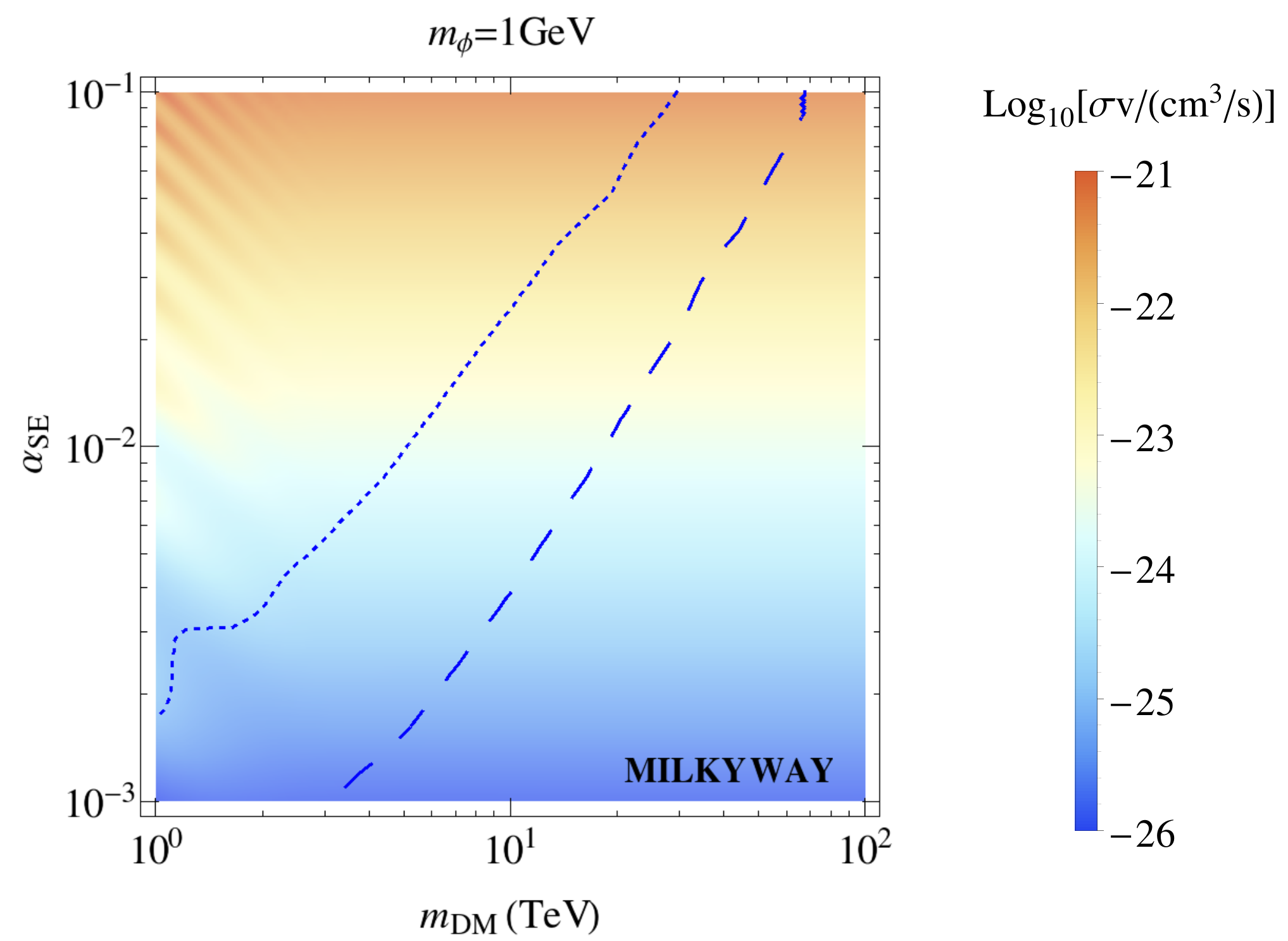}
\includegraphics[width=0.495\textwidth]{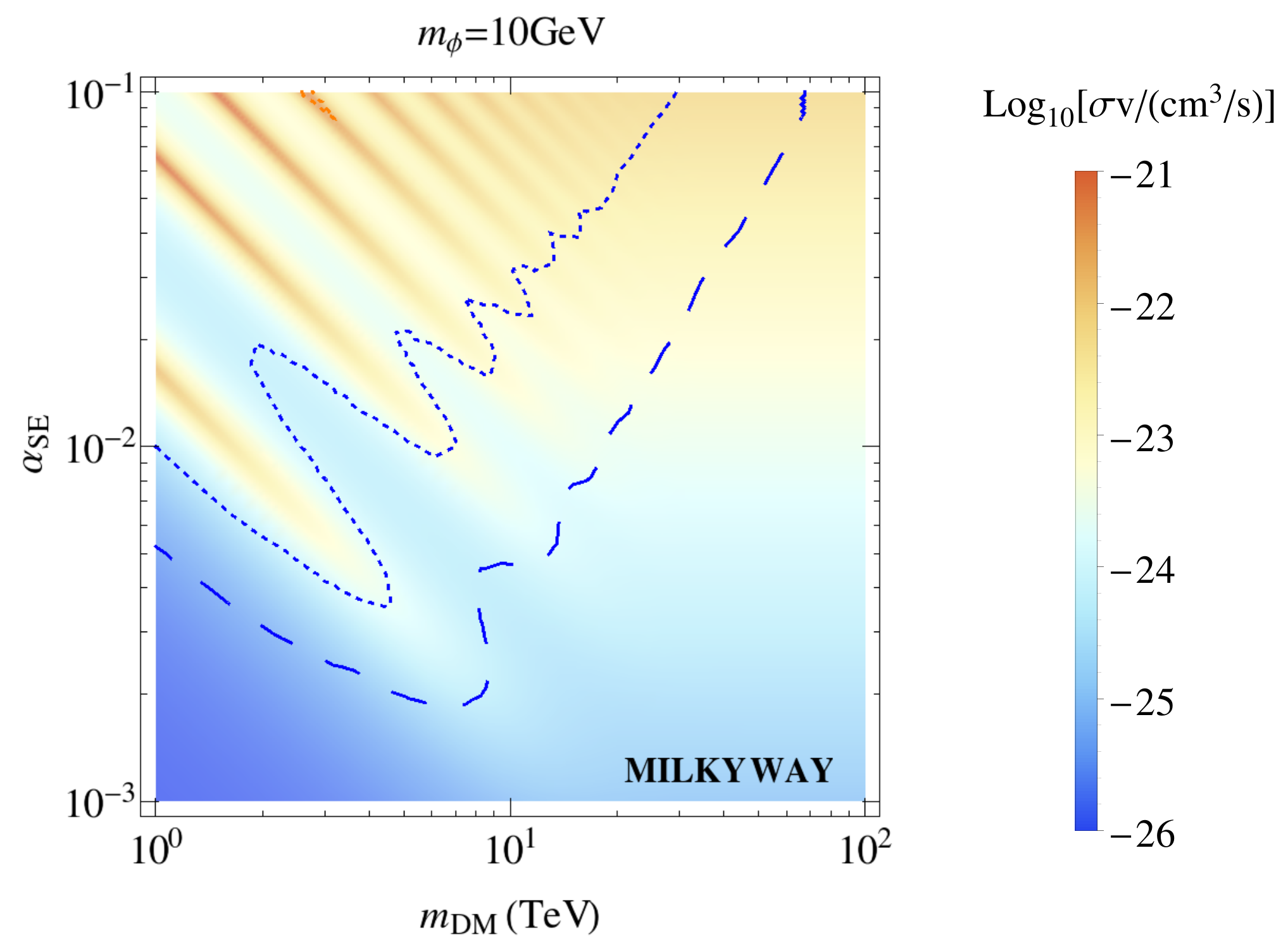}\\
\includegraphics[width=0.495\textwidth]{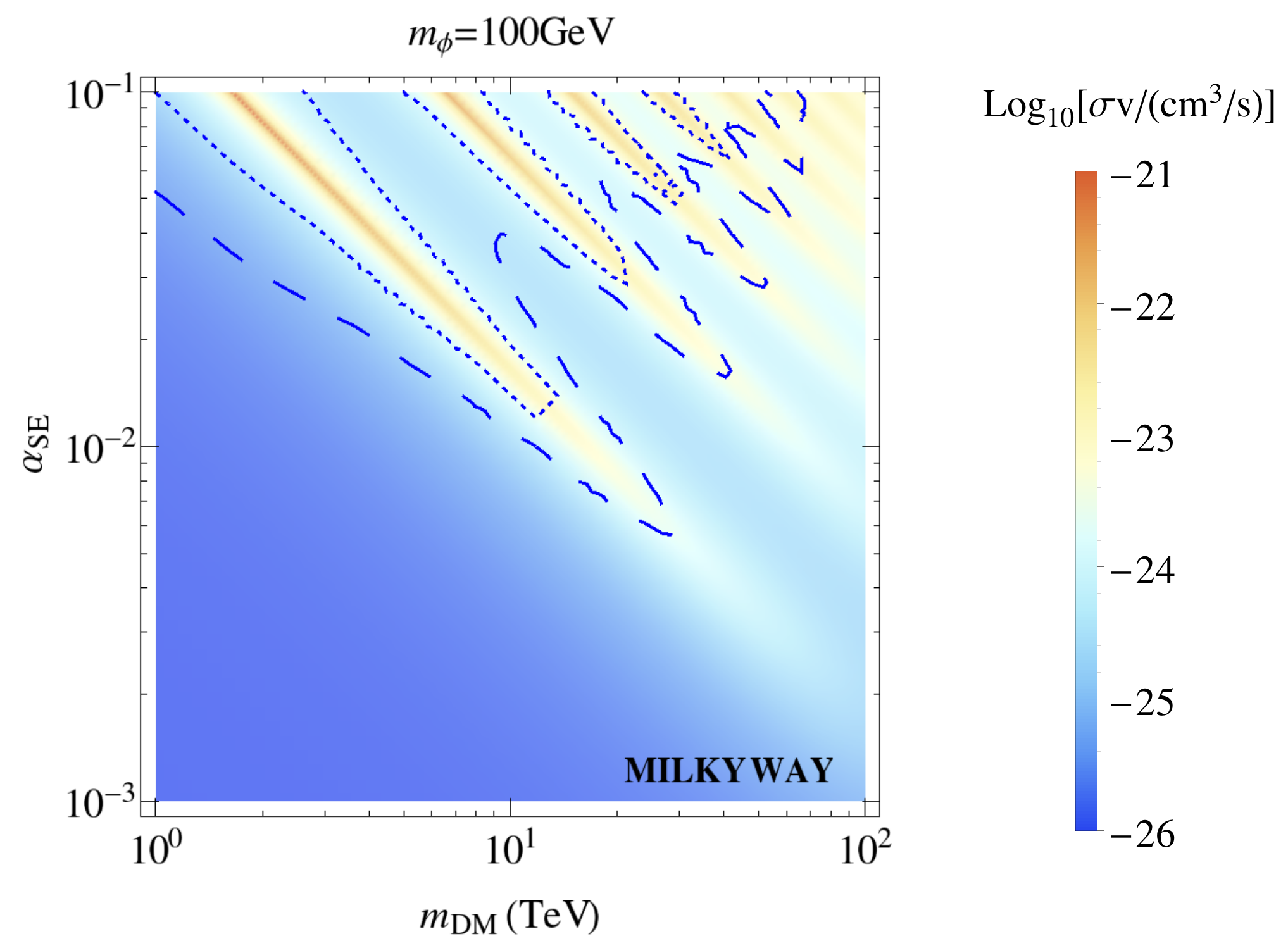}
\includegraphics[width=0.495\textwidth]{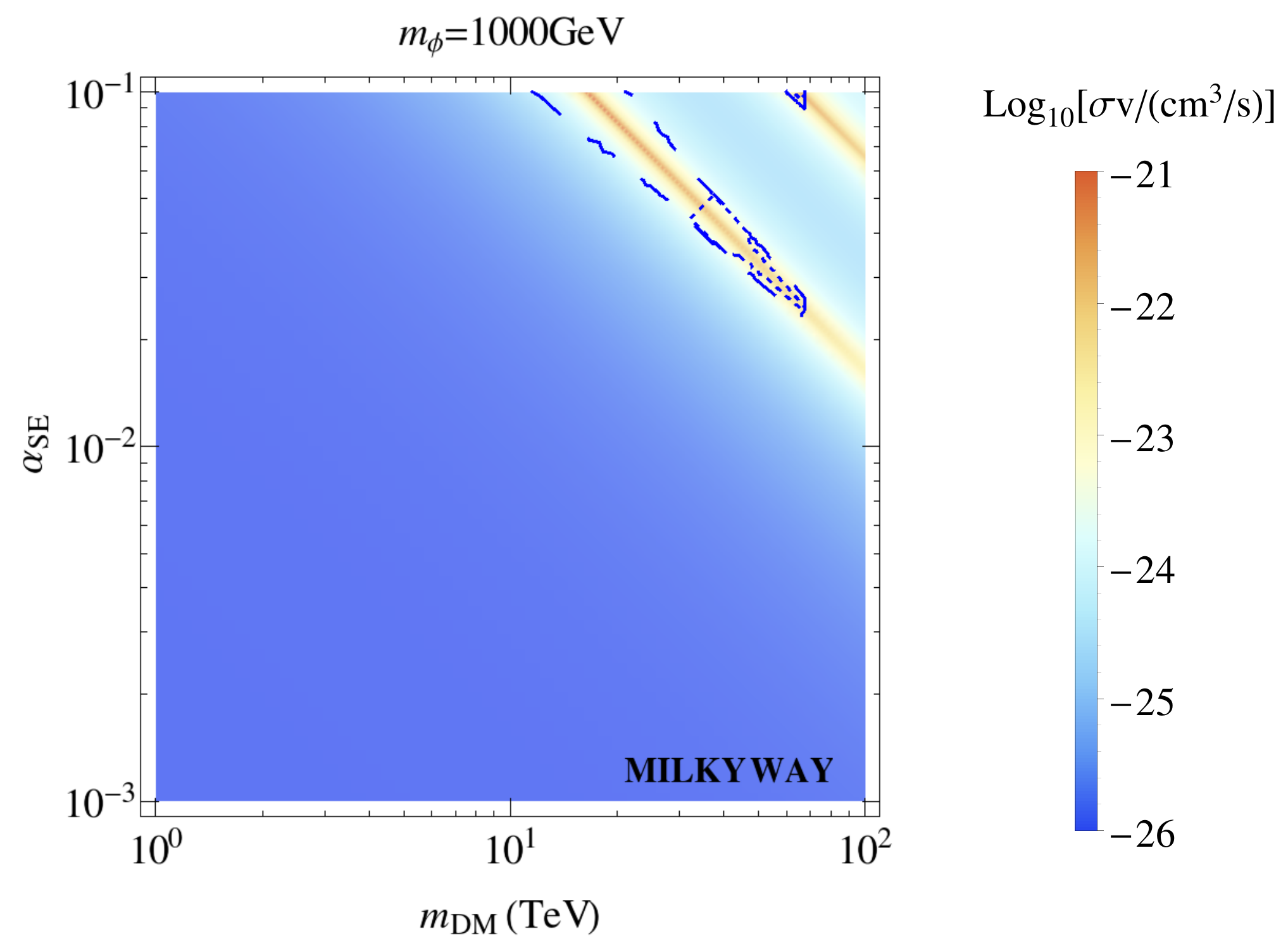}
\caption{
DM annihilation cross section into neutrinos of all flavors for model $F_4$ as a function of the DM mass and the coupling inducing the Sommerfeld effect. For these models, this cross section also gives the total annihilation cross section into charged leptons.  This cross section is here given for an annihilations in the Milky Way, assuming a DM relative velocity of $v\approx 2\times10^{-3} c$~\cite{Bertone:2010zza}. The blue dotted (dashed) lines are the limits for $\mu^+\mu^- $  ($\tau^+\tau^-$) final states from HESS~\cite{Abdallah:2016ygi} under the assumption of a NFW profile and correspond to ten years of observation of the galactic center.  All the limits in this figure have been rescaled to account for the fact that we are considering Dirac DM.
}
\label{fig:nuFluxF3F4}
\end{figure*}

In Fig.~\ref{fig:nuFluxF3F4}, we plot for model $F_4$ the total annihilation cross section into neutrinos (or charged leptons) as a function of $\alpha_\text{SE}$ and the mediator mass $m_\phi$. For model $F_3$ the plot is  the same except that the neutrino cross section reported in this figure has to be divided by 3, due to the fact that a $Z'$ produces neutrinos in a flavor-democratic way at the source. 
As for  $F_{1,2}$ , the models $F_{3,4}$  predict an equal production of charged leptons and neutrinos. Therefore,  the observation of a neutrino line is only possible for a neutrino cross section below the limits on charged leptons.
Here too, this implies  that an observation of a line below  1 TeV in the near future is not expected in these models, given the stringent limits on charged leptons at these low energies. 
But, unlike models $F_{1,2}$, below the charged-lepton limits such an observation is in principle possible for any DM mass above 1 TeV. This results from the fact that here the mass of the Sommerfeld mediator and its coupling strength are not fixed.  By varying these two parameters, the positions of the peaks and the strength of the Sommerfeld effect can vary, as shown in Fig.~\ref{fig:nuFluxF3F4}. This leads us to the conclusion that, in presence of a new Sommerfeld mediator, an observation of a neutrino line is basically possible everywhere for cross sections below the charged-lepton limits. (Notice that this would also be possible for model $F_{1,2}$ if we added a new Sommerfeld mediator.)

We would like to mention that model $F_4$ also gives rise to $\DM\DM \to \nu\nu$, as shown in Table~\ref{table:models}. An explicit calculation 
shows that  $\sigma v(\DM \DM \to \nu_\alpha \nu_\beta)/\sigma v(\DM \overline{\DM} \to \nu_\alpha \overline{\nu}_\beta) < \lambda_5^2 v_\text{EW}^4/32m_\DM^4$.  Therefore, the $\nu\nu$ final states are only relevant below the electroweak breaking scale, where we find that $\sigma v(\DM \DM \to \nu_\alpha \nu_\beta) \sim \sigma v(\DM \overline{\DM} \to \ell_\alpha \overline{\ell}_\beta)$. Since the latter are severely constrained,  $\nu\nu$ as the dominant annihilation channel in model $F_4$ is not an option.

\FloatBarrier

\subsection{Models $S_1$ and $S_2$}
\label{model:S1S2}

As discussed in Section \ref{sec:classification}, models $S_1$ and $S_2$ lead to annihilation cross sections that are suppressed by powers of $v_\text{EW}/m_\DM$. Therefore, they are viable models for neutrino lines only below the TeV scale. These models, as $\SSS_1$ and $\FF_1$, have the interesting property of not leading to an equal production of charged leptons and neutrinos. For models $S_{1}$ and $S_{2}$, this is in fact a necessary condition because, as already said above,  charged-lepton limits at low scales reach sensitivities that neutrino telescopes will not reach before long. 

\begin{figure*}[h]
\includegraphics[trim=0cm 0.0cm 0cm 0.1cm,clip,height=0.14\textheight]{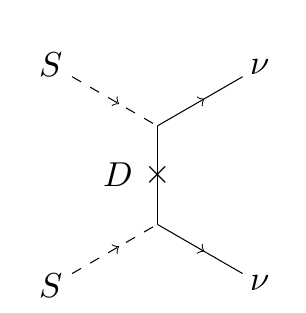}
\includegraphics[trim=0cm 0.0cm 0cm 0.0cm,clip,height=0.14\textheight]{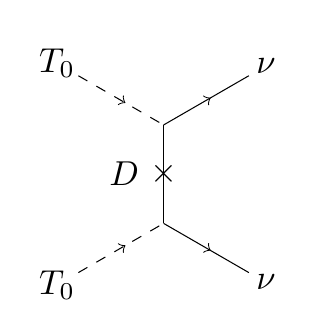}
\caption{  DM annihilation into neutrinos for model $S_1$ (left) and model $S_2$ (right). We use the notation of Table~\ref{table:models}.}. 
\end{figure*} 

\FloatBarrier

At such low masses, the electroweak Sommerfeld effect does not take place, even for model $S_2$ whose DM candidate belongs to a $SU(2)_L$ triplet. 
Thus, as for  models $F_{3,4}$, an additional light mediator is required to induce the Sommerfeld effect. Such a mediator can be a scalar or a vector boson. In both cases, DM annihilations into the mediator proceed via the s-wave and they must therefore be considered at freeze-out and in DM halos. Moreover, if the mediator dominantly decayed into charged fermions or photons, the annihilation into monochromatic neutrinos would not be  the prime signature of these scenarios.

\begin{figure*}[t]
\centering{Model $S_1$ }\\
\includegraphics[width=0.495\textwidth]{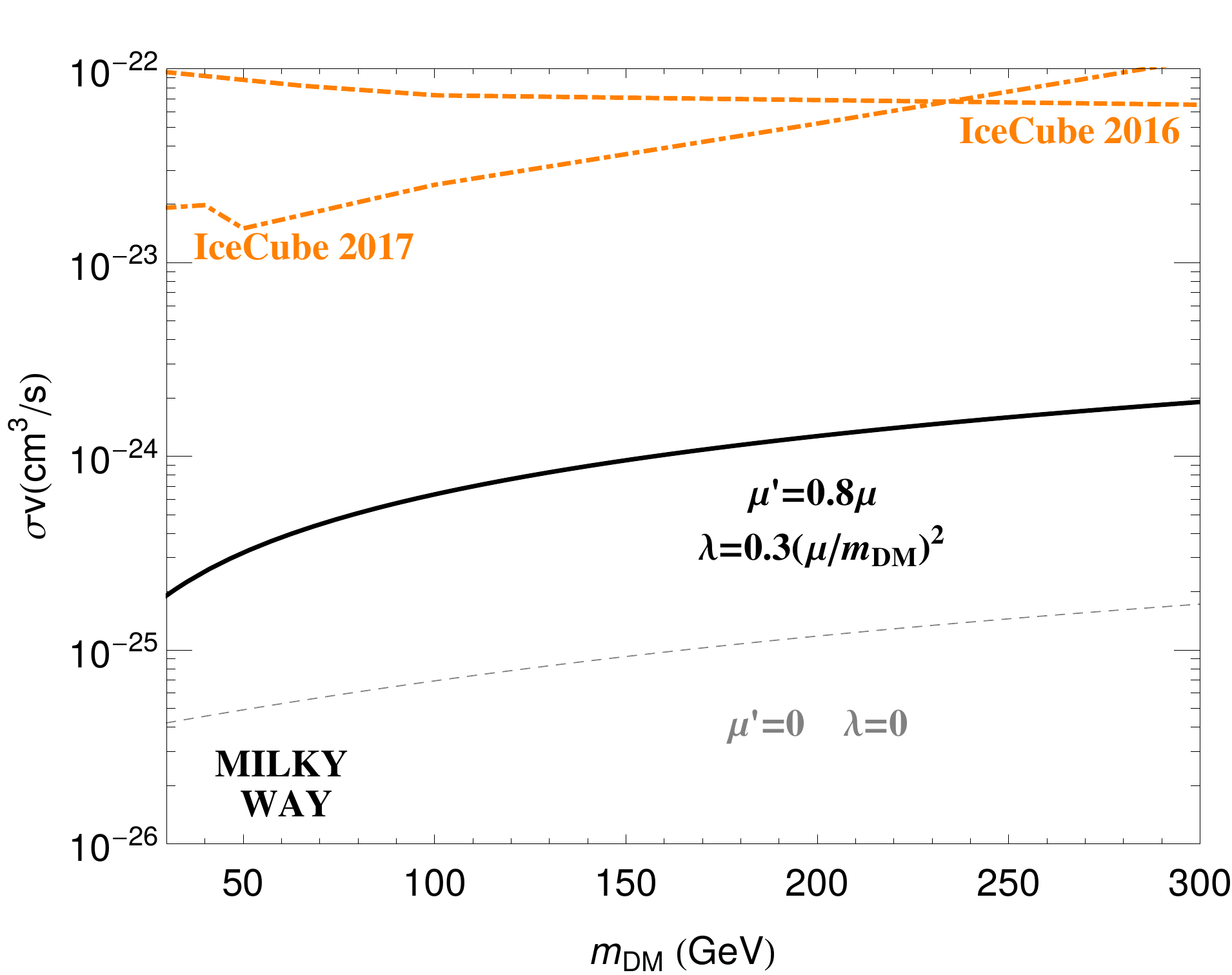}
\includegraphics[width=0.495\textwidth]{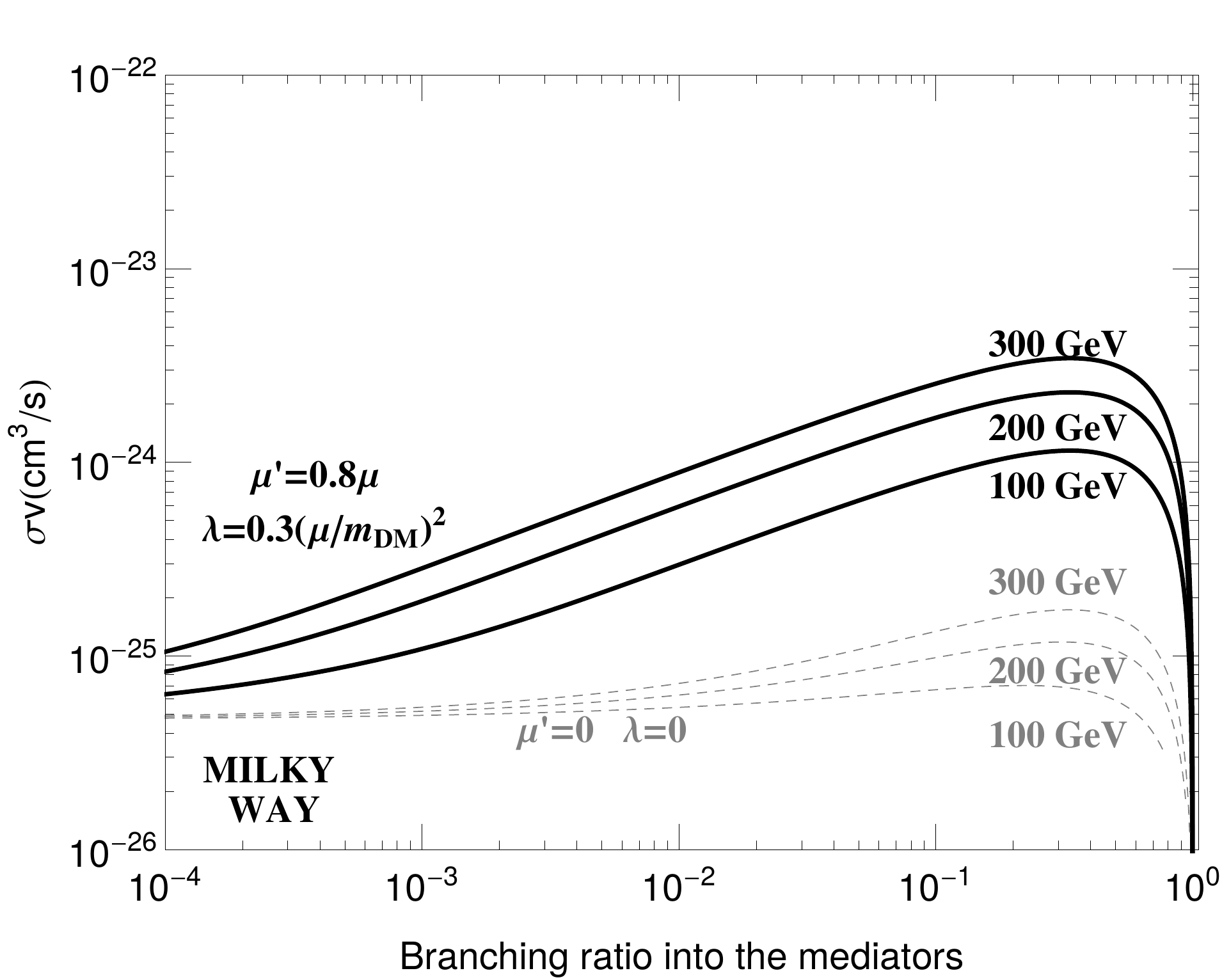}\\
\caption{\emph{Left panel:} Total annihilation cross section into neutrinos { and antineutrinos of all flavors} in model $S_1$ for $m_\phi=\unit[1]{MeV}$ and two different choices of couplings in Eq.~\eqref{eq:LS1}. In this model, there is no associated  annihilation into charged leptons. A branching ratio of 30 \% is assumed for the annihilation channel $\overline{\DM} \DM \to \phi\phi$. The IceCube limits correspond to those of Refs.~\cite{Aartsen:2017ulx, Aartsen:2016pfc} after rescaling for the fact that DM is not self-conjugate in model $S_1$.
\emph{Right Panel:} Same as the left panel, but varying the branching ratio for fixed values of $m_\DM$.  
}
\label{fig:nufluxS1}
\end{figure*}

Let us first discuss model $S_1$ where DM is a complex scalar singlet $\Phi$ coupling to a fermionic doublet $\chi_D$. Furthermore, we assume that DM couples to a real scalar $\phi$ that mostly  decays into neutrinos and whose mass is $m_\phi\lesssim\unit[1]{MeV}$. The relevant piece of the Lagrangian for our purposes is
\begin{equation}
{\cal L}_\phi \supset ( y_\alpha \Phi \overline{L_\alpha} \chi_D +h.c.)-(\mu\,\phi+\lambda\, \phi^2  )\Phi^* \Phi - \mu'\, \phi^3  \,.
\label{eq:LS1}
\end{equation}
{To ensure that this model does not generate neutrino masses at one loop, we must forbid terms such as $\Phi^2$ (by invoking, for example, a $U(1)'$ symmetry).}   
The cubic interaction between the DM and the mediator induces a Yukawa potential of the form $V(r)= - \mu e^{-m_\phi r}/(16\pi m_\DM^2 r)$~\cite{Garcia-Cely:2015khw}. Furthermore, the other interactions lead to $\Phi^* \Phi\to \nu\nu$ and $\Phi^* \Phi\to \phi\phi$. For the sake of simplicity, we neglect the Sommerfeld enhancement in the early universe. We then have 
\begin{equation}
\left[\sigma v(\Phi^* \Phi^*\to \overline{\nu}\, \overline{\nu})+\sigma v(\Phi \Phi\to \nu\nu)+\sigma v(\Phi^* \Phi\to \phi\phi)\right]_\text{Freeze-out}=\unit[4.7\times10^{-26}]{cm^3/s}\,, 
\label{eq:S1FO}
\end{equation}
while in DM halos
\begin{equation}
\sigma v (\Phi^* \Phi^*\to \overline{\nu} \,\overline{\nu})=\sigma v (\Phi \Phi\to \nu\nu) =
  \text{BF}\,\times (1-\text{Br}_{\phi\phi}) \,\left(\unit[2.35\times10^{-26}]{cm^3/s}\right)\,.
\end{equation}
where $\text{BF}$ is the Sommerfeld boost factor. As before, we calculate it using the Hulth\'en potential approximation~\cite{Cassel:2009wt}. Here, $\text{Br}_{\phi\phi}$ is the branching ratio into the mediators.
If we fix this branching ratio as well as $\mu'$ and $\lambda$, we can calculate the coupling $\mu$  by means of Eq.~\eqref{eq:S1FO} and therefore the Sommerfeld boost factor BF, which allows to calculate the neutrino signal today. This is illustrated in the left panel of Fig.~\ref{fig:nufluxS1}, where we show the annihilation cross section into neutrinos in the Milky Way  for two different combinations of  $\mu'$ and $\lambda$, $\text{Br}_{\phi\phi}=0.3$ and $m_\phi= \unit[1]{MeV}$. In the corresponding right panel, we show the dependence of $\sigma v$ on the branching ratio. For low $ \text{Br}_{\phi\phi}$, the Sommerfeld boost is small. For large $ \text{Br}_{\phi\phi}$, the signal into neutrinos gets naturally suppressed in the early universe and therefore also today. In the figure, we also show current limits by {IceCube~\cite{Aartsen:2017ulx, Aartsen:2016pfc}}. Even  after including the Sommerfeld effect by an additional mediator, the cross section predicted by model $S_1$ is still more than one order of magnitude below the  current sensitivity.

As for model $S_2$, which involves  a DM triplet, it  turns out that it is excluded in the region below 1 TeV, where it can produce neutrino lines. For {$m_\DM > m_W$}, DM annihilations into $WW$ are too fast in the early universe and can not account for the observed relic density. Below this $W$-threshold, it can be checked that this model leads to annihilations into $\gamma\gamma$ arising from $W$-boson loops, which are already excluded by current limits on photon-line production~\cite{Ackermann:2015lka}. 

Thus, to summarize this section, model $S_2$ is excluded, and model $S_1$, unlike models $F_{1-4}$, could lead to an observable neutrino line below 1 TeV provided that the neutrino-telescope sensitivities improve by at least one order of magnitude. 

\section{Flavor structure of the lines }
\label{sec:flavor}

So far, our discussion has focused on the line feature because it may be used to better discriminate a signal from the astrophysical background. In this sense, indirect detection with neutrinos and photons are similar. Nonetheless, there is an important difference between the two cases. In contrast to photons, neutrinos carry flavor, which can be used as an indicator of their origin.

\begin{figure*}[t]
\includegraphics[width=0.75\textwidth]{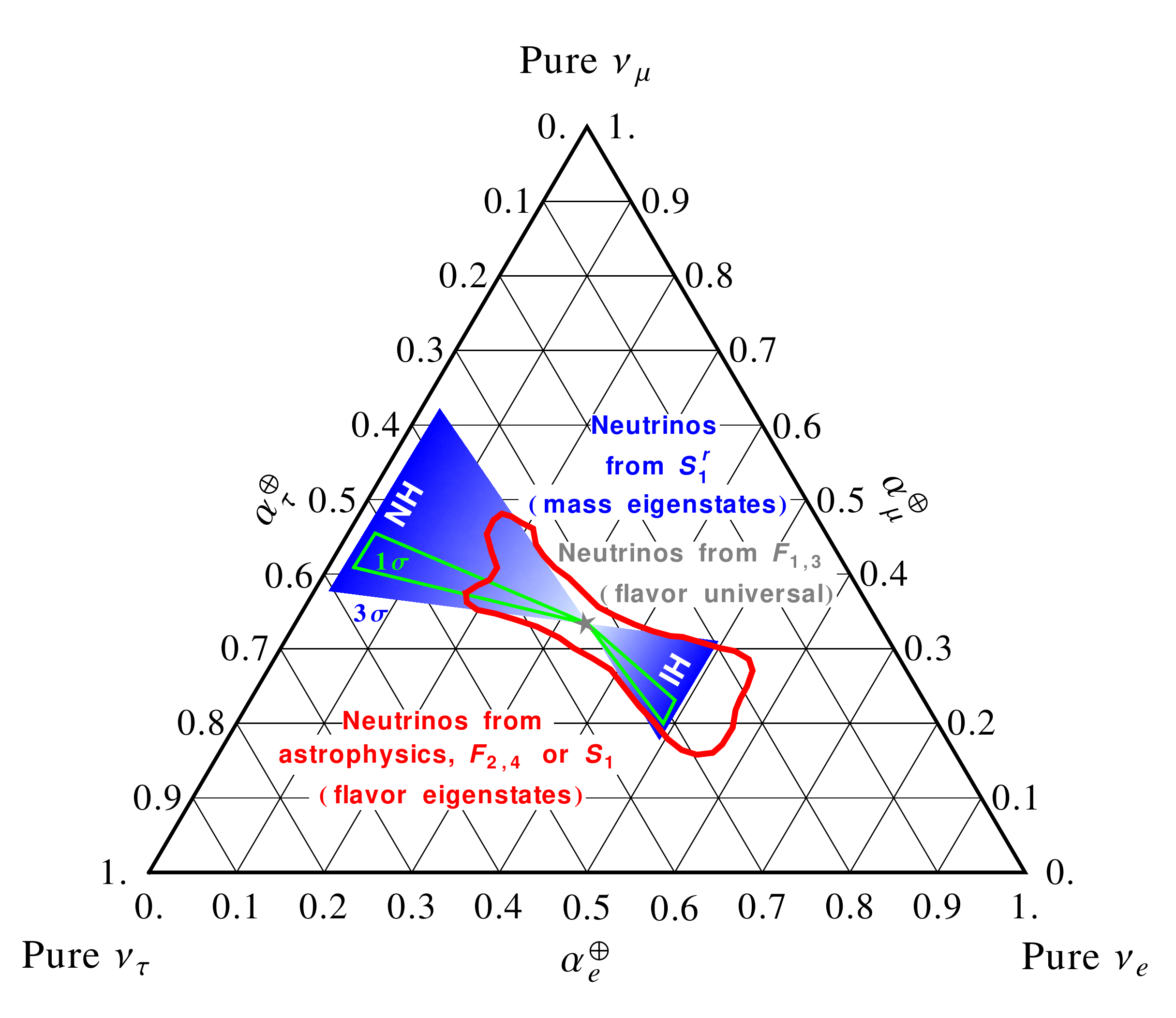}
\caption{ Flavor composition of the neutrino lines in the detector for the different models under consideration in this work. In the case of model $\SSS_1$, the flavor composition depends on the neutrino parameters that we take from Ref.~\cite{Esteban:2016qun} at $3\sigma$ ($1\sigma$) in blue (green). In addition, the blue-color gradient refers to the mass of the lightest neutrino (the darker the color, the lighter the mass). 
}
\label{fig:triangle}
\end{figure*}

Thus, the prediction of the flavor ratios of the models of Table~\ref{table:models} is of particular relevance. Doing so requires to account for neutrino oscillations because they average out the flavor ratios when the neutrinos are produced at very large distances from  Earth.  Before presenting the predictions of each model, and for the sake of comparison and illustration, we first discuss the case of neutrinos produced in astrophysical processes (not including DM annihilations or decays), such as cosmic-ray spallations. These are produced as flavor eigenstates with  particular ratios~\cite{Bustamante:2015waa}, but even for an arbitrary flavor composition  at the source, $\alpha^S_e:\alpha^S_\mu:\alpha^S_\tau $, neutrino oscillations will average out this composition in such  a way that on Earth  $\alpha^\oplus_e:\alpha^\oplus_\mu:\alpha^\oplus_\tau$ gets closer to $1:1:1$. More precisely~\cite{Vincent:2016nut,Bustamante:2015waa}, 
$\alpha^\oplus_\ell \simeq \sum_{\ell'} P_{\ell \ell'} \,\alpha_{\ell'}^S$, with $P_{\ell \ell'} = \sum_{i} |U_{\ell i}|^2  |U_{\ell' i}|^2 $. Taking into account the most recent data on neutrino parameters, one finds that,  for an arbitrary composition at the source, the flavor ratios in the detector lies within the red contour depicted in Fig.~\ref{fig:triangle}.  

Having mentioned astrophysical neutrinos, let us consider model $F_1$ or $F_3$. According to Eqs.~\eqref{eq:BrF1} and \eqref{eq:BrF3F4}, they predict flavor-universal branching ratios. This is because DM annihilations into neutrinos are mediated by a $Z'$ boson. Therefore,  the flavor ratios  are $1:1:1$ at the annihilation site, for instance at the galactic center. Neutrino oscillations will not affect this pattern and the flavor ratios in the detector are thus  $\alpha^\oplus_e:\alpha^\oplus_\mu:\alpha^\oplus_\tau \sim 1:1:1$. This is depicted in  Fig.~\ref{fig:triangle} as the gray star. 

Let us now consider models $F_2$, $F_4$ and $S_1$. In that case, DM is coupled to the neutrinos by means of flavored Yukawa interactions,  which give rise to neutrinos with arbitrary flavor branching ratios (see Eqs.~\eqref{eq:BrF2}, \eqref{eq:BrF3F4} and \eqref{eq:LS1}). As a result, in these models the neutrino composition of the lines also lies  within the red contour, see the flavor triangle in Fig.~\ref{fig:triangle}.

In contrast, for the case where the couplings $Y^L$ of Eq.~(\ref{eq:LSm7Fm7}) are the only source of neutrino masses, model $\SSS_1$ produces lines that are \emph{mass eigenstates}. They  do not oscillate by definition.  The mass-eigenstate ratios of the lines are determined by Eq.~\eqref{eq:BrSr7}. The corresponding flavor composition is simply given by performing a transformation from the mass to the flavor basis, which leads to
$
\alpha^\oplus_\ell = \sum_{j=1} m_j^2 |U_{\ell j}|^2/\sum_{j=1} m_j^2
$. Taking into account  the  $3\sigma$ ($1\sigma$) fit of the neutrino parameters presented in Ref.~\cite{Esteban:2016qun}, we calculate such flavor composition and show it as the blue (green) region in Fig.~\ref{fig:triangle}. There, we differentiate between inverted and normal hierarchy. For the former case, one obtains a flavor composition in the detector that neutrinos of astrophysical origin can also give. On the contrary, in the case of normal hierarchy, the flavor composition does not mimic astrophysical neutrinos in most cases. All this discussion is completely analogous to the case of a majoron decaying into neutrinos, as recently studied  in Ref.~\cite{Garcia-Cely:2017oco}.

\section{Beyond the basic picture}
\label{sec:caveats}

So far, to determine the list of models that could lead to a large emission of monochromatic neutrinos today, we have made a series of standard minimal  assumptions on the structure of the DM model, see (i)--(iii) in Section \ref{sec:classification}. In this section we discuss how, by relaxing these assumptions---that is to say by complicating the DM scenario---one could enlarge the possibilities of having an observable flux.

 \subsection{Beyond triplets}
 
In the basic setup considered above,  assumption (ii) allows for DM and mediator multiplets up to triplets.
It is not difficult to generalize this to higher representations. In Table~\ref{table:models}, any DM or mediator that transforms as a doublet can be replaced by a 4-plet, 6-plet, etc. Similarly, any triplet can be traded for a 5-plet, 7-plet, etc. For instance, one possible candidate is a hyperchargeless 5-plet with either a 4-plet or a 6-plet mediator, or a 7-plet with either a 6-plet or a 8-plet mediator. 

A $2n$-plet (with $n \in \mathbb{N}$) necessarily has hypercharge, implying that its components must be split in mass to evade direct detection constraints. This in turn means that only the final state $\nu\nu$ can take place via the s-wave. However, this channel breaks lepton number and  the corresponding neutrino-mass constraints put a severe upper bound on the corresponding  annihilation cross section into neutrinos. For instance, DM coming from a $2n$-plet  can not produce neutrino lines in the t-channel because of the reasons presented in Section~\ref{sec:classification} for doublets.  In the s-channel, the doublet models $\SSS_1$ and $\FF_1$  can be  generalized to a $2n$-plet but require fine-tuning as explained above.

Nonetheless, DM as a Dirac $(2n+1)$-plet  could constitute a viable option. Here, we will discuss the case in which it is  coupled to a $Z'$ boson, that is, the generalization of model $F_1$. For simplicity, we will neglect  the contribution of pure gauge interactions to the annihilation matrix. Because of this, and because the $Z'$ boson couples to all the components of the multiplet in the same way,
Eqs.~\eqref{eq:GammaF1Gaugev2} and \eqref{eq:GammaF1Gaugev3} generalize to 
\begin{eqnarray}
\label{eq:GammaF1Gaugev2gen}
\Gamma^{S=1}_{\nu_\alpha\bar{\nu}_\beta} &=&
\Gamma^{S=1}_{\ell_\alpha\bar{\ell}_\beta} =
\frac{1}{9}\,\sigma v_0\,
\begin{psmallmatrix}
1&1&\cdots& 1\\ 
1&1&\cdots& 1 \\
\vdots&&\ddots&\\
1&1&\cdots&1
\end{psmallmatrix} \delta_{\alpha \beta} \,,   
\end{eqnarray}
where $\sigma v_0$ is a global factor chosen so that the total annihilation cross section into neutrinos, as given by Eq.~\eqref{eq:SEsigmav_text}, is 
\begin{equation}
\sum_{\alpha \beta}\sigma v \left(\psi^0 \overline{\psi^0}\to \nu_\alpha\overline{\nu_\beta}\right) = \sigma v_0 \bigg|\sum_{i=-n}^n d_i\bigg|^2\,. 
\end{equation}
The last factor in this equation is  the Sommerfeld enhancement for the $(2n+1)$-plet for the annihilation matrices in Eq.~\eqref{eq:GammaF1Gaugev2gen}. In order to calculate $\sigma v_0$, we first notice that Eq.~\eqref{eq:GammaF1Gaugev2gen} only has one eigenvector with non-zero eigenvalue, which corresponds to the isospin singlet. This is expected because the annihilation proceeds via the s-channel exchange of a singlet particle. Then, using the notation of Appendix~\ref{sec:AppSE}, we have $\Gamma_{I=0,S=1}=2(2n+1)\sigma v_0/3 $. Similarly, the potential associated to singlet state\footnote{The potential is $V(r) = \alpha_2 I_1 \cdot I_2/r$, where $I_1$ and $I_2$ are the isospin operators of the incoming particles in the annihilation. From the equalities $I_1\cdot I_2= \frac{1}{2}\left((I_1+I_2)^2-I_1^2-I_2^2\right)$, $I_1^2= I_2^2 = n(n+1) {1\!\!1} $ and $(I_1+I_2)^2=0$ (for the isospin singlet state), the relation  $I_1\cdot I_2= - n(n+1) {1\!\!1}$ holds.} is $V(r) =  -\alpha_2 n(n+1)/r$. The relic density constraint of Eq.~\eqref{eq:relic0} is therefore 
\begin{equation}
\sigma v_0 =\left(\frac{2n+1}{1+ \sqrt{\pi x_f}n(n+1)\alpha_2} \right)\left(\unit[2.35\times 10^{-26}]{\frac{cm^3}{s}}\right)\,.
\label{eq:relicF1gen}
\end{equation}

\begin{figure*}[t]
\includegraphics[width=0.60\textwidth]{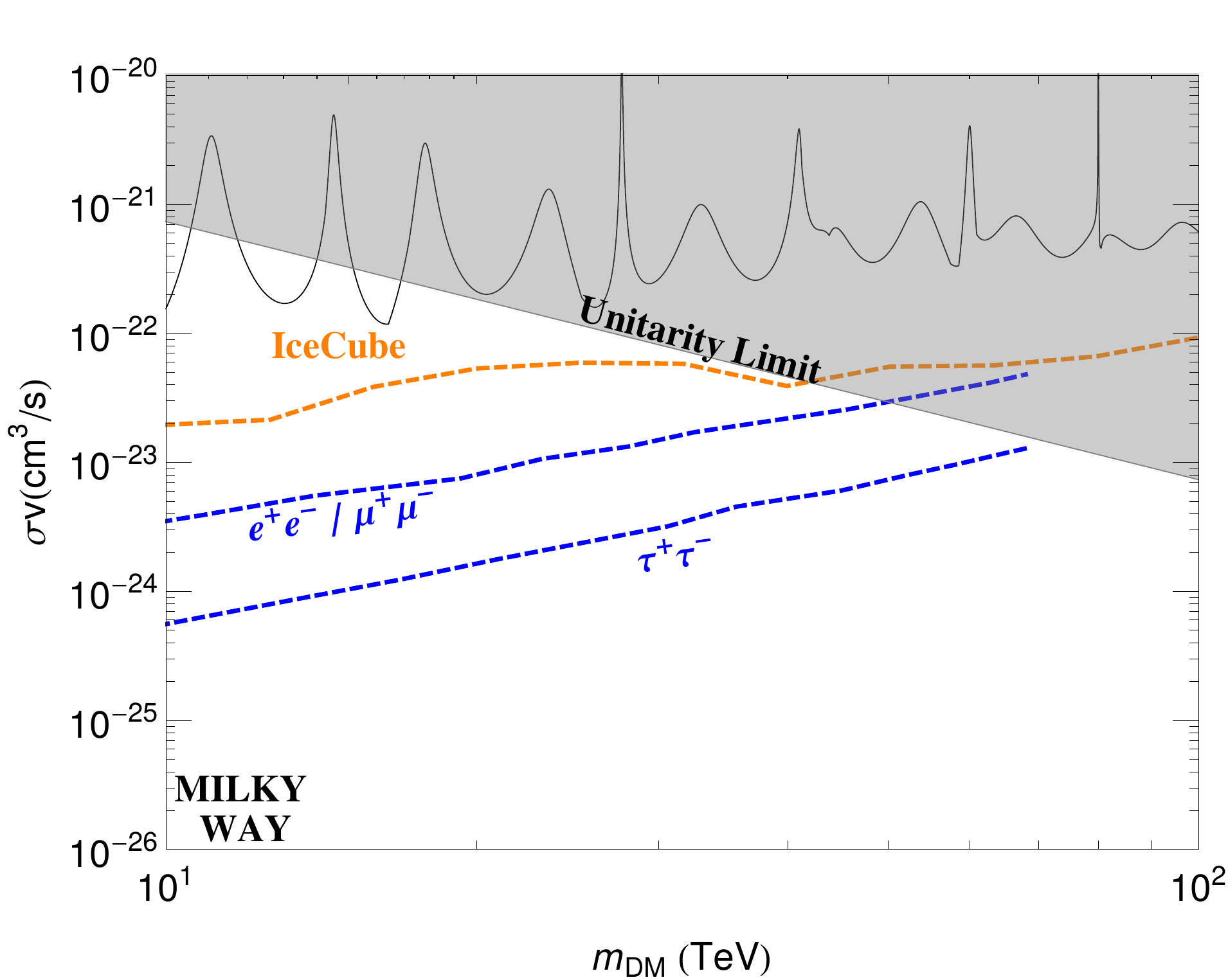}
\caption{DM annihilation cross section into neutrinos of all flavors (black) for  5-plet DM coupled to a $Z'$ boson. In this scenario, this also gives the total annihilation cross section into charged leptons. This cross section holds for annihilations in the Milky Way with a DM relative velocity of $v\approx 2\times10^{-3} c$~\cite{Bertone:2010zza}. The limits are the same as those presented in the left panel of Fig.~\ref{fig:nuFluxF2} 
}
\label{fig:nuFluxF15plet}
\end{figure*}

Using the results from Ref.~\cite{Garcia-Cely:2015dda} (see also Refs.~\cite{Cirelli:2007xd, Cirelli:2009uv,Cirelli:2015bda}), we can calculate the total annihilation cross section into neutrinos and present the case of a quintuplet in Fig.~\ref{fig:nuFluxF15plet}. 
The cross sections obtained in this case are above $\unit[10^{-22}]{cm^3/s}$, which is not only greater than the upper bounds from IceCube data but also violates the unitarity limit~\cite{Griest:1989wd}. Thus,  quintuplet DM  is not a good candidate for observing a neutrino line. Such a large value, as compared to the case of a triplet, is due to a larger multiplicity. Based on an estimate similar to that of footnote~\ref{ft:potential}, we have explicitly checked that larger representations give rise to even larger cross sections and are therefore excluded as well. We thus conclude that larger DM representations do not lead to more possibilities for observable neutrino lines.

\subsection{Freeze-out from extra interactions}

\begin{figure*}[t]
\includegraphics[width=0.495\textwidth]{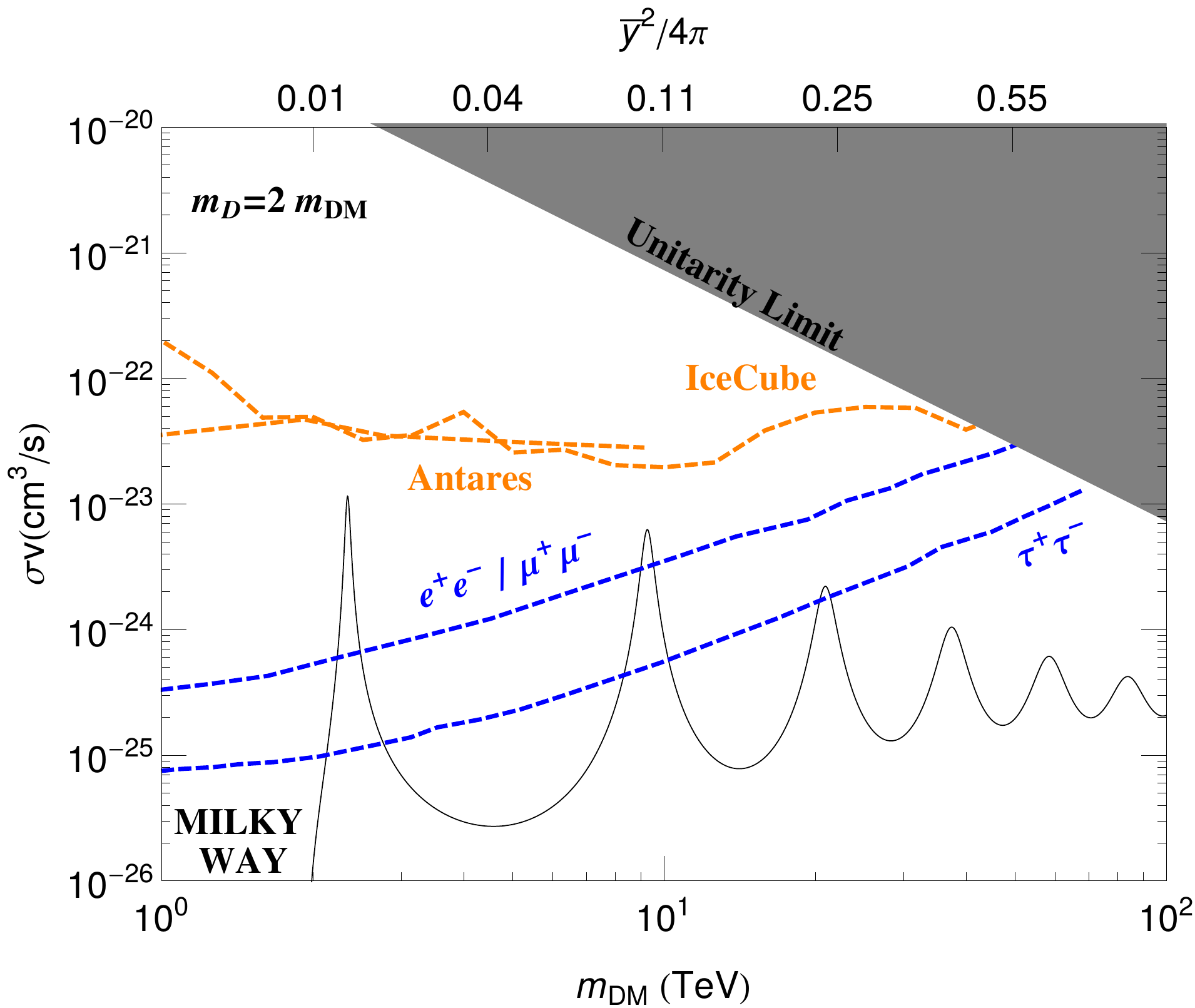}
\includegraphics[width=0.495\textwidth]{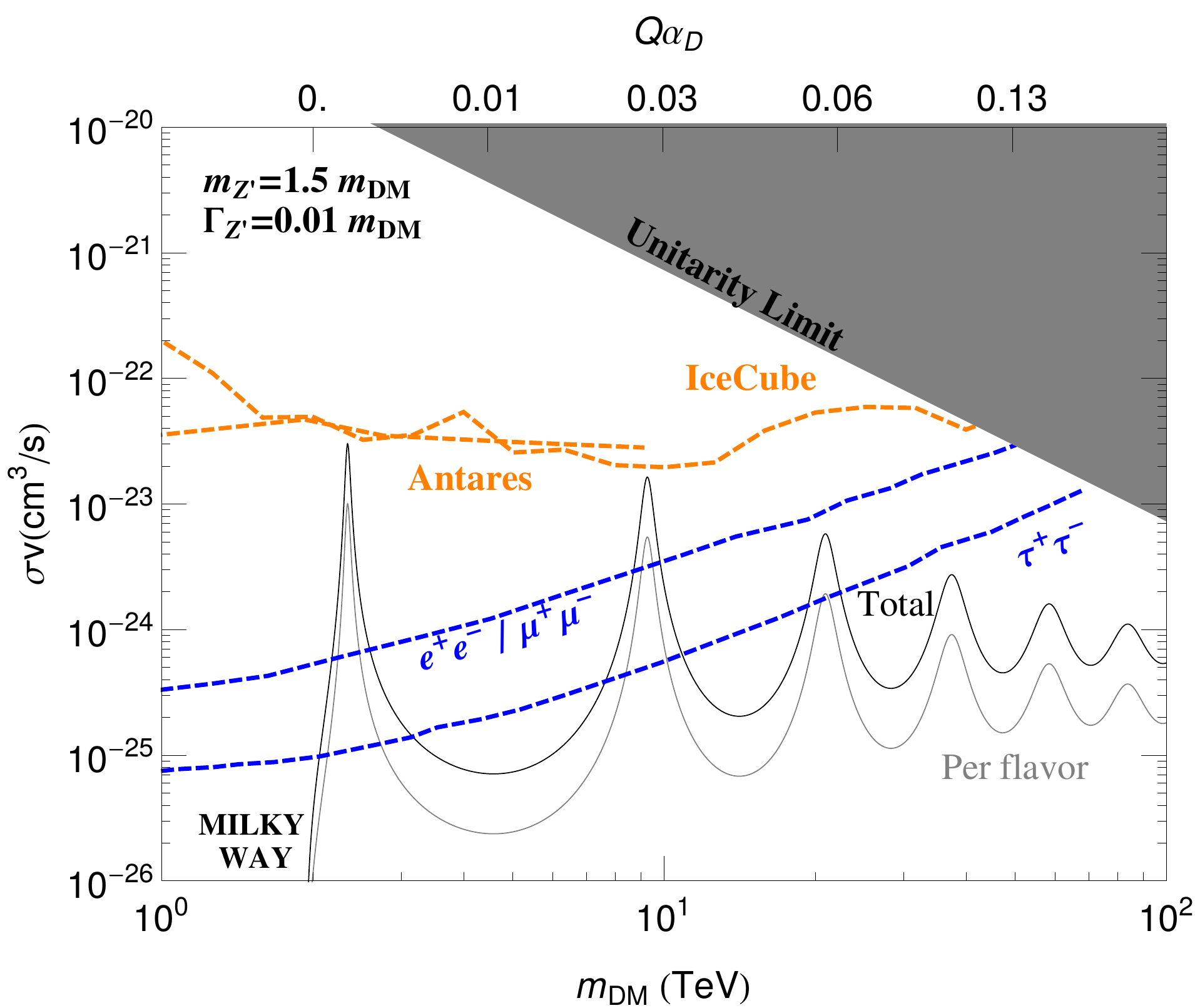}
\caption{ Same as left column of Fig.~\ref{fig:nuFluxF2} except that neutrino interactions are only responsible for 5\% of the annihilation cross section at freeze-out. }
\label{fig:nuFluxF2inv}
\end{figure*}

If  one still assumes that the DM is made of a single particle species but allows for extra annihilation channels (on top of those induced by the mediator in Table~\ref{table:models} and on top of the possible Sommerfeld effect mediator), one can in some cases considerably increase the parameter space leading to an  observable neutrino line. 
For instance, for the $F_{1,2}$ models one can assume an annihilation into neutrinos twenty times slower, in which case the  prediction for the neutrino cross section in Fig.~\ref{fig:nuFluxF2} decreases by a factor twenty, so that the corresponding cross sections are now allowed by charged-lepton constraints for almost any DM mass above $\unit[2]{TeV}$, except close to the peaks. We present such a scenario in Fig.~\ref{fig:nuFluxF2inv}. 
 
Of course, this still requires that neutrino telescopes can reach a  sensitivity below the charged lepton limits. For instance, above 10 TeV, this requires a sensitivity  around $\unit[3\times 10^{-24}]{ cm^3/s}$ ($\unit[5\times10^{-25}]{ cm^3/s}$) in the $e$ and $\mu$ channels ($\tau$ channel).
On the model-building side, this implies the assumption of extra interactions inducing new annihilation channels leading to $\Omega_{DM}\simeq 26\%$ through thermal freeze-out.

Under this assumption, one has to make sure that these new interactions do not induce too large fluxes of cosmic rays. For Dirac candidates, this can be done by considering annihilations into scalars, which are p-wave suppressed. More generally, this could also be done by considering new annihilation channels into new dark sector  states. We will not consider explicit examples of the latter here.

\subsection{Multi-component DM}

Another possibility is to give up the assumption that DM is made out of a single particle species.  For instance, one bosonic  and one fermionic component can annihilate into a neutrino and a gauge boson, possibly giving rise to two different sorts of lines such as those discussed in Ref.~\cite{Aisati:2015ova} for decaying DM. A comprehensive study of such multi-component scenarios is beyond  the scope of this work. We nevertheless discuss what happens when the scenarios of Table~\ref{table:models} give a sub-dominant component of  DM.

On the one hand, if the DM component producing neutrinos (and charged leptons) has no other equally important annihilation channels,  one must consider  annihilation cross sections much larger than in the single-component case, otherwise  DM would  be produced in the early universe much more. By doing this, i.e. decreasing the amount of DM particles producing neutrinos, but increasing the corresponding cross section,  one is left with the same neutrino flux today as in the case of a single component. As a result, the overall picture does not change significantly.

The conclusion is different  if the DM component producing neutrinos also have other annihilation channels, naturally diminishing  its abundance. This is the case of  models with Dirac triplets with masses below $\unit[2]{TeV}$, where DM significantly annihilates into gauge bosons via its s-wave. Thus, the annihilation rate into neutrinos at freeze-out can be assumed to be below the thermal value. 
In this way, the neutrino flux that is produced---which equals that of charged leptons---can easily go below the limits on charged lepton production. Note nevertheless that below $\unit[2]{TeV}$ these limits require a sensitivity below $\unit[3 \times 10^{-25}]{ cm^3/s}$ for the $e$ and $\mu$ channels and below $\unit[6\times 10^{-26}]{ cm^3/s}$ in the $\tau$ channel. This appears to be quite challenging for neutrino telescopes.

\subsection{Multiplet mixing in the initial state} 

In sections \ref{sec:classification}-\ref{sec:SEnoEW}, we have allowed for one DM and one mediator multiplets to be involved in the annihilation into neutrinos. Another option is to add an extra multiplet,  whose neutral component mixes with the DM particle through electroweak symmetry breaking. For instance, before annihilating into neutrinos, a singlet DM particle can mix with the neutral component(s) of a doublet or  a triplet.
In this case, for what concerns the annihilation into neutrinos, the model is like the ones where the DM is the neutral component of this doublet or triplet, up to the fact that the cross section is multiplied by mixing factors. For DM representations with hypercharge, this allows to avoid a too large cross section on nucleons, as this one is also suppressed by a mixing factor. Models of this kind can be found in 
Ref.~\cite{Lindner:2010rr} and also---in the context of the MSSM---in Ref.~\cite{Arina:2015zoa}. We will not consider this possibility further. These models are suppressed by powers of $v_\text{EW}/m_\text{DM}$, so that they cannot perturbatively lead  to a large production of neutrinos above the TeV scale. At lower scales they require, on top of the multiplet mediating the interaction and the ones mixing in the initial state, yet an extra light mediator leading to the necessary Sommerfeld enhancement.
Moreover, for DM multiplets with hypercharge, if the direct detection constraint is satisfied thanks to this mixing, a suppression from the latter will also apply to the production of neutrinos.

\subsection{Spin-1 DM}

We did not consider spin-1 DM in this work because it cannot  be easily studied along simple criteria of minimality similar to those considered in Section~\ref{sec:classification} for lower spins.
Nevertheless, two annihilating spin-1 particles can easily form states with total spin larger than zero. As a result, s-wave annihilations into neutrinos are not so restricted, unlike scalar or Majorona DM. {For an example} in the context of Kaluza-Klein DM see e.g. Refs.~\cite{Blennow:2009ag,Hooper:2002gs}. Annihilation into monochromatic neutrinos is however not a generic feature of spin-1 DM particles as the simple case of Hidden Vector DM~\cite{Hambye:2008bq} shows. In that scenario, DM only communicates with the SM via the Higgs portal, which leads to (at least) d-wave suppressed DM annihilations into neutrino pairs.  In fact, elaborate model building is required in order to obtain s-wave annihilations into neutrinos for spin-1 DM.  

\subsection{Beyond tree level}

Above, we have made a systematic study of the DM models leading to annihilation into monochromatic neutrinos at tree level\footnote{In fact, by including electroweak Sommerfeld enhancement we have also considered  some of the loop-induced annihilations involving electroweak gauge bosons. See e.g.~\cite{Hisano:2002fk} for a study of the connections between the Sommerfeld effect and multiloop annihilation processes.}.  
In principle, the annihilation could also proceed at loop level. In order for such a loop-induced cross section to be {around} the freeze-out value, one would expect that the  underlying interactions are non-perturbative, or close to it. Even though a more compelling statement  requires a systematic study that is beyond the scope of this work,   we would like to mention that many aspects of the  study above remain true at loop level.
 
On the one hand, the helicity  arguments that lead to discarding models giving rise to p-wave or d-wave suppression are valid to any  order of the loop expansion because they only depend on angular momentum or CP conservation. In particular, Majorana or scalar DM annihilations into $\nu\bar{\nu}$ are always at least p-wave suppressed. Likewise, the rate for the processes giving rise to  final state $\nu\nu$ is  suppressed by four powers of the electroweak vev, unless the DM multiplet carries hypercharge, which---because of direct detection constraints---requires to break the corresponding neutral component into self-conjugate fields with a mass splitting.

\section{Beyond lines: other sharp spectral features}
\label{sec:beyondlines}

In the previous sections, we made a systematic study of monochromatic neutrinos arising from DM annihilations. As is well known, this is not the only possible  sharp spectral feature, even if it is the most striking one. 
Two other possibilities are  box-shaped and bremsstrahlung (annihilations into 3-body final states) spectra. For a review of such features in the context of gamma-ray indirect searches, see Ref.~\cite{Bringmann:2012ez}, or Ref.~\cite{Aisati:2015vma} for  decaying DM. 
 In this section, we do not provide a systematic classification of every possible spectra within these two categories. We make general considerations on the shape of these neutrino spectra and discuss two representative examples.
Note that the IceCube and Antares collaboration have not provided so far any limits on spectra of these kinds.
This is why we will also derive limits on the production of neutrinos along this type of spectral features at the end of this section.

\subsection{Box-shaped spectra}

If DM annihilates into particles, say a pair of ``$\phi$" particles, which subsequently decay in-flight into photons or neutrinos, the corresponding spectrum exhibits a spectral feature confined in a kinematical box. If the intermediary particle is a scalar, the spectrum is flat~\cite{Ibarra:2012dw}. In the case of arbitrary spin, the shape within the box is determined by angular momentum conservation and depends on the polarization of the intermediary particles. These spectra have been comprehensively  studied for neutrinos in Ref~\cite{Garcia-Cely:2016pse}  (see also Ref.~\cite{Ibarra:2016fco}).

For the sake of illustration,  and because we are focusing on neutrino fluxes at the reach of current telescopes, we will discuss  the case in which the intermediary particle decaying into neutrinos also induces a Sommerfeld effect\footnote{For similar setups in the context of self-interacting DM models, see Refs.~\cite{Bringmann:2016din,Aarssen:2012fx}.}. Therefore, we assume that the intermediary $\phi$ is either a scalar or vector field. We further assume  that  DM  is not self-conjugate  and that the s-wave annihilation into a pair $\phi\phi$  fixes the relic density\footnote{For a vector boson $\phi$, this assumption means that  DM must be a complex scalar or a Dirac particle, while for a scalar $\phi$, it implies that DM must be a complex scalar.}.
Then, today,
\begin{equation}
\sigma v \left(\DM \overline{\DM}\to \phi \phi\right) \approx \text{BF} \,\times \left(\unit[4.7\times10^{-26}]{cm^3/s}\right)\,,
\label{eq:boxes}
\end{equation}
 where BF is the boost factor due to Sommerfeld enhancement.  Notice that we are neglecting the Sommerfeld effect in the early universe and that there are no co-annihilation channels. 
Proceeding through the various steps of the calculation as for the other models above, it turns out that the cross section into $\phi\phi$ obtained in the Milky Way is the same as the cross section into neutrino and charged leptons for model $F_4$ above {(Fig.~\ref{fig:nuFluxF3F4})}. 
This stems from the fact that Eq.~\eqref{eq:boxes} and Eq.~\eqref{eq:relicF3F4} are identical. 
However the neutrino energy spectra distributions are obviously not the same. 

\begin{figure}[t]
\includegraphics[height=0.435\textwidth]{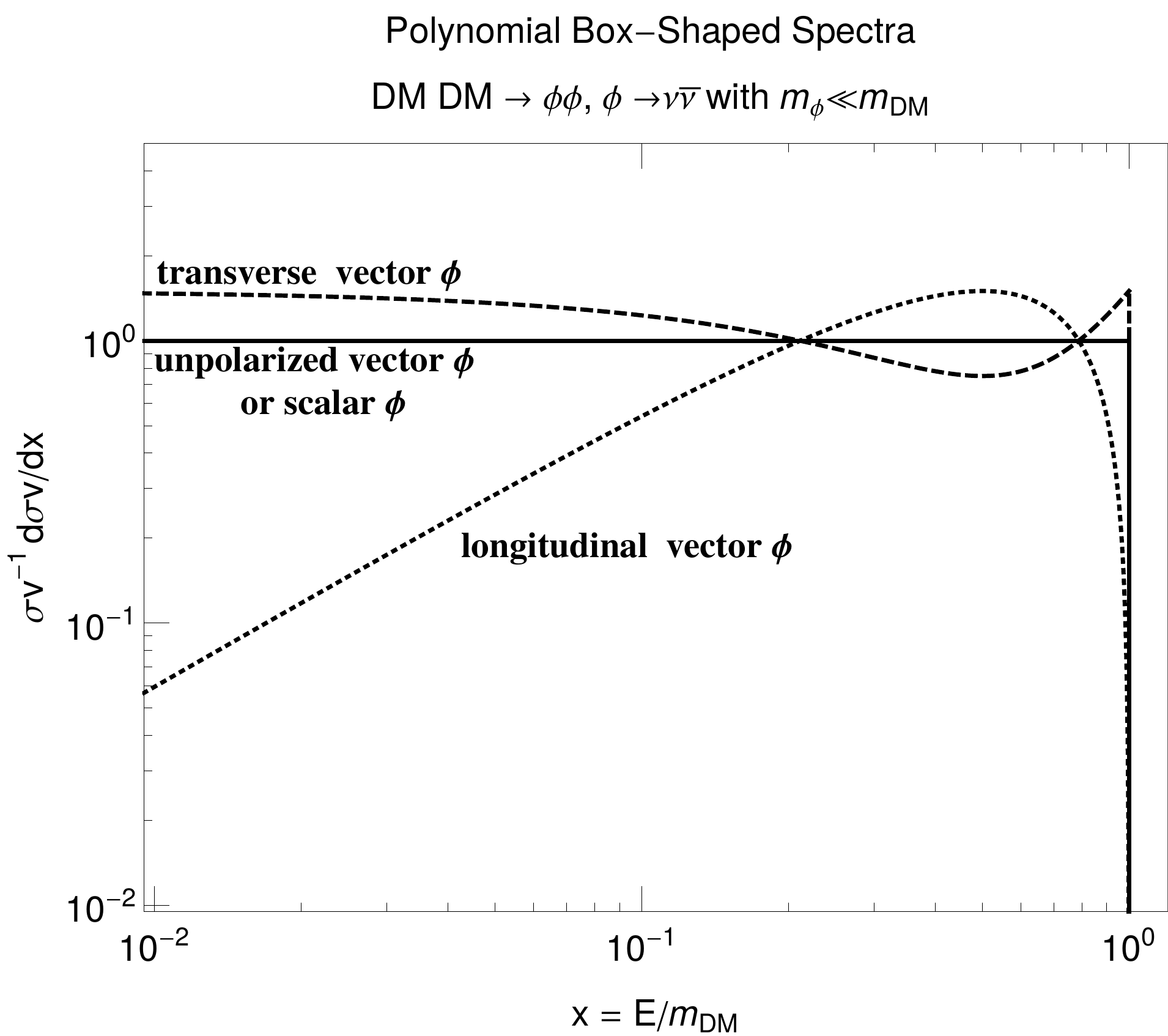}
\includegraphics[height=0.435\textwidth]{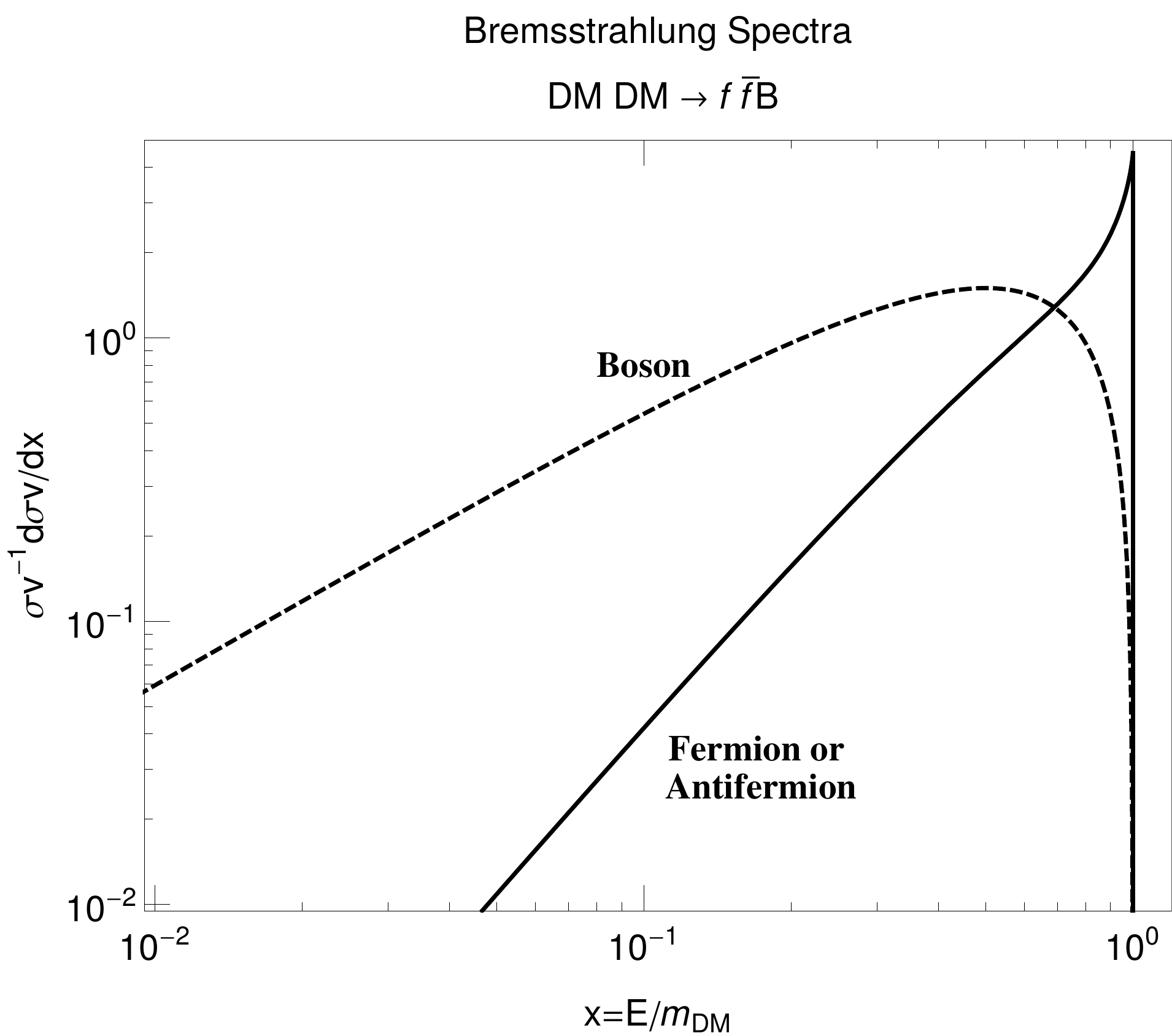}
\caption{\emph{Left panel:} Box-shaped spectra for vector bosons decaying in flight into neutrinos according to their polarization~\cite{Garcia-Cely:2016pse}. The spectra for scalars and that of the unpolarized vector boson are the same.  \emph{Right panel:} Bremsstrahlung spectra from scalar doublet DM for fermions and bosons, as given by Eqs.~\eqref{eq:bremsstrahlung_fermions} and Eqs.~\eqref{eq:bremsstrahlung_bosons}, respectively. }
\label{fig:dNdx}
\end{figure}

For the case in which the $\phi$ boson is a massive vector boson much lighter than the DM, and depending on its polarization,  the neutrino energy spectrum is~\cite{Garcia-Cely:2016pse}
\begin{align}
\frac{1}{\sigma v}\frac{\sigma v}{dx}=\Theta(x)\Theta(1-x)\times
\left\{
\begin{array}{ll}
1\,,&\text{Unpolarized }\\
\frac{3}{2}(1-2x+2x^2)\,,&\text{Transverse}\\
6(1-x)x\,,&\text{Longitudinal}\\
\end{array}
\right. \,,
\label{eq:dNdxboxes}
\end{align}
where $x$ is the neutrino energy normalized to the DM mass. For a scalar boson, as well as for an unpolarized vector boson, the spectrum is flat.
These spectra are shown in the left panel of Fig.~\ref{fig:dNdx}. 

As for neutrino lines, a neutrino box-shaped spectrum may also be accompanied by a similar spectrum of photons or by a flux of cosmic rays associated to charged-lepton channels. For the case under consideration, this depends on the properties of the mediator. For instance,  if the mediator only decays into neutrinos, this does not happen. Nevertheless, that is challenging from the model-building point of view.

The decay of a $Z'$  cannot produce a diphoton spectrum ~\cite{Landau,Yang:1950rg} but will produce a pair of charged leptons with the same rate as for neutrinos (unless kinematically forbidden). Instead, a scalar triplet with $Y=2$ only produces neutrinos, but the associated charged triplet components, if produced too, will decay into charged particles. 

In this regard, one could also consider Majoron mediators generating neutrino boxes. Their decay into photons or charged leptons is parametrically independent from the decays into neutrinos~\cite{Garcia-Cely:2017oco} and one can assume that the latter dominate. Nevertheless, 
one must keep in mind that the Sommerfeld effect for pseudo-scalar mediators is generally much more involved than for scalar or vector particles. See Refs.~\cite{Bedaque:2009ri,Bellazzini:2013foa} for a detailed discussion in the case of fermion DM.

\subsection{Bremsstrahlung}
\label{sec:bremsstrahlung}

The emission of an additional photon out of two-body DM annihilations---the so-called bremsstrahlung process---typically has a hard gamma-ray spectrum with a line-like feature. Interestingly, the latter can mimic monochromatic photons with the energy resolutions of current telescopes~\cite{Bergstrom:1989jr,Flores:1989ru,Beacom:2004pe,Bergstrom:2004cy,Bergstrom:2005ss,Bringmann:2007nk,Garcia-Cely:2013zga}. More generally, a hard spectrum for the additional particle that is emitted is also induced when this  particle is a gauge boson~\cite{ Garny:2011cj,Garny:2011ii, Garny:2015wea,Barger:2011jg} or even a Higgs boson~\cite{Luo:2013bua,Kumar:2016mrq}. For a recent comprehensive study on this subject, see Ref.~\cite{Bringmann:2017sko}.  
For the case where the   two-body final state is a neutrino pair emitting a SM gauge boson that then produces secondary photons, see Ref.~\cite{Queiroz:2016zwd}.
In this work, instead of focusing on the production of secondary photons out of the neutrinos, we will directly study the properties of the neutrinos, i.e.~the neutrino spectra that such processes lead to. This is fully relevant at neutrino telescopes.

The appearance of a spectral feature for certain bremsstrahlung processes is particularly interesting when, on the basis of a symmetry, these processes naturally dominate over the corresponding two-body processes.
For the case of neutrinos  in the two-body final state, as argued in Section~\ref{sec:classification}, a significant number of DM scenarios gives rise to DM annihilations that are p-wave or d-wave suppressed because rotational invariance or the CP-symmetry forbids the s-wave. 
In this case, the two-body processes are suppressed in DM halos but the three-body ones have no reason to be suppressed in the same way.
Consequently,  studying the bremsstrahlung process is fully relevant here. In practice this means that if there are scenarios where the s-wave annihilation into $\nu\overline{\nu}$  is forbidden, the corresponding annihilation into $\nu\overline{\nu}Z$ or $\ell \overline{\nu}W$ is not~\cite{Bringmann:2007nk,Bergstrom:1989jr}.  
Another scenario where the bremsstrahlung production can naturally be dominant arises if the two-body neutrino production proceeds through the vev of a scalar field.
In this case, for DM masses above roughly 10 TeV, the three-body process involving the corresponding scalar field dominates. This possibility has been studied in Ref.~\cite{Aisati:2015ova} for decaying DM (see also~\cite{Dudas:2014bca}).
In this work, we will focus on the cases in which the two-body processes  are p-wave or d-wave, i.e.~forbidden by symmetry.

Before illustrating the previous remarks with a concrete example, let us first discuss the channel involving charged leptons. The arguments of Section~\ref{sec:classification} show that, whenever the process DMDM $\to\nu\overline{\nu}Z$ exists, $SU(2)_L$ invariance (more precisely, invariance under the  $e^{i \pi T_2}$ transformation)  guarantees that there is another process with a similar rate with final state $\ell^-\ell^+Z$. For DM belonging to singlet or triplet representations with $Y=0$, the initial state of the latter process is also a pair of DM particles because  $e^{i \pi T_2}\,$DM$=$DM. Then, the constraints on charged lepton production will compete with limits on neutrino spectral features, similarly to what happens for models $F_{1-4}$. In contrast, for doublets with $Y=1$, the initial state of the process with the charged leptons is not a  DM pair. This is because, in that case, DM is not invariant under the $e^{i \pi T_2}$ transformation.  More precisely, let us consider the case where the  DM candidate, $H^0$, belongs to a doublet $\phi_D= (H^+,\frac{H^0+iA^0}{\sqrt{2}})^T$. 
Then
\begin{align}
\begin{pmatrix}
H^+\\ \frac{H^0+iA^0}{\sqrt{2}}
\end{pmatrix}
 &\to 
\begin{pmatrix}
 \frac{H^0+iA^0}{\sqrt{2}}\\-H^+
\end{pmatrix} = e^{ i \pi \, T_2}
\begin{pmatrix}
H^+\\ \frac{H^0+iA^0}{\sqrt{2}}
\end{pmatrix}
 \,.
\end{align}
Hence, the process $H^0 H^0\to\nu \overline{\nu}Z$ is related by $SU(2)_L$ invariance to the process $H^+ H^-\to\ell^- \ell^+Z$. Consequently, for the case of doublets, the limits on charged leptons may not directly constrain neutrino features.

Motivated by that, we discuss scalar doublet DM as an example of a bremsstrahlung process\footnote{The bremsstrahlung out of diboson final states with DM belonging  to a scalar doublet was studied in Refs.~\cite{Garcia-Cely:2013zga,Garcia-Cely:2015khw}.}.  In order to obtain neutrinos in the final state,  we postulate the presence of a Dirac singlet $\psi$ that couples to DM by means of the Lagrangian
\begin{eqnarray}
{\cal L}_D &\supset& y_\alpha\, \overline{\psi } P_L L_{\alpha} \phi_D+ h.c.\,.
\label{eq:LIB}
\end{eqnarray}
\begin{figure*}[t]
\includegraphics[width=0.495\textwidth]{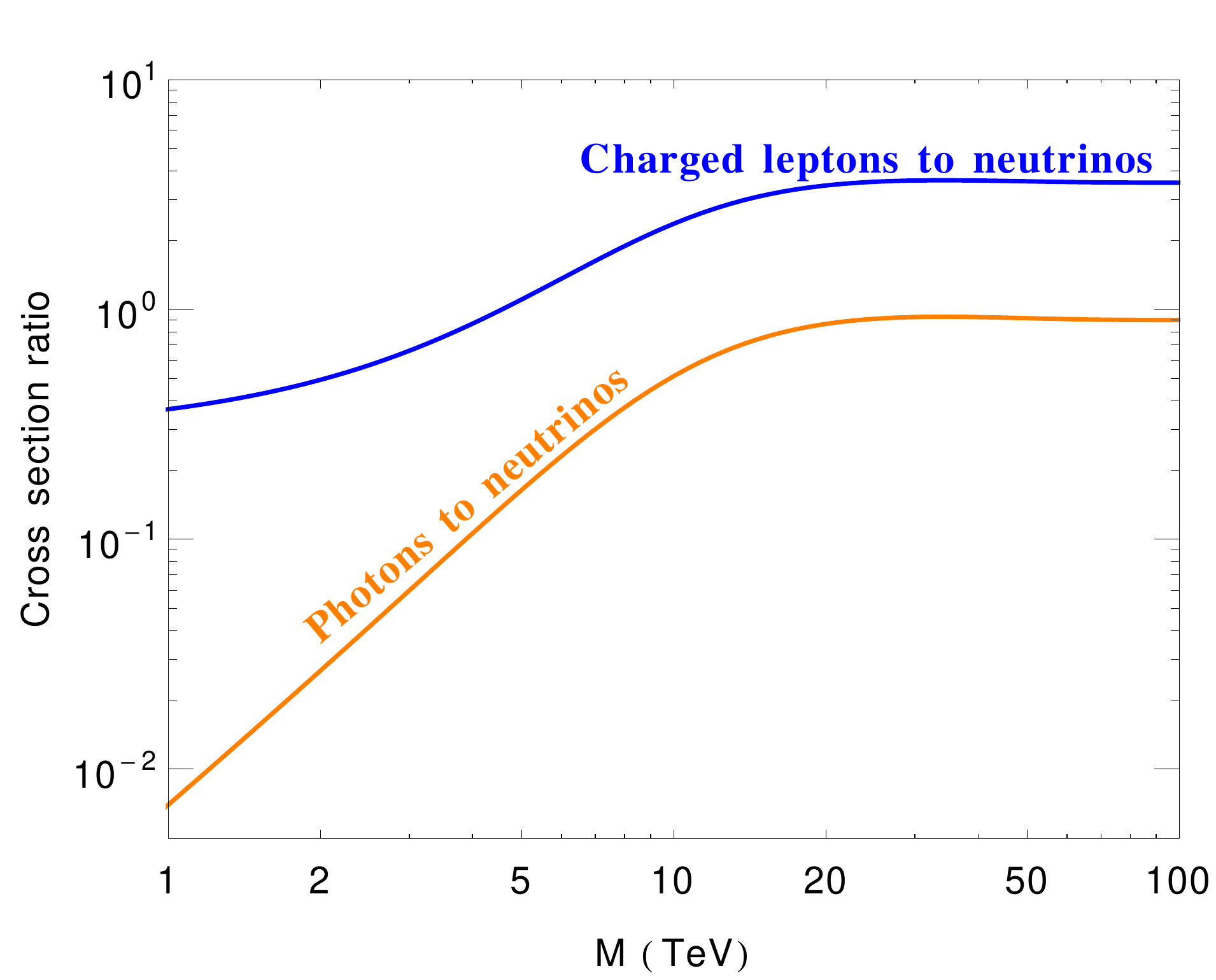}
\caption{Ratio of the indicated cross sections  for a relative velocity of $v\approx 2\times 10^{-3} c$ under  the assumption that ${M_{H^\pm}}-M={M_{A^0}}-M= \unit[1]{GeV}$ for the particles in the doublet. 
}
\label{fig:IBratio}
\end{figure*} 

This scenario naturally gives neutrino lines at tree level, $H^0 H^0 \to \nu \overline{\nu}$, and does not produce $\ell^-\ell^+$ final states. Nevertheless, as explained at length in Sec.~\ref{sec:classification}, the corresponding cross section has no s-wave piece. In fact, it first happens via its d-wave\footnote{Two identical scalars can not be arranged in the $L=1$ configuration that  is antisymmetric. See Ref.~\cite{Giacchino:2013bta} for a related discussion.}. As expected, the emission of a gauge boson removes such a suppression. We find the following differential s-wave cross sections
\begin{eqnarray}
\frac{1}{\sigma v_{f_1 \overline{f_2} B}} \frac{d\sigma v_{f_1\overline{f_2} B}}{dx}&=&\frac{9}{2} \left( 8x -7 x^2+4 (x-2) (x-1) \log (1-x)\right)\,,\label{eq:bremsstrahlung_fermions}\\
\frac{1}{\sigma v_{f_1 \overline{f_2} B}} \frac{d\sigma v_{f_1\overline{f_2} B}}{dx'}&=&6 x'(1-x')\,,
\label{eq:bremsstrahlung_bosons}
\end{eqnarray}
with $x=E_{f_1}/M$ and $x'=E_{B}/M$. In the final state ${f_1 \overline{f_2} B}$,  $B$ is $\gamma, W$ or $Z$  and $f_i$ is a lepton. The corresponding spectra are shown in the right panel of Fig.~\ref{fig:dNdx}. Interestingly, one observes that, even though both spectra exhibit a spectral feature, those of the leptons are far more prominent. 

For the total cross sections,  we find:

\begin{align}
\sigma v_{\nu_\alpha \overline{\nu}_\beta Z}= \sigma v_0\,\,\big|d_1+d_2\big|^2\,,&&
\sigma v_{\nu_\alpha \overline{\ell}_\beta W}+
\sigma v_{\overline{\nu_\alpha }\ell_\beta W}
 =& c_W^2 \,\sigma v_0\,\,|d_1+d_2+\sqrt{2}d_3|^2\nonumber\\
\sigma v_{\ell_\alpha \overline{\ell}_\beta Z}=2c_{2W}^2 \,\sigma v_0\,\,|d_3|^2\,,&&
\sigma v_{\ell_\alpha \overline{\ell}_\beta \gamma}=& 2 s_{2W}^2\,\sigma v_0\,\,|d_3|^2 \,,&
\label{eq:IBcross}
\end{align}
where we define
\begin{equation}
\sigma v_0 \equiv \frac{|y_\alpha|^2|y_\beta|^2  \alpha_2 }{288 \pi^2 c_W^2  m_\text{DM}^2 r_D^4}\,,
\end{equation}
and where $d_1$,  $d_2$ and $d_3$ are the Sommerfeld factors introduced in Appendix~\ref{sec:AppSEIDM}. 
If  the Sommerfeld effect is negligible, $d_3\simeq0$, and  $\sigma v_{\ell_\alpha \overline{\ell}_\beta Z} \simeq0$ as expected from the discussion above. Nevertheless, in the presence of the Sommerfeld effect, we can not ignore $\sigma v_{\ell_\alpha \overline{\ell}_\beta Z}$. The reason for this is simple. The Sommerfeld enhancement, which encodes the non-perturbative effects associated to the exchange of $W^\pm$ boson, creates transitions from $H^0H^0$ to $H^+H^-$, and the latter do annihilate into  ${\ell_\alpha \overline{\ell}_\beta Z}$.

For the particular case when ${m_{H^\pm}}-m_{H^0}={m_{A^0}}-m_{H^0}= \unit[1]{GeV}$, Fig.~\ref{fig:IBratio} shows the ratio of the cross section of all channels producing charged leptons to the corresponding quantity for neutrinos (blue). From the figure, we conclude that the observation of neutrino spectral features also implies  the production of charged leptons with a hard spectrum, specially well above the electroweak scale.   In the same figure, we show the ratio of the cross section $\sigma v_{\ell_\alpha \overline{\ell}_\beta \gamma}$ to that of neutrinos (orange).  Again, we conclude that the neutrino feature appears simultaneously with  a softer photon feature, whose rate is approximately the same.  Moreover, 
in the electroweak symmetric limit, that is for $m_\text{DM} \gg v_\text{EW}$, regardless of the mass splittings we find that $d_2\to -d_1$ and $d_3\to \sqrt{2} d_1$. According to Eqs.~\eqref{eq:IBcross}, this means that  the ratio of  photons to neutrinos is $4 s_W^2\approx 0.9$, while the ratio of charged leptons to neutrinos is  $1+2/c_W^2\approx 3.5$. These values are in perfect agreement with the asymptotic behavior observed in Fig.~\ref{fig:IBratio}.

In summary, the emission of gauge bosons removes the d-wave suppression of the annihilations into $\nu \bar{\nu}$. Also, even though we choose a DM representation in which the annihilation into charged leptons and one boson is absent at tree level, such channel is  present after accounting for the Sommerfeld effect and makes the limits on charged leptons relevant and competitive to those of the neutrino lines.

\begin{figure*}[t]
\includegraphics[width=0.65\textwidth]{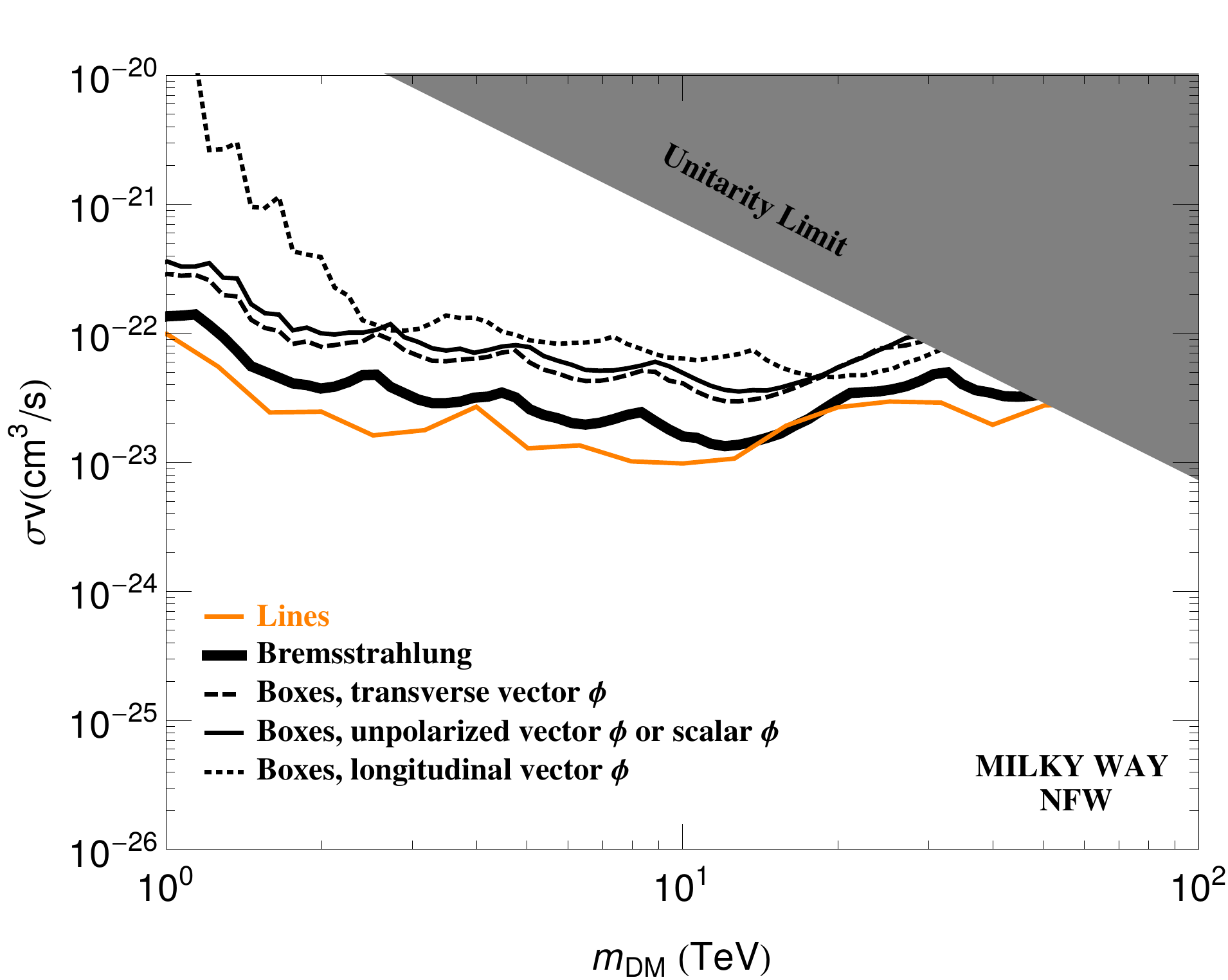}
\caption{Limits on the DM annihilation cross section for neutrino box-shaped spectra (for three different types of intermediary particles, see Eq.~\eqref{eq:dNdxboxes}) and for  the neutrino bremsstrahlung spectrum of Eq.~\eqref{eq:bremsstrahlung_fermions}. For comparison we also show the limits on neutrino lines from Fig~\ref{fig:dNdxline}.
}
\label{fig:IB-box-limits}
\end{figure*}

\subsection{IceCube limits on box-shaped and bremsstrahlung spectra}

The IceCube and Antares collaborations have not provided any limits on box-shaped or bremsstrahlung spectra from DM  decays or annihilations.
For decaying DM, this has been derived in Ref.~\cite{Aisati:2015vma}. From the very same sample and methodology as that considered in this reference (which have also been used  to get the neutrino-line limit of Fig.~\ref{fig:dNdxline}), we derive a limit on the neutrino production cross section for  box-shaped and bremsstrahlung spectra. The results are given in Fig.~\ref{fig:IB-box-limits} for the box-shaped and bremsstrahlung spectra under consideration in this section, taking into account all the possible neutrino flavors in the final state.
As expected, the limits are slightly below those applying to a line. This is because the spectral features considered here are less sharp.

\section{Conclusions}

\label{sec:conclusions}

The sensitivities to monochromatic neutrinos are expected to improve in the near future. 
We illustrated this by deriving, from a public IceCube data sample, a new limit on the  production of such signal from DM annihilations in our galaxy (see Fig.~\ref{fig:dNdxline}).

Motivated by this fact, in this work we have systematically studied  the possibility of observing a neutrino line from DM annihilations. In Section~\ref{sec:classification},  as a first step towards this goal,
  we have listed the set of models that give rise to  unsuppressed rates into monochromatic neutrinos under certain criteria of simplicity. 
This list is given in Table~\ref{table:models}. We have found that neutrino-mass constraints are extremely important in this classification. 
In fact, only eight models satisfy all these criteria. Models $F_{1-4}$
do it  
without any problem. They all have the DM in a hyperchargeless Dirac representation (singlet or triplet).
Two other models, $\SSS_1$ and $\FF_1$, where DM consists of a real scalar or Majorana fermion sitting in a doublet, pass all the constraints at the price of fine-tuned cancellations occurring in the neutrino masses. The last two models, $S_1$ and $S_2$, where DM consists of a complex scalar out of a hyperchargeless multiplet (singlet or triplet), lead to unsuppressed cross sections into neutrinos for  DM masses below 1 TeV. However, $S_2$ is excluded from the fact that a DM triplet below the electroweak scale produces too many monochromatic photons. 
All these models conserve lepton number except  $\SSS_1$ and $\FF_1$.

In general, we expect the viable models in Sec.~\ref{sec:classification} to produce lines with a cross section of the order of $\unit[10^{-26}]{cm^3/s}$ because we assume that DM is produced via the standard freeze-out mechanism.
Since neutrino telescopes are not expected to reach such a sensitivity in the near future,  as a second step in our analysis, we have studied in detail the possibility of having a boost in the annihilation cross section from the Sommerfeld effect. 

Our findings in Sections~\ref{sec:SE} and~\ref{sec:SEnoEW} suggest that neutrino lines can be observed in the near future, with cross sections of the order of $\unit[10^{-24}]{cm^3/s}$. Four models---namely, $F_{1}$, $F_{2}$, $S_1^r$ and $F_1^m$---undergo large electroweak Sommerfeld enhancement for DM masses above 2 TeV.  $F_1$ and $F_2$ lead to cross sections which  are already excluded by neutrino telescopes around the Sommerfeld peaks. Elsewhere (and above $\sim 3$~TeV), the predicted fluxes are reachable by neutrino telescopes. This is also true for models $S_1^r$ and $F_1^m$ .

The other three models, $F_{3}$, $F_{4}$ and $S_1$ do not have any electroweak Sommerfeld effect,  either because they involve a DM singlet or because they are only viable below the TeV scale.  As discussed in Section~\ref{sec:SEnoEW}, they can nevertheless take advantage of another source of Sommerfeld enhancement, if they couple to an extra light mediator. For $F_{3}$ and $F_{4}$, depending on the mediator mass and on the strength of the mediator interaction with DM, an observable signal is possible for any value of the DM mass (see Fig.~\ref{fig:nuFluxF3F4}), thanks to the fact that the DM annihilation into the mediator is p-wave suppressed (unlike for model $S_1$, see Fig.~\ref{fig:nufluxS1}).

Nevertheless, neutrino lines do not appear alone. They are accompanied with cosmic-ray signals at a similar production rate. These cosmic rays are mostly charged leptons (see Figs.~\ref{fig:nuFluxF2} and \ref{fig:nuFluxF3F4}). A continuum of  gamma rays is also produced from other annihilation channels (see Fig.~\ref{fig:nuFluxSm7Fm7}). Therefore, charged-lepton and gamma-ray limits  on DM annihilations indirectly constrain neutrino lines. For instance,  models $F_{1-4}$ predict an equal flux of charged leptons and neutrinos. Thus, the annihilation cross sections into neutrinos must be below the existing upper bounds  on the corresponding cross section for charged leptons.  In fact, for  $F_{1}$ and $F_2$, 
the possibility of observing  a neutrino line could be excluded by a slight improvement of the charged lepton bounds. Alternatively, a neutrino line can potentially be discovered if neutrino telescopes reach the charged-lepton sensitivity. This seems feasible for DM masses above few TeV. {If a neutrino line is observed {there, the}} unavoidable {corresponding} flux of charged leptons must {also} be observed.
A similar conclusion arises for models $F_{3}$ and $F_{4}$. In contrast,  models $S_1^r$ and $F_1^m$ as well as the low-scale scenario  $S_{1}$  do not lead to a flux of charged leptons equal to that of neutrinos, and are therefore less constrained.

From the results obtained, it turns out that only very few models with specific mass and
interactions could lead to an observable neutrino line from DM annihilations (unlike for decaying DM~\cite{Aisati:2015ova, Aisati:2015vma}). In this sense, one could argue that we would be lucky to observe 
such a signal in the near future. On the other hand, as there are very few models and since they lead to different annihilation cross sections into neutrinos and cosmic rays, the observation of a
neutrino line would point out towards DM candidates with  specific properties.

Complementary information to discriminate the origin of the neutrino line is provided by the flavor composition  of the neutrino flux. 
The models studied in this work produce lines of three different types: flavor democratic (associated to a $Z'$ boson in the s-channel), flavor eigenstates (such as those associated to the t-channel exchange of a mediator) and mass eigenstates (from the s-channel exchange of a scalar with $Y=2$). In particular, in the latter case and 
for a normal hierarchy, neutrino lines may possess  a unique flavor composition that astrophysical neutrinos cannot.  This is illustrated in Fig.~\ref{fig:triangle}. 

Even though our conclusions are fairly general, they are still model dependent because they are based on a set of simple assumptions.
In Section~\ref{sec:caveats}, we have gone beyond our minimal assumptions and discussed more general cases. 
 Finally, we have also considered  line-like features in Section~\ref{sec:beyondlines}. In particular, we have discussed bremsstrahlung and boxed-shaped features, which mimic neutrino lines, given the current sensitivity of neutrino telescopes. In Fig.~\ref{fig:IB-box-limits}, we have presented limits on annihilation processes with these spectra  using IceCube data.

As a closing remark, we would like to encourage experimentalists to study the  multi-TeV region. There is no reason not to consider DM annihilations above 10 TeV,  although it has often been the case. In fact, as far as annihilating DM is concerned, it is in this region that simple models leading to  neutrino lines can be probed best with neutrino telescopes. Furthermore,  for such masses colliders can not probe the models presented in this work. For instance, for those where DM belongs to a  triplet, the most important bound is currently given by ATLAS, which only excludes DM masses below $\unit[0.27]{TeV}$~\cite{Aad:2013yna}.

\section*{Acknowledgments}
The authors would like to thank Michael Gustafsson, Julian Heeck and Sergio Palomares-Ruiz  for  useful discussions. In particular, we are indebted to Michael Gustafsson for cross-checking our  derivation of the limit presented in Section \ref{sec:limits}. This work is supported by the FNRS, the ``Probing dark matter with neutrinos" ULB-ARC grant, the FRIA and the Belgian Federal Science Policy through the Interuniversity Attraction Pole P7/37. 

\newpage

\appendix

\section{Sommerfeld Enhancement for Dirac triplets}
\label{sec:AppSE}

\vspace{0.3cm}
\begin{centering}
\noindent{\small\textbf{Annihilations in DM halos}}\\
\end{centering}

\vspace{0.6cm}

The Sommerfeld enhancement  in DM annihilations  is a non-perturbative effect arising when the DM particles are non-relativistic and are able to exchange lighter particles that induce a long range force that modifies their wave function, and therefore their annihilation cross sections. In this appendix, we discuss such effect in detail for Dirac triplets at the TeV scale. In that case, DM belongs to the $SU(2)_L$ triplet representation  $\psi= (\psi_1^+, \psi^0,  \psi_2^-)^T $, more precisely, $\DM = \psi^0$. The exchange of $W^\pm$ bosons converts the pairs $\psi^0 \overline{\psi^0}$ into $\psi_1^+\psi_1^-$ or  $\psi_2^+\psi_2^-$. Consequently, the potential $V(r)$ describing the long-range force is a $3\times3$ matrix in the basis  $(\psi^+_1\psi^-_1, \psi^0\overline{\psi^0}, \psi^+_2\psi^-_2 )^T$ given by
\begin{align}
V(r)=
-\frac{\alpha_2}{r}
\begin{psmallmatrix}
 s_W^2+c_W^2 e^{-m_Z r} & e^{-m_W r} & 0 \\
 e^{-m_W r} & 0 & e^{-m_W r} \\
 0 & e^{-m_W r} &s_W^2+ c_W^2 e^{-m_Z r} \\
\end{psmallmatrix}\,.
\label{eq:V}
\end{align}
with $\alpha_2 = \alpha/s_W^2$.
This potential is valid for any spin configuration of the annihilating DM pair, that is, for both $S=0$ and $S=1$. This is in sharp contrast to self-conjugate triplets, also called Winos, because in that case the state symmetrization induces different potentials for $S=0$ and $S=1$~\cite{Hisano:2006nn}.

The Sommerfeld effect is encoded in a matrix $g(r)$ that satisfies the Schr\"odinger equation~\cite{Hisano:2004ds}
\begin{align}
-\frac{1}{m_\DM} g''(r) +\left(-\dfrac{1}{4} m_\DM v^2 {1\!\!1} +2 \delta m + V(r) \right) g(r) =0 \,,
\label{eq:SoDE}
\end{align}
where $v$ is the relative velocity between DM particles and $m_\DM$ is their mass. Moreover, $\delta m $ is a diagonal matrix whose entries are the mass splittings between the charged particles and the DM. For Dirac triplets, the splittings between the different components appear radiatively after electroweak symmetry breaking. For TeV DM~\cite{Cirelli:2005uq} 
$m_{\psi_{1,2}}-m_\DM= \alpha_2 m_W (1-c_W)/2 \simeq  \unit[166]{MeV}$. Therefore, $\delta m=( \unit[166]{MeV})\,\text{diag}(1,0,1)$.

For $M\lesssim \unit[100]{TeV}$ and typical relative velocities in DM halos, the kinetic energy of the DM is smaller than the mass splitting between pairs, which implies that only the pair $\psi^0\bar{\psi^0}$  can exist  at large distances in the scattering process. Thus, one of the boundary conditions of Eq.~(\ref{eq:SoDE}) is that
\begin{align}
g(r) \longrightarrow  e^{ i
m_\DM vr /2 }\begin{psmallmatrix}
  0   & 0 & 0 \\
  d_+ & d_0 & d_- \\
 0 & 0 & 0 
 \end{psmallmatrix}\text{ for }r\to \infty\,.
\end{align}
The other boundary condition is $g(0)={1\!\!1}$. A rigorous derivation of these boundary conditions is given in Ref.~\cite{Hisano:2004ds}.
Here $d \equiv (d_+, d_0, d_-)$ is the only non-zero row of the matrix $g(r)$ after factorizing out the out-going wave describing the DM pairs at infinity. Notice that the potential in Eq.~\eqref{eq:V} gives rise to $d_+ = d_-$. In terms of $d$ factors, 
the s-wave cross section for a pair of DM particles in the configuration with total spin S annihilating into a two-body final state $ab$ reads
\begin{equation}
\sigma v^S \left(\psi^0 \overline{\psi^0}\to a b\right)  =  \,d^*\, \Gamma^S_{ab}\, d^T\,,
\label{eq:SEsigmav}
\end{equation}
where $\Gamma_{ab}$ are the annihilation matrices. Electroweak interactions alone induce  annihilations into SM particles, whose corresponding annihilation matrices are  given by 
\begin{eqnarray}
\label{eq:GammaGauge}
\Gamma^{S=0}_{WW}=  \frac {\alpha_2^2 \pi }{2 m_\DM^2} 
\begin{psmallmatrix}
  1   & 2 & 1 \\
 2 & 4 & 2 \\
 1 & 2 & 1
\end{psmallmatrix}
\,, & &
\Gamma^{S=0}_{\gamma\gamma}= \tfrac{s_W^2}{2 c_W^2} \Gamma^{S=0}_{\gamma Z} =  \tfrac{s_W^4}{ c_W^4}  \Gamma^{S=0}_{ZZ} =  \frac {\alpha_2^2 s_W^4 \pi }{m_\DM^2} 
\begin{psmallmatrix}
  1   & 0 & 1\\
  0 & 0 &0\\
 1 & 0 & 1
\end{psmallmatrix}
\,,\\
\label{eq:GammaGaugev2}
\Gamma^{S=1}_{WW}= \Gamma^{S=1}_{Zh}  = \frac {\alpha_2^2 \pi }{ 48 m_\DM^2} 
\begin{psmallmatrix}
 1 & 0 & -1 \\
 0 & 0 & 0 \\
 -1 & 0 & 1
\end{psmallmatrix}
,&&
\Gamma^{S=1}_{f_{\alpha}\overline{f_\beta}} = 2 N_f\, \Gamma^{S=1}_{WW}\delta_{\alpha \beta}\,. 
\label{eq:Gammasff}
\end{eqnarray}
The annihilation matrices for each spin due to electroweak interactions  are then
\begin{eqnarray}
\Gamma^{S=0}_{\text{EW}}=  \frac {\alpha_2^2 \pi }{2 m_\DM^2} 
\begin{psmallmatrix}
  3   & 2 & 3 \\
 2 & 4 & 2 \\
 3 & 2 & 3
\end{psmallmatrix}
\,, & &
\Gamma^{S=1}_{\text{EW}}= \frac {25\alpha_2^2 \pi }{24 m_\DM^2} 
\begin{psmallmatrix}
  1   & 0 & 1\\
  0 & 0 &0\\
 1 & 0 & 1
\end{psmallmatrix}
\,.
\label{eq:Gammatot}
\end{eqnarray}

\vspace{0.3cm}
\begin{centering}
\noindent{\small\textbf{Relic density in the electroweak symmetric phase}}\\
\end{centering}

\vspace{0.6cm}
Calculating the relic density  also  requires to take into account the Sommerfeld effect in the early universe. This was first shown  for supersymmetric triplets in Ref.~\cite{Hisano:2006nn} (see also Refs.~\cite{Cirelli:2007xd,Beneke:2016ync,Beneke:2014hja,Beneke:2014gja,Beneke:2012tg,Hellmann:2013jxa}). A priori, such an effect can play an important role for very heavy DM candidates, well above the electroweak scale. In such case, it is justified to work in the electroweak symmetric phase\cite{Cirelli:2007xd, Cirelli:2009uv}, in which $v_\text{EW}$ is set to zero and therefore there is no mass splitting in Eq.~\eqref{eq:SoDE}, nor gauge boson masses in Eq.~\eqref{eq:V}.  In fact, in that limit, Eqs.\eqref{eq:V},\eqref{eq:GammaGauge} and \eqref{eq:GammaGaugev2} can be simultaneously diagonalized by the matrix 
\begin{equation}
U=
\begin{psmallmatrix}
  \frac{1}{\sqrt{3}}  &  \frac{1}{\sqrt{3}}& \frac{1}{\sqrt{3}} \\
 -\frac{1}{\sqrt{2}} & 0 & \frac{1}{\sqrt{2}} \\
 \frac{1}{\sqrt{6}} & -\sqrt{\frac{2}{3}} & \frac{1}{\sqrt{6}}
\end{psmallmatrix}\,.
\end{equation}
Concretely,
\begin{align}
U V(r) U^T = \frac{\alpha_2}{r}\begin{psmallmatrix}
  -2 &  0&0\\
 0  & -1 & 0 \\
 0& 0& 1
\end{psmallmatrix}\,, &&
U \Gamma^{S=0}_{\text{EW}}  U^T = \frac{\pi\alpha_2^2}{m_\DM^2}\begin{psmallmatrix}
 4 &  0&0\\
 0  & 0 & 0 \\
 0& 0& 1
\end{psmallmatrix}\,,&&
U \Gamma^{S=1}_{\text{EW}}  U^T = \frac{\pi\alpha_2^2}{m_\DM^2}\begin{psmallmatrix}
 0 &  0&0\\
 0  & \frac{25}{12} & 0 \\
 0& 0& 0
\end{psmallmatrix}\,.
\label{eq:MatSym}
\end{align}
The matrix $U$ describes the change of basis from annihilating pairs with definite electric charge into the states of total isospin $I$. In fact, 
\begin{equation}
\begin{pmatrix}
 |m_I=0, I=0\rangle\\
 |m_I=0, I=1\rangle\\
 |m_I=0, I=2\rangle
\end{pmatrix}
=
U\begin{pmatrix}
 \psi^+_1\psi^-_1\\
 \psi^0\overline{\psi^0}\\
 \psi^+_2\psi^-_2
\end{pmatrix}
\label{eq:rotQ0}\,.
\end{equation}

As a consequence of Eqs.~\eqref{eq:MatSym}, the Schr\"odinger  Eq.~\eqref{eq:SoDE} in the isospin basis reduces to one-dimensional equations with a Coulomb potential. The Sommerfeld effect is known analytically for those cases and is given by
\begin{equation}
\sigma v_{I,S}= {\cal S} (2 \pi \alpha_I/v)\Gamma_{I,S}\,, \hspace{10pt}\text{with}\hspace{10pt} {\cal S}(t) = t/(e^t-1)\,.
\end{equation}
Here, $\Gamma_{I,S}$ and $\alpha_I/r$  are respectively defined as the eigenvalues of the annihilation and potential matrices for a given total spin $S$ and isospin $I$. For instance, if only electroweak interactions are present, according to Eqs.~\eqref{eq:MatSym}, $\alpha_{I=0} =-2 \alpha_2$ $\alpha_{I=1}=-\alpha_2$ and $\alpha_{I=2}=\alpha_2$. Similarly, $\Gamma_{I=0,S=0} =4\pi \alpha_2^2/m_\DM^2$, $\Gamma_{I=1,S=1} =25\pi \alpha_2^2/12m_\DM^2$ and $\Gamma_{I=2,S=0} =\pi \alpha_2^2/m_\DM^2$. Notice that they must not depend on $m_I$ because of isospin invariance. 
These definitions are particularly useful in the early universe, where the effective cross section that takes into account co-annihilations simply reads
\begin{equation}
\sigma v_\text{eff} = \frac{1}{9}\sum_{I,S}  (2I+1) (2S+1)\sigma v_{I,S}\,.
\end{equation} 
The factor of $9=3\times 3 =\sum I = 1+3+5$ in the denominator is the total number of co-annihilating pairs.  We will now consider the thermal average of $\sigma v_\text{eff}$, which is given by 
\begin{equation}
\langle\sigma v_\text{eff}\rangle =\frac{\int^\infty_0 dv v^2 e^{-\frac{Mv^2}{4T}} \sigma v_\text{eff}}{\int^\infty_0 dv v^2 e^{-\frac{Mv^2}{4T}}}\,.
\end{equation}

The thermal average for $S(2\pi \alpha_I/v)$ can be calculated analytically by expanding in $\alpha_I$. We find 
\begin{equation}
\langle\sigma v_\text{eff}\rangle \approx \frac{1}{9}\sum_{I,S}  (2I+1) (2S+1) \left(1-\frac{1}{2} \sqrt{\pi x}\, \alpha_I\right)\Gamma_{I,S}\,,
\label{eq:SEGamma}
\end{equation}
where $x = m_\DM/T$.  With this, we can now consider 
the Boltzmann equation describing the abundance of DM particles that are \emph{not} self-conjugate~\cite{Gondolo:1990dk}
\begin{equation}
\frac{dY}{dx} =- \frac{1}{2} \frac{s\langle \sigma v_\text{eff}\rangle }{H\,x} (Y^2 -Y^2_\text{eq})\,,
\label{eq:BE}
\end{equation}
where $Y=n_\DM/s$ is the yield associated to  the total number of DM particles and anti-particles which are  assumed to be present in equal amounts. Notice the factor of $1/2$ which is not present for the case of real scalar or Majorana DM. In this work we solve this equation by means of the freeze-out approximation, which yields 

\begin{align}
\Omega h^2 \approx  \frac{\unit[1.03\times 10^9]{GeV^{-1}}}{g_\star^{1/2} M_\mathrm{Pl} } \left(\int^\infty_{x_f} \frac{\langle \sigma v \rangle_\mathrm{eff}}{2x^2} \dd x \right)^{-1}  = 0.11 \left(\frac{x_f}{27}\right)
\frac{  (\unit[4.7\times 10^{-26}]{cm^3/s})}{x_f\int^\infty_{x_f} x^{-2}\langle \sigma v \rangle_\mathrm{eff}\dd x}.
\label{eq:Omega}
\end{align}
We take the number of relativistic degrees of freedom, $g_\star \approx 106$, because we are focusing on multi-TeV  DM candidates that freeze out at $x_f\approx 27$, that is to say with a freezout temperature above $\sim$100~GeV.
If DM annihilates with a constant s-wave effective cross section, according to Eq.~\eqref{eq:Omega}, the observed DM abundance of $\Omega h^2 \approx 0.11$ is obtained for $x\approx27$ and $\langle\sigma v_\text{eff}\rangle \approx\unit[4.7\times 10^{-26}]{cm^3/s} $. This is in agreement with Ref.~\cite{Steigman:2012nb}. In the case of the Dirac triplets, the cross section is not exactly constant because of  the Sommerfeld effect, as given in Eq.~\eqref{eq:SEGamma}. In that case, the relic density constraint is
\begin{equation}
\frac{1}{9}\sum_{I,S}  (2I+1) (2S+1) \left(1- \sqrt{\pi x}\, \alpha_I\right)\Gamma_{I,S}\approx \left(\frac{x_f}{27}\right) (\unit[4.7\times 10^{-26}]{cm^3/s})\,.
\label{eq:relic0}
\end{equation}
If only electroweak interactions are present, the observed  relic density is obtained for 
\begin{equation}
\frac{\pi \alpha_2^2}{12\,m_\DM^2}( 37+29 \sqrt{\pi x} \alpha_2) =  \left(\frac{x_f}{27}\right) \left(\unit[4.7\times 10^{-26}]{cm^3/s}\right)\,,
\label{eq:relic}
\end{equation}
which gives $m_\DM\approx\unit[2]{TeV}\approx \unit[2.8/\sqrt{2}] {TeV}$.  This must be compared with Wino DM (a Majorana triplet at 2.8 TeV) when the Sommerfeld effect is taken into account in the early universe~\cite{Hisano:2006nn}. 

As a further example, we now consider model $F_2$ in which not only electroweak interactions contribute to the annihilation matrices but also those induced by Eq.~\eqref{eq:LF2}. The total annihilation matrix into leptons $\sum_{\alpha \beta}(\Gamma^{S=1}_{\nu_\alpha\bar{\nu}_\beta}+\Gamma^{S=1}_{\ell_\alpha\bar{\ell}_\beta})$  (see Eq.~\eqref{eq:GammaF2Gaugev2}) can be brought to the weak isospin eigenvector basis by means of Eq.~\eqref{eq:rotQ0}, which gives
\begin{align}
\Gamma_{I=0,S=1}=\frac{9 \overline{y}^4 m_\DM^2}{16 \pi (m_\DM^2+m_D^2)^2}\,,&&
\Gamma_{I=1,S=1}= \frac{3\overline{y}^4 m_\DM^2}{8 \pi (m_\DM^2+m_D^2)^2} + \frac{ \alpha_2  \overline{y}^2}{2 (m_\DM^2+m_D^2)} \nonumber\,.
\end{align}
 
If we neglect the last term in $\Gamma_{I=1,S=1}$,  corresponding to the interference and generally subdominant, we obtain $ \Gamma_{I=0,S=1}=2\sigma v_0 $ and $\Gamma_{I=1,S=1}=4\sigma v_0 /3$, where $\sigma v_0$ was defined in Eq.~\eqref{eq:BrF2}. This allows to obtain the formula

\begin{eqnarray}
\sigma v_0\bigg|_{F_2}
\approx \dfrac{3}{6+8\sqrt{\pi x_f}\alpha_2}\left(\frac{x_f}{27} \left(\unit[4.7\times 10^{-26}]{\frac{cm^3}{s}}\right)-\frac{\pi \alpha_2^2 ( 37+29 \sqrt{\pi x_f} \alpha_2)}{12\,m_\DM^2}\right).
\label{eq:relicF2}
\end{eqnarray}

In spite of  the many details that go into this formula, the physical picture is clear. The cross section $\sigma v_0$ is given, up to an overall factor related to co-annihilations, by the typical effective cross sections at freeze-out for Dirac DM~\cite{Steigman:2012nb}   minus the DM annihilation cross section into other states induced by the electroweak interactions of the triplet.

Similarly, for $F_1$ we find 

\begin{eqnarray}
\sigma v_0 \bigg|_{F_1} \approx \frac{3}{2+4\sqrt{\pi x_f}\alpha_2}\left( \frac{x_f}{27} \left(\unit[4.7\times 10^{-26}]{\frac{cm^3}{s}}\right)-\frac{\pi \alpha_2^2( 37+29 \sqrt{\pi x_f} \alpha_2)}{12\,m_\DM^2}\right)\,.
\label{eq:relicF1}
\end{eqnarray}

\section{Sommerfeld Enhancement for scalar doublets}
\label{sec:AppSEIDM}

In this Appendix, we review the properties of scalar doublets, focusing on the TeV scale where the Sommerfeld effect is relevant.

Consider a scalar doublet with hypercharge $Y=1$ and odd parity under an unbroken $Z_2$ symmetry. 
Its Lagrangian is 
\begin{eqnarray}
 {\cal L}_{\phi_D}&& = 
  (D_\mu \phi_D)^\dagger  (D^\mu \phi_D)-m_D^2\phi_D^\dagger \phi_D\nonumber - \lambda_2(\phi_D^\dagger \phi_D)^2- \lambda_3(H^\dagger H)(\phi_D^\dagger \phi_D) \label{eq:L}\\
&& - \lambda_4(H^\dagger \phi_D)(\phi_D^\dagger H) 
-\dfrac{1}{2} \left( \lambda_5(H^\dagger \phi_D)(H^\dagger \phi_D)+{\rm \text{\small h.c.}}\right) \;,
\label{eq:IDML}
\end{eqnarray}
where  $H$ is the SM doublet, and $\phi_D$ is the inert doublet which can be written as
\begin{align}
\phi_D= \begin{pmatrix} H^+ \\ \frac{H^0+ i A^0}{\sqrt2} \end{pmatrix}\;,
\label{eq:fieldcompIDM}
\end{align}
where the  scalar components are two charged states $H^\pm$, one CP-neutral state $H^0$ and one CP-odd neutral state $A^0$. After electroweak symmetry breaking, the masses of these particles are split according to 
\begin{align}
m^2_{H^+} &= m^2_{H^0} -\frac{1}{2}(\lambda_4+\lambda_5)v_\text{EW}^2\,,& 
m^2_{A^0} &= m^2_{H^0} -\lambda_5 v_\text{EW}^2 \;.&
\label{eq:IDMmasses}
\end{align}
Without loss of generality, we can assume that $m_{A^0} >m_{H^0}$ and, in order to have a neutral DM candidate, we can assume $m_{H^+} > m_{H^0}$. In this case $H^0$ is a viable DM candidate~\cite{Deshpande:1977rw,Barbieri:2006dq,Ma:2006km,LopezHonorez:2006gr,Cirelli:2007xd,Hambye:2009pw}. It couples to the $Z$ boson because  it belongs to a multiplet with non-zero hypercharge. Concretely,  ${\cal L}_{\phi_D} \supset  (g/2c_W)  (H^0 Z \cdot\partial A^0 -A^0 Z \cdot\partial H^0 )$. As discussed at length in Sec.~\ref{sec:classification}, unless $A^0$ is sufficiently heavier than $H^0$, the exchange of a $Z$ boson leads to direct-detection cross sections much larger than currently allowed by experiments. According to Eq.~\eqref{eq:IDMmasses}, it is therefore necessary for $\lambda_5$ to be different from zero. More precisely, since $m_{A_0}>m_{H_0}$, one has
\begin{equation}
-\lambda_5 > 2\left(\frac{ m_{H^0}}{v_\text{EW}}\right)\left( \frac{m_{A^0}-m_{H^0}}{v_\text{EW}}\right) \approx 3 \cdot 10^{-6} \left( \frac{m_{H^0}}{\unit[1]{TeV}}  \right) \left(\frac{m_{A^0}-m_{H^0}}{\unit[100]{keV}}\right)\,.
\label{eq:DD_IDM}
\end{equation}
The Sommerfeld effect is relevant for doublets at the TeV scale (see Ref.~\cite{Garcia-Cely:2015khw}). Here we just summarized the main aspects. The discussion is similar to the one given for Dirac triplets in the previous appendix, with two  differences. First, the Sommerfeld effect is very small in the early universe because of the doublet nature of the DM candidate~\cite{Garcia-Cely:2015khw, Hisano:2006nn}. This allows us to focus on the effect in DM halos such as the Milky Way. Second, in this case the DM is  self-conjugate, and Eq.~\eqref{eq:SEsigmav} that relates the annihilation matrices and the cross sections gets an extra factor of two
\begin{equation}
\sigma v \left(H^0 H^0\to a b\right)  = 2 \,d^*\, \Gamma_{ab}\, d^T\,.
\label{eq:SEsigmavIDM}
\end{equation}
Here, $d\equiv (d_1,d_2,d_3)$ are the Sommerfeld factors obtained by solving Eq.~\eqref{eq:SoDE} with $m_\DM=m_{H^0}$ and the boundary conditions
\begin{align}
g(r) \longrightarrow  e^{ i
m_\DM vr /2 }\begin{psmallmatrix}
  d_1   & d_2 & d_3 \\
  0 & 0 & 0 \\
 0 & 0 & 0 
 \end{psmallmatrix}\text{ for }r\to \infty\,, && g(0)={1\!\!1}\,.
\end{align}
We are working in the basis $(H^0H^0, A^0A^0,H^+H^-)$, where the potential reads 

\begin{align}
V(r)=
-\frac{\alpha_2}{r}\begin{psmallmatrix}
0 & \frac{  e^{-m_Z r}}{4 c_W^2 }& \frac{  e^{-m_W r}}{2 \sqrt{2} }\\
\frac{ e^{-m_Z r}}{4 c_W^2 }   & 0 & \frac{e^{-m_W r}}{2 \sqrt{2} } \\
\frac{ e^{-m_W r}}{2 \sqrt{2} }  & \frac{ e^{-m_W r}}{2 \sqrt{2} } &  s_W^2 +\frac{(1-2 c_W^2)^2 e^{-m_Z r}}{4 c_W^2 }
\end{psmallmatrix}\,.
\label{eq:V_IDM}
\end{align}
We would like to remark that there is another piece of the potential associated to the exchange of scalar bosons. We neglect such contribution here because it is proportional to $v_\text{EW}^2/m_\DM^2$ and is therefore subdominant for $m_\DM\gg 1 $ TeV.

An interesting aspect of  scalar doublets is that the mass splittings entering in Eq.~\eqref{eq:SoDE} are not fixed and change as a function of the quartic couplings according to Eq.~\eqref{eq:IDMmasses}. In section~\ref{sec:results}, we discuss the implications of that for the position of the Sommerfeld peaks.

In this work, we use Eq.~\eqref{eq:SEsigmavIDM} in order to calculate cross sections . For the case of neutrino final states, we report the relevant annihilation matrices in the text, while for the case of final states with SM bosons ($WW$, $ZZ$, $\gamma Z$, $hh$) we use the expressions reported in Ref.~\cite{Garcia-Cely:2015khw}.

The only model in Table~\ref{table:models} involving doublet scalar DM and passing all the constraints is $\SSS_1$.  As for Dirac triplets, we can work in the symmetric phase in order to calculate the relic density constraint. For doublets we can neglect the Sommerfeld effect~\cite{Hisano:2006nn,Garcia-Cely:2015khw} but not the co-annihilations.  Moreover,  we note that Eq.~\eqref{eq:GammaSm7Fm7} has only two non-vanishing eigenvectors with the same eigenvalue, which correspond to the isospin triplets with hypercharge $Y=\pm2$, as expected for the exchange of $T_2$ on the s-channel. Then, generalizing the notation of Appendix~\ref{sec:AppSE} (or following Ref.~\cite{Garcia-Cely:2015khw}), we have $\Gamma_{I=1,Y=\pm1}=\sigma v_0 $. The relic density constraint of Eq.~\eqref{eq:relic0} is therefore 
\begin{equation}
\frac{2}{4^2}(2\times1+1)(\Gamma_{I=1,Y=-1}+\Gamma_{I=1,Y=+1}) =\unit[2.35\times 10^{-26}]{\frac{cm^3}{s}}\,.
\label{eq:relicSm7Fm7}
\end{equation}
Here, the factor $4^2$ comes from the fact that there are four particles annihilating $H^0, A^0$ and $H^\pm$, and the global factor of 2 is analogous to the one in Eq.~\eqref{eq:SEsigmavIDM}. This implies
\begin{equation}
\sigma v_0\Bigg|_{\SSS_1}\approx\frac{4}{3}\left(\unit[2.35\times 10^{-26}]{cm^3/s}\right)\,.
\end{equation}

\bibliographystyle{utcaps_mod}
\bibliography{BIB}
\end{document}